\documentclass[sn-basic]{sn-jnl}

\usepackage{graphicx}%
\usepackage{multirow}%
\usepackage{amsmath,amssymb,amsfonts}%
\usepackage{amsthm}%
\usepackage{mathrsfs}%
\usepackage[title]{appendix}%
\usepackage{xcolor}%
\usepackage{textcomp}%
\usepackage{manyfoot}%
\usepackage{booktabs}%
\usepackage{algorithm}%
\usepackage{algorithmicx}%
\usepackage{algpseudocode}%
\usepackage{listings}%

\usepackage{stmaryrd}
\usepackage{enumerate}
\usepackage{ulem}
\usepackage{comment}
\usepackage{bold-extra}

\usepackage{xspace}
\newcommand{\vide}{\textsc{VIDE}\normalfont\xspace}
\newcommand{\spark}{\textsc{Sparkling}\normalfont\xspace}
\newcommand{\zobov}{\textsc{ZOBOV}\normalfont\xspace}

\graphicspath{ {./plots/} } 

\usepackage[nameinlink,noabbrev]{cleveref}
\Crefname{equation}{Eq.}{Eqs.}
\Crefname{section}{Sect.}{Sects.}
\Crefname{figure}{Fig.}{Figs.}
\crefname{equation}{Equation}{Equations}
\crefname{section}{Section}{Sections}
\crefname{figure}{Figure}{Figures}
\creflabelformat{equation}{#2#1#3}

\begin{document}

\title[The era of precision cosmology with voids]{The era of precision cosmology with voids}


\author[1]{\fnm{Sofia} \sur{Contarini}}\email{contarini@mpe.mpg.de}
\equalcont{These authors contributed equally to this work.}

\author[2,3]{\fnm{Giovanni} \sur{Verza}}\email{gverza@flatironinstitute.org}
\equalcont{These authors contributed equally to this work.}

\author[4,5]{\fnm{Alice} \sur{Pisani}}\email{pisani@cppm.in2p3.fr}
\equalcont{These authors contributed equally to this work.}

\affil[1]{Max Planck Institute for Extraterrestrial Physics, Giessenbachstrasse 1, 85748 Garching, Germany}

\affil[2]{Center for Computational Astrophysics, Flatiron Institute, 162 5th Avenue, 10010, New York, NY, USA}

\affil[3]{ICTP, International Centre for Theoretical Physics, Strada Costiera 11, 34151, Trieste, Italy}

\affil[4]{Aix-Marseille Universit\'e, CNRS/IN2P3, CPPM, Marseille, France}

\affil[5]{Department of Astrophysical Sciences, Peyton Hall, Princeton University, Princeton, NJ 08544, USA}


\abstract{
Cosmic voids, the large underdense regions of our Universe, have emerged over the past decade as powerful cosmological laboratories: their simple dynamics, sensitivity to local gravitational effects and cosmic expansion, and ability to span large volumes, make them uniquely suited to test fundamental physics. Fueled by advances in theory, simulations, and observations, void science has matured into a precision tool for constraining the parameters of the standard cosmological model and its possible extensions. In this review, we provide a comprehensive description of the statistical tools developed to characterize voids, the theoretical models that link them to cosmological parameters, and the methodologies used to extract information from survey data. We highlight the growing synergy between void-based observables and other cosmological probes, and showcase the increasingly stringent constraints derived from voids measured from current survey data and expected from future missions. With the advent of the next generation of galaxy surveys, voids are poised to play a central role in the future of cosmology, turning what was once regarded as emptiness into one of the most promising frontiers of fundamental science.
}

\keywords{Cosmology, Voids, Large-scale structure, Theory, Surveys}


\maketitle
\tableofcontents
\newpage

\section{Introduction}\label{sec1}

Cosmology is an ambitious science: it aims to understand our Universe as a whole, by investigating its evolution and composition, while observing from our tiny, remote spot in space and time. Modern data already allow us to achieve a good understanding of the Universe's fundamental properties. We know that the Universe is expanding at an accelerated rate, and that the existence of dark energy is postulated to explain such observed acceleration \citep{Riess_1998,Perlmutter_1999}. The nature of dark energy, accounting for $\sim69\%$ of our Universe remains, however, elusive. Alongside with the mystery of dark energy, cosmologists and particle physicists struggle to understand dark matter, accounting for $\sim26\%$ of the energy content of the Universe. 

Cosmic voids, the vast underdense regions observed in the galaxy distribution, feature the capability to provide tight constraints on the most relevant challenges in modern cosmology. Voids form in the spaces between the dense structures of the cosmic web---the intricate network of clusters, filaments, and walls shaped by the gravitational evolution of matter in the Universe. 
The remarkable sensitivity of voids to cosmological parameters can be explained by considering the aspects related to their unique properties \citep{Pisani_2019, Moresco_2022, Cai_Neyrinck_2025}.

Firstly, since the matter content of voids is low, dark energy dominates voids. It is therefore reasonable to expect that voids will be sensitive to the properties of dark energy, such as its time evolution or total contribution \citep{Bos_2012,Lavaux_Wandelt_2010,Pisani_2015a,Verza_2024b}.
While the nature of dark energy is still unknown, it is possible to constrain its equation of state, for example considering the Chevallier--Polarski--Linder (CPL) parameterization \citep{Chevallier_Polarski_2001,Linder_2003},
$w(a) = w_0 + (1-a) \, w_a$, where $w_a$ accounts for the dark energy time evolution and $w_0$ its value at the current time. Cosmic voids allow us to constrain $w_0$ and $w_a$, as well as the dark energy density parameter $\Omega_\Lambda$.

Secondly, if dark energy is not invoked to explain the acceleration of the expansion of the Universe, then such acceleration can be explained by modifications of the laws of gravity. In the latter case, the impact on voids would be stronger, since screening mechanisms are typically not effective in voids \citep[see e.g.][]{Barreira_2015,Baker_2018,Giocoli_2018,Davies_2019,Perico_2019,Contarini_2021,Maggiore_2025}. Typically void sensitivity would allow us to e.g. constrain the growth rate of structure $f$, whose value can be predicted within the $\Lambda$-cold dark matter (CDM) standard model of cosmology. 

Thirdly, voids are sensitive to the sum of neutrino masses, therefore being highly relevant in understanding these elusive particles, e.g. by constraining their total mass \citep[see e.g.][]{Massara_2015,Sahlen_2019,Bayer_2021,Contarini_2022,Kreisch_2022}. Since neutrinos are a diffuse component, they may impact matter fluctuations on the typical size of voids, as well as the formation of low mass galaxies, that in turn will impact the properties of cosmic voids. Also, since voids contain less matter, it is reasonable to expect a stronger relative influence of neutrinos on void properties.

Last but not least, the evolution of cosmic voids plays a fundamental role in shaping the Universe’s large-scale structure and is also relevant to astrophysical studies. Voids provide a unique framework for investigating the evolution of the cosmic web, the interplay between galaxies and their environments, and the underlying baryonic physics~\citep{Sutter_2014b,Cautun_2015,vandeWeygaert_2016,Wojtak_2016,Paillas_2017,Adermann_2018,Valles-Perez_2021,Bromley_Geller_2025,Schuster_2025}. Thanks to the weak impact of astrophysical processes and their simple dynamics---close to the linear regime---voids offer a clean environment for precision tests. For example, they have been shown to be sensitive to the properties of dark matter \citep{Arcari_2022}. While this may seem contradictory, as there is little matter in voids, the simpler environment of voids actually provides a higher signal-to-noise to probe various dark matter models, given that dark matter signals are usually buried in strong astrophysical backgrounds \citep{Arcari_2022}. Nevertheless, in this review our emphasis is placed on the use of cosmic voids as cosmological probes rather than on their astrophysical or dynamical roles within the cosmic web.

The sensitivity of voids to cosmology and physics is further enhanced by the fact that voids span different scales (their sizes span from a few to hundreds of megaparsecs). This adds sensitivity to scale-dependent effects for the considered cosmological observables.
Nowadays, one of the strongest caveats to extract cosmological information from the distribution of galaxies is related to the presence of nonlinearities that become dominant at smaller scales. While these scales are rich of information, methodologies such as modeling the redshift-space distortions of the two-point galaxy correlation function will suffer of model break down below 20--30 $h^{-1}\, \mathrm{Mpc}$. For voids, however, the simplistic single-stream motion of galaxies in their interior, paired with the milder galaxy-astrophysics dominated environment, allows a more direct connection to cosmology \citep{Schuster_2024}, a connection that promises to hold down to galaxy properties \citep{Wang_Pisani_2024,Curtis_2024}. 

Given the promise in constraining power that voids provide thanks to their privileged environment, it is not surprising that they are expected to also enlighten the landscape of recent tensions arising in cosmology: the value of $H_0$---showing a tension of up to $\geq 4 \sigma$ between supernovae and cosmic microwave background (CMB) based measurements \citep{DiValentino_2021b,DiValentino_2021a,Verde_2024}, depending on the considered scenario; and the value of $\sigma_8$ measured from the CMB and other techniques such as weak gravitational lensing or cluster abundances\footnote{It should be noted that recent results from the Kilo-Degree Survey (based on state-of-the-art modeling with a careful treatment of systematic errors and a blinded set-up for the analysis) suggest a release of this tension \citep{KiDS_2025_cosmo}.} \citep{DiValentino_2021c,Amon_2022}. 

In the era of precision cosmology, it is therefore all the more important to address tensions using independent and complementary probes, as well as to rely on theoretical models that perform well at various scales. Void statistics fulfill both these requirements and are expected to powerfully complement our understanding of cosmology in the landscape of tensions \citep{Contarini_2024}. 

With this review, we aim to provide a comprehensive description of the use of cosmic voids as cosmological probes. We begin by introducing the main void observables and discussing the richness of void-based statistics (\Cref{sec:observables}), along with an overview of the most common void-finding algorithms used to identify underdensities in both simulations and galaxy surveys. The core of the review focuses on the theoretical modeling of void statistics (\Cref{sec: Theoretical models}), with dedicated sections on the void size function, the void-galaxy cross-correlation function, velocity profiles, void ellipticity, CMB cross-correlations, and void lensing. In \Cref{sec:constraints_from_surveys} we highlight the sensitivity of these statistics to cosmological parameters and their complementarity with other probes. We then focus on the application of the main void summary statistics to real data catalogs, describing how models are extended to account for observational effects and summarizing current cosmological constraints from voids. Finally, we provide an overview of the role of voids in recent, ongoing and future surveys (\Cref{sec:modern_surveys}), outlining the prospects for high-precision void cosmology in the coming years.

\section{Voids: one tool, many statistics}\label{sec:observables}

Observing voids in the large-scale distribution of cosmic tracers allows us to define many statistics, each with its own theoretical set up, measuring features and impacting systematic errors. Before diving into describing such statistics, we ought to ask ourselves the question of how to define and find voids from a distribution of mass tracers (e.g. dark matter particles, dark matter halos, galaxies and cluster of galaxies). We then explain how to extract different types of void-summary statistics and describe their typical properties. 

To provide the reader with illustrative examples on this topic, we derived key void statistics from a subset of the \textsc{AbacusSummit}\footnote{Publicly available at \url{https://abacusnbody.org/}.} cosmological $N$-body simulation suite \citep{Maksimova_2021}.
We used one realization from the baseline $\Lambda$CDM simulation set, which is characterized by a \textit{Planck} 2018 cosmology \citep{Planck_2020} with $\Omega_\mathrm{cdm}=0.2645$, $\Omega_{\rm b}=0.0493$, $\Omega_\mathrm{de}=0.6862$, $h=0.6736$, $A_{\rm s}= 2.0830 \times 10^9$ and $n_{\rm} = 0.9649$. This simulation includes approximately 330 billion dark matter particles with a mass of $\sim2 \times 10^9 \ h^{-1} \, M_\odot$, distributed within boxes of side length $2 \ h^{-1} \, \mathrm{Gpc}$.
Depending on the considered void statistic, we prepared catalogs composed either of dark matter particles or dark matter halos. The former are obtained by selecting a 5\% random sub-sample of the public catalogs (already provided in a diluted form that retains 3\% of the original particles), while the latter consist of a cleaned sample of halos identified with \textsc{CompaSO} \citep[see][]{Hadzhiyska_2022}, filtered to have halos containing at least 300 particles (corresponding to $\sim6 \times 10^{11} \ h^{-1} \, M_\odot$). We considered three redshifts: $z = 0.2$, $0.5$, and $1.1$, extracting the 3D positions and velocities of each tracer. At each redshift, the dark matter catalog contains roughly half a billion particles, corresponding to a mean inter-particle separation of $\sim 2.53 \ h^{-1} \ \mathrm{Mpc}$, which provides enough spatial resolution to detect small underdensities and to robustly trace the inner regions of deep cosmic voids. For dark matter halos, we only show the results $z = 0.2$, where the mean separation is $~ 5.40 \ h^{-1} \, \mathrm{Mpc}$.
While it is beyond the scope of this review to test and compare different void finders, we pedagogically select two different types of void finders (a tessellation based and a spherically based) as illustration.
Using these void finders, two different void catalogs were then constructed from the \textsc{AbacusSummit} data, in order to illustrate the impact of the identification method on the resulting summary statistics.

\subsection{Void definition and identification}\label{sec: Void finders}
Since their discovery \citep{Gregory_1978,Joeveer_1978,Kirshner_1981,deLapparent_1986,Thompson_2011}, various methodologies have been tested to detect cosmic voids from both simulations and real data. While cosmic voids generally refer to underdense regions, a unique, specific and universal definition does not exist. This situation closely parallels that of dark matter halos: although a halo is typically considered as a bound and collapsed clump of dark matter, no unique definition of such an object or its extent exists, and numerous halo-finding algorithms have been developed~\citep[see e.g.][]{Knebe_2011,Onions_2012}. Similarly, each void definition corresponds to a specific void-finding algorithm, often optimized for a particular application. In cosmological analyses, the key aspect is therefore not to adhere to a single void definition, but rather to understand the used void definition and its implications, to construct a robust void catalog and to maximize the cosmological information that can be extracted from it.

When found in simulations, voids are detected in the dark matter, halo or galaxy distribution; while finding voids from survey data usually relies on the galaxies' positions (photometric or spectroscopic redshifts, jointly with right ascension and declination) or on 2D maps. 
Typically, void finders can be divided into three main classes \citep{Lavaux_Wandelt_2010, Colberg_2008}:
\begin{itemize}
    \item \textit{Topological}: algorithms that identify voids as geometrical underdense structures composed of cells or polyhedra, such as \zobov \citep{Neyrinck_2008} and void finders based on it, e.g. \vide \citep{Sutter_2015} and \textsc{Revolver} \citep{Nadathur_2019c}. These algorithms rely on methods such as the Voronoi tessellation to estimate the density field, followed by techniques to merge the Voronoi cells, e.g. the watershed transform \citep{Platen_2007}.
    \item \textit{Spherical}: algorithms that search for voids as spheres, or ensembles of spheres, embedding a given integrated density contrast\footnote{The density contrast is defined as $\delta \equiv (\rho-\overline{\rho})/\overline{\rho}$, with $\rho$ and $\overline{\rho}$ being the local density and the mean density of the Universe, respectively.}, like \spark \citep{Ruiz_2015,Ruiz_2019}, \textsc{PopCorn} \citep{Paz_2023}, the void finder of the library \textsc{Pylians} \citep{Villaescusa-Navarro_2018}, \textsc{DIVE}\footnote{While we place \textsc{DIVE} in this class, it actually belongs to a slightly separate category, since it considers empty spheres that are constrained by four elements of a point set, relying on the Delaunay tessellation. As such, it corresponds, in fact, to performing a volume statistics analysis of the field \citep{Chan_Hamaus_2021}.} and \textsc{VoidFinder} (from \citet{El-Ad_Piran_1997} and \citet{Hoyle_Vogeley_2002}, also implemented in the \textsc{VAST} toolkit \citep{Douglass_2022}.
    \item \textit{Dynamical}: algorithms based on the reconstruction of the cosmic velocity field, which define voids as regions from which matter is evacuated \citep[see e.g.][]{Lavaux_Wandelt_2010,Elyiv_2015}.
\end{itemize}
A summary of the publicly available algorithms that are widely used to identify cosmic voids is provided in Appendix~\ref{app:A}, where we include a brief description of each code, together with the relevant references and links necessary for their use.

In recent years, cosmological studies have mostly relied on topological void finders, but also on algorithms that characterize all cosmic web structures, such as \textsc{DisPerSE} \citep{Sousbie_2013} and \textsc{NEXUS+} \citep{Cautun_2013}, which have been used, for example, for the characterization of other cosmic web environments such as filaments. Moreover, particularly in the context of void lensing analyses, it becomes important to consider the 2D projection of voids. This can be done by selecting 3D voids in redshift slices and projecting their signal \citep{Cautun_2018,Fang_2019,Davies_2021}, or by identifying voids directly in 2D lensing maps \citep{Sanchez_2017,Maggiore_2025}.
 
Finally, variations can be considered with respect to the assumptions made by each void finder; for example, the choice of the void identification threshold (including considering empty spheres), the criterion for void merging or overlap, or the method used to assign the void center or define the void radius.
Overall, the great variety of algorithms can penalize an easy, clean and coherent physical interpretation and result comparison. In the last decade an accurate comparison of results, paired with the widespread use of topological void finders for recent studies, allowed the extraction of robust cosmological constraints from a number of data sets. 

While all the void finders are sensitive to the physical signal of the underdensities, there is agreement in the community that some definitions may be more suitable than others based on \textit{a}) the considered void statistics \textit{b}) the specific features of the data set \citep{Colberg_2008, Cautun_2018, Veyrat_2023}. This typically depends on the fact that systematic errors (such as the impact of spurious voids and redshift errors) vary across definitions and that enhancing the signal-to-noise ratio is crucial for a stronger sensitivity to cosmology. A deep understanding of the void finder features and wide testing on simulations and mocks is crucial for robust applications, as well as the possibility to link the void finder definition with the corresponding theoretical description of the considered statistics, and the study of the impact of observational systematic errors. 

As anticipated, in this review we apply two void finders to the catalogs extracted from the \textsc{AbacusSummit} simulation. 
From the class of topological void finders, we consider the Void IDentification and Examination toolkit \citep[\textsc{VIDE,}][]{Sutter_2015}, a parameter-free watershed-based algorithm that can be applied to both periodic simulation boxes and observational survey data. \textsc{VIDE} identifies underdense regions as basins in the density field, which is estimated through a Voronoi tessellation of the tracer distribution. The center of each void is defined as the volume-weighted barycenter of the Voronoi cells composing it. As is common for topological methods, the effective radius of a void is defined as that of a sphere with the same total volume as the sum of its constituent Voronoi cells \citep{Neyrinck_2008}. By construction, the resulting void catalog fills the entire volume traced by the mass distribution, since every Voronoi cell is assigned to a void.

From the class of spherical finders, we consider \textsc{Sparkling} \citep{Ruiz_2015,Ruiz_2019}, an algorithm designed for the identification of underdense spherical regions, optimized to operate efficiently in both $N$-body simulations and galaxy surveys. The algorithm begins by reconstructing a continuous density field through a Voronoi tessellation of the tracer distribution. The minima of this field are then used as initial candidates for void centers. Starting from these positions, spherical regions are iteratively expanded until they enclose a given integrated density contrast, $\Delta_{\rm v}$\footnote{In this review, we will use the notation $\Delta(r)$ when the density contrast is explicitly computed by averaging its local value inside spheres of radius $r$, which mathematically corresponds to integrating over the radial coordinate up to $r$. Consequently, we use $\Delta_{\rm v}$ to indicate the threshold used in void finders based on the integrated density contrast.}, which is set by the user and we fix to $-0.8$. To refine the void centers, random displacements are tested, and the position yielding the largest possible expansion is retained. In the final step, overlapping spheres are discarded, keeping only the largest ones to produce a clean, non-overlapping void catalog.

\begin{figure}[ht]
    \centering
    \includegraphics[height=6.32cm]{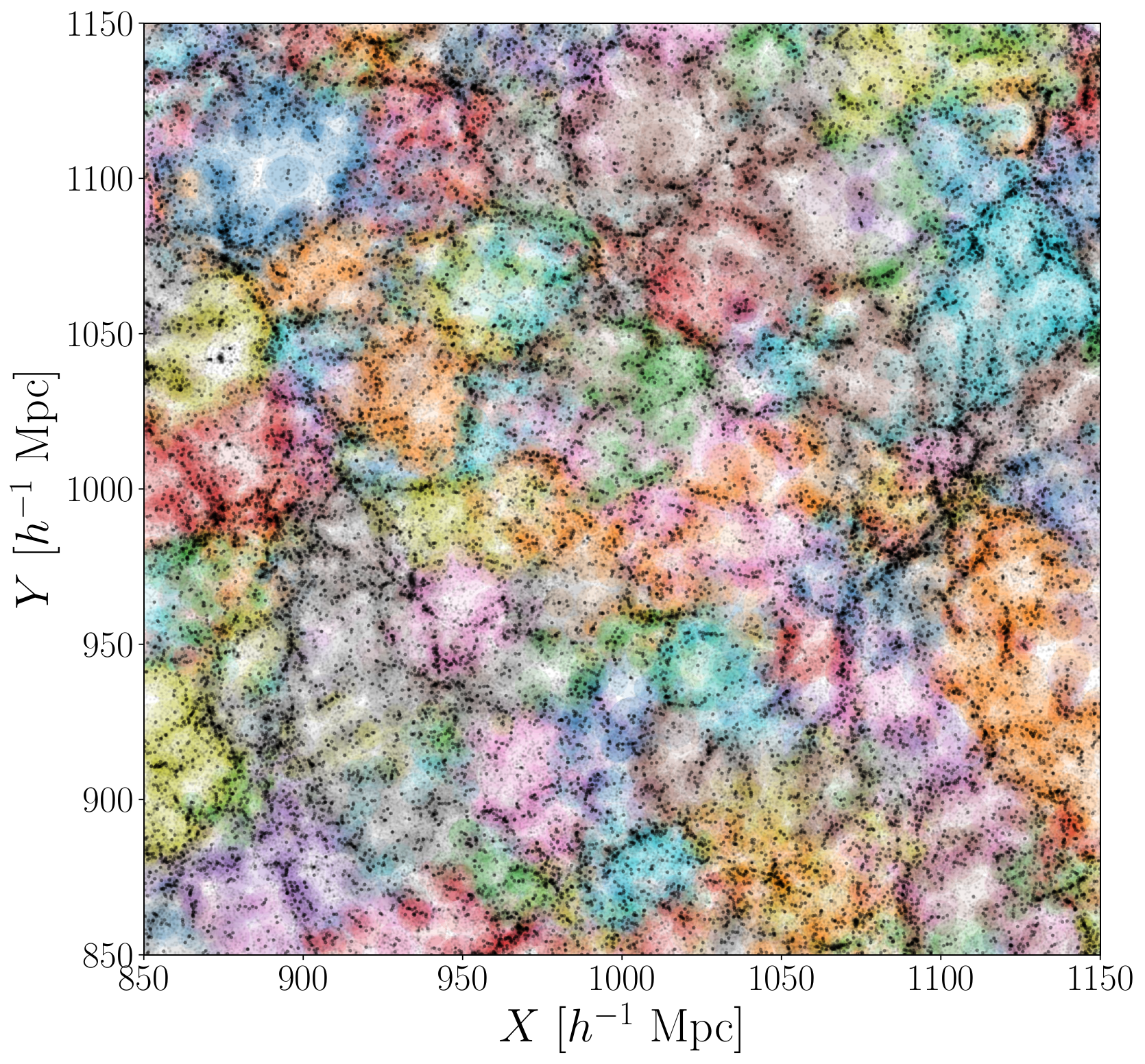}
    \includegraphics[height=6.25cm]{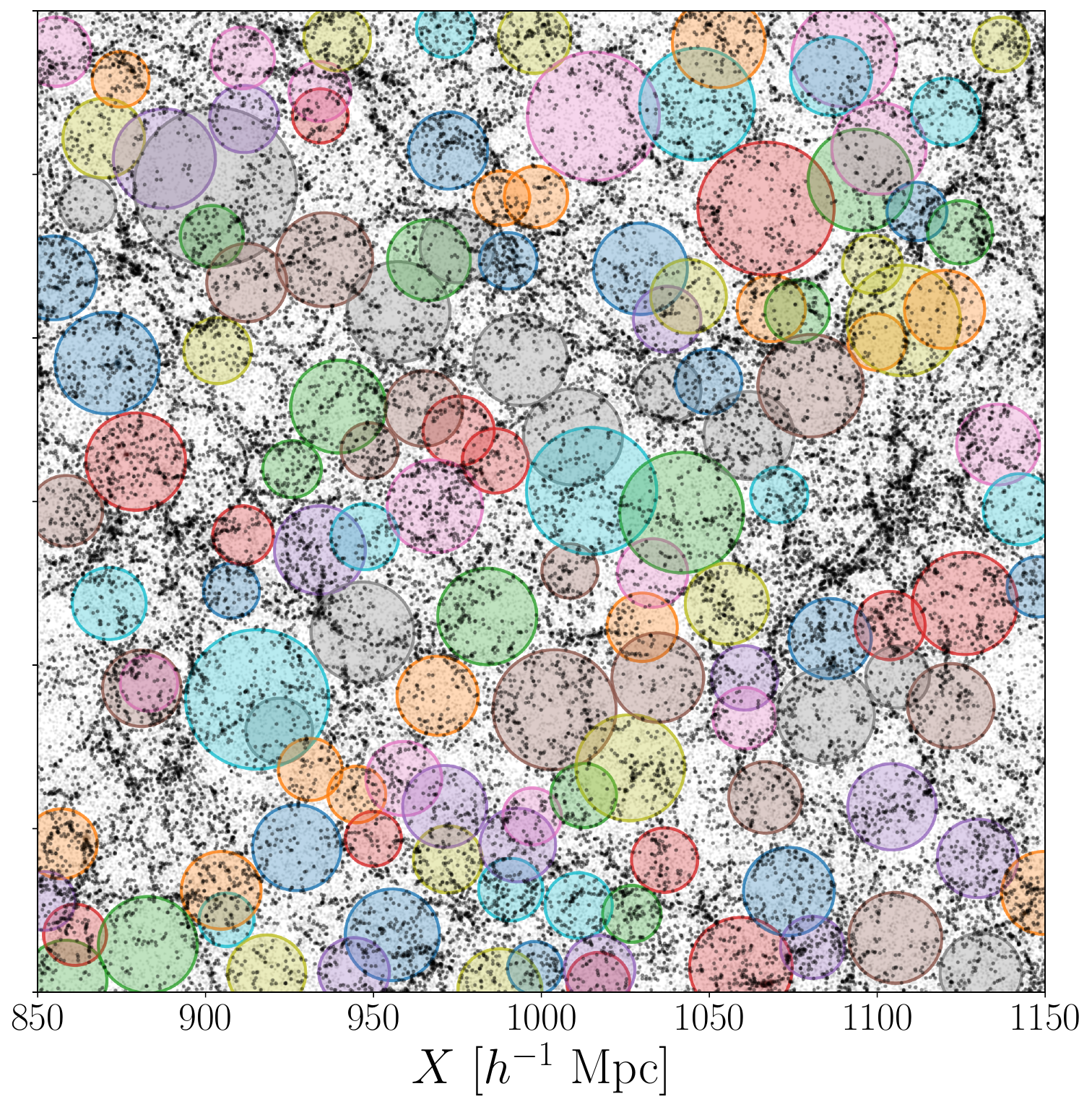}
    \caption{Visual representation of cosmic voids found with \vide (left) and \spark (right). We show with different colors the voids identified by the two algorithms in the distribution of dark matter halos (black dots) extracted from the \textsc{\textsc{AbacusSummit}} simulation suite at $z=0.2$. Both panels cover the central part of the box and represent the 2D projection of a $30 \ h^{-1} \, \mathrm{Mpc}$ slice along the Z-axis.}
    \label{fig:void_visual_comparison}
\end{figure}

We present now the representation of the void catalogs obtained when applying these two finding algorithms. Visualizing cosmic voids can be challenging due to their intrinsic emptiness and often irregular 3D shapes, as well as projection effects. Nevertheless, in \Cref{fig:void_visual_comparison}, we provide a visual representation by showing the 2D projection of voids identified within a $30 \ h^{-1} \, \mathrm{Mpc}$-thick slice of the $\Lambda$CDM \textsc{AbacusSummit} simulation box at $z = 0.2$. We consider all the voids intersecting the selected Z-axis range, and show only the portions lying within the slice. In this way, the colors indicate particles associated with different voids in a manner consistent with the Z-axis projection. As expected, by representing the voids identified with \textsc{VIDE}, the cosmic web appears divided in different sub-structures, each composed of an underdensity surrounded by dense ridges (i.e. filaments). Voids identified with \spark are instead spherical non-overlapping spheres with the same internal density contrast; any apparent overlap of the circles in \Cref{fig:void_visual_comparison} is caused by projection effects along the Z-axis. 
We will demonstrate later that, despite appearing visually different, the two void samples we have just presented trace regions of the Universe characterized by the same global properties. For each void, the center and the effective radius can be used to build different summary statistics, as described in the next section.

\subsection{Void statistics \label{sec: Void statistics}}
The measurement of cosmic voids provides void centers, void radii and, usually, information on galaxies' membership to voids (the latter only when voids are not empty spheres). With these it is possible to build various statistics, such as: the void size function, the void-galaxy cross-correlation function, the void auto-correlation function, void ellipticity, etc. If galaxy velocities are also available, it is possible to measure velocity profiles, while additional CMB or lensing measurements further allows us to measure the void-CMB cross-correlation or the lensing signal by voids. 

The various void statistics mentioned above are not all equally developed for applications in cosmological analyses. For example, statistics such as the void size function and the void-galaxy cross-correlation function can be modeled satisfactorily and their use already provides constraints from survey data, while others are still at development stage, such as the ellipticity. In the following sub-sections we briefly introduce the various void statistics, leaving the detailed discussion of their theoretical modeling to the section that follows (\Cref{sec: Theoretical models}). We also refer the interested reader to Appendix \ref{app:A} for a list of public codes that can be used to manipulate void catalogs and easily measure several of these statistics.

\subsubsection{Void size function}
The void size function (VSF) is a first-order statistic in cosmology that quantifies the number density of voids as a function of their size, commonly characterized by their effective radius or equivalent volume. It describes the abundance of underdense regions in the universe, which is a property dependent on the overall matter distribution and its evolution. Due to the hierarchical nature of cosmic structures, smaller voids are typically far more numerous than larger ones. As a result, the VSF decreases rapidly with increasing void radius---at least on scales not dominated by statistical noise. The maximum measurable void size depends on the sampled volume: larger survey volumes increase the likelihood of detecting rare and extended voids. Conversely, the minimum measurable size is determined by the mean separation of the tracers used to identify voids, as smaller voids become challenging to detect with sparse or low-resolution data sets. Indeed, when the void size approaches the spatial resolution of the tracer catalog, it becomes more likely to find voids that are unphysical, since empty areas in sparse data sets simply correspond to regions where tracers of low mass cannot be detected. These are commonly known as spurious voids or Poisson voids \citep{Neyrinck_2008,Pisani_2015b,Cousinou_2019}.

To measure the VSF for cosmological analyses, the sample of identified voids is binned according to their radius, $R_\mathrm{v}$, producing a distribution $N(R_\mathrm{v})$, which represents the number of voids in each size bin. This can then be normalized by the survey volume, to obtain the number density of voids, $n(R_\mathrm{v})$. Moreover, it is important to correct for possible selection effects, such as survey geometry and observational systematic effects, and include the covariance matrix to account for uncertainties and correlations between bins.

\begin{figure}[t]
    \centering
    \includegraphics[height=5.5cm]{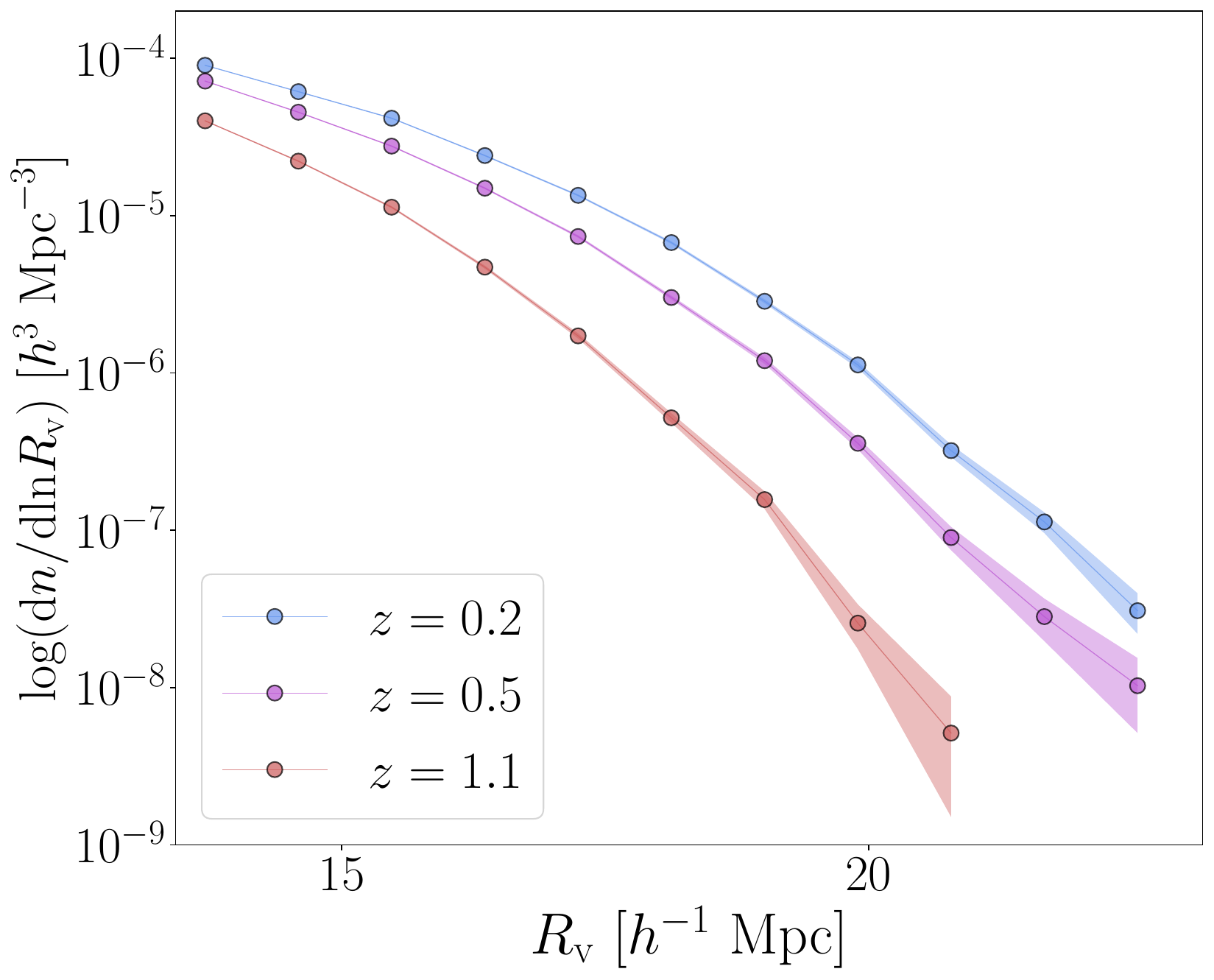}
    \includegraphics[height=5.5cm]{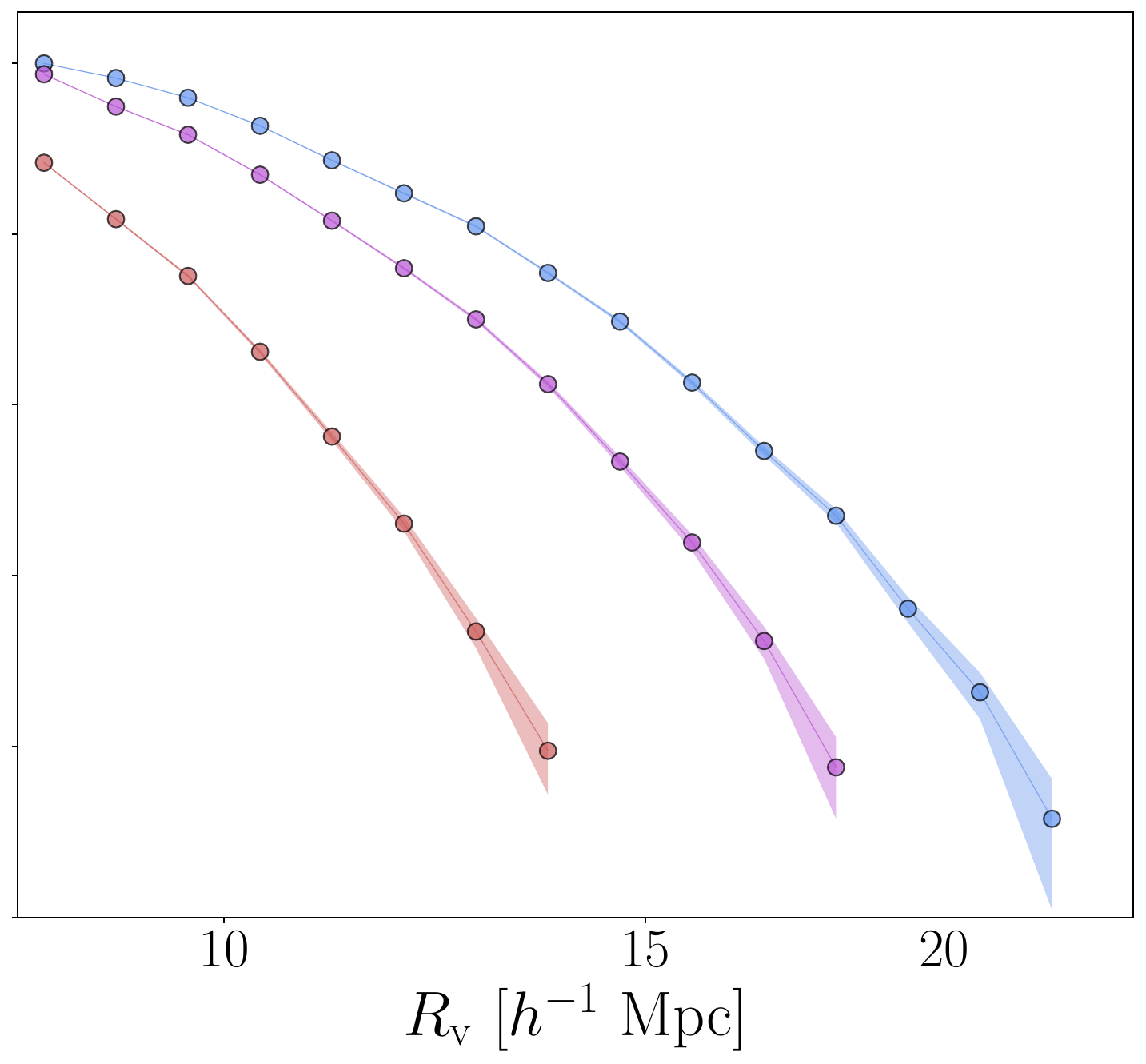}
    \caption{Comparison between the VSFs measured using cosmic voids identified with \vide (left) and \spark (right) in the \textsc{AbacusSummit} simulation suite, using dark matter particles as tracers. In both panels, different colors indicate the VSF at different redshifts: $z = 0.2$ in blue, $z = 0.5$ in purple, and $z = 1.1$ in red. Shaded regions represent the uncertainty in each bin, derived from the Poisson error as $\sqrt{N}/(V \, \Delta\ln R_{\rm v})$, where $N$ is the counts of voids in the finite logarithmic bin $\Delta \ln R_{\rm v}$ and $V$ the volume occupied by the cosmic tracers.
    }
    \label{fig:VSF_comparison}
\end{figure}

In \Cref{fig:VSF_comparison}, we present a comparison of the VSFs obtained by applying the finders \vide and \spark to the sub-sampled dark matter particles of the \textsc{AbacusSummit} simulation suite at different redshifts ($z = 0.2$, $0.5$, and $1.1$). As these finders adopt different definitions for the void radius---\vide typically assigning larger effective radii with respect to \spark---we use in this figure different selections for the minimum radius. We chose the size range where the two catalogs showcase the same order of magnitude in void number counts, while excluding the small scales where the shot noise becomes significant. The overall shape of the VSFs is similar for both catalogs, and their redshift evolution also follows the same trend, with voids increasing in size and number over time in the selected radius ranges\footnote{It is important to clarify that voids tend to expand over time as they evacuate matter from their interiors. At the same time, neighboring underdense regions can merge as the cosmic web evolves, forming larger voids. Although the total volume occupied by voids may remain approximately conserved, the number of large voids naturally increases as a consequence of both expansion and merging processes.}. This points to a coherent physics-driven behavior among definitions. We finally emphasize that the shapes of the void size functions identified with \spark and \vide are remarkably similar, even when dark matter halos are used as tracers. No clear redshift evolution can however be seen when voids are traced with a biased sample selected with the same minimum halo mass at all redshifts. This happens because the dark matter density contrast used to define voids, $\Delta_{\rm v, DM}(z)$, evolves over time proportionally to the growth factor, while the linear bias of the tracers, $b(z)$, evolves inversely with it. As a result, the halo density contrast within a given void, $\Delta_{\rm v, tr}(z) \equiv \Delta_{\rm v, DM}(z) \, b(z)$, is by construction almost constant in time, effectively leading to a void size function that is nearly independent of redshift.

\subsubsection{Void density profiles} 

We start this section by highlighting that the void density profile is the equivalent of the void-tracer cross-correlation function. In observations, in fact, this statistic is commonly measured using galaxies as tracers, hence it corresponds the so-called void-galaxy cross-correlation function (VGCF), $\xi_\mathrm{vg}(\bf{r})$, which is the excess probability of finding a galaxy at a relative position $\bf r$ with respect to the void center. This statistic {\it de facto} corresponds to measuring the density contrast profile of voids \citep[see e.g.][]{Hamaus_2014c}. In an isotropic universe, the VGCF depends only on the relative void center--galaxy distance $r$ and not on its orientation. The VGCF approaches the value $-1$ from above at small $r$, as the probability of finding galaxies near the void center is low. At larger distances outside of voids, the VGCF approaches $0$, as the galaxy density reaches the mean field value far from the void center. However, the exact VGCF shape depends on the considered void finder. 
The VGCF usually presents a positive peak around the average void radius \citep{Hamaus_2014b}. This feature is called ``compensation wall'' \citep{Sheth_van_de_Weygaert_2004,Hamaus_2014c,Hamaus_2014a} and corresponds to the overdensities formed by galaxies surrounding each void. The presence of the compensation wall is particularly useful for cosmological analyses (see \Cref{sec: Theoretical models}), therefore it is common to enhance it using the stacking technique \citep{Lavaux_Wandelt_2012,Sutter_2012,Sutter_2014a}, which maintains the characteristic shape of the void profile while reducing the statistical noise. The stacking procedure typically consists of considering voids of similar size or behavior, averaging their signal after rescaling the distance from the center by the void radius. This is also crucial for cosmological analyses, since individual voids cannot be used statistically.

\begin{figure}[t]
    \centering
    \includegraphics[height=5.4cm]{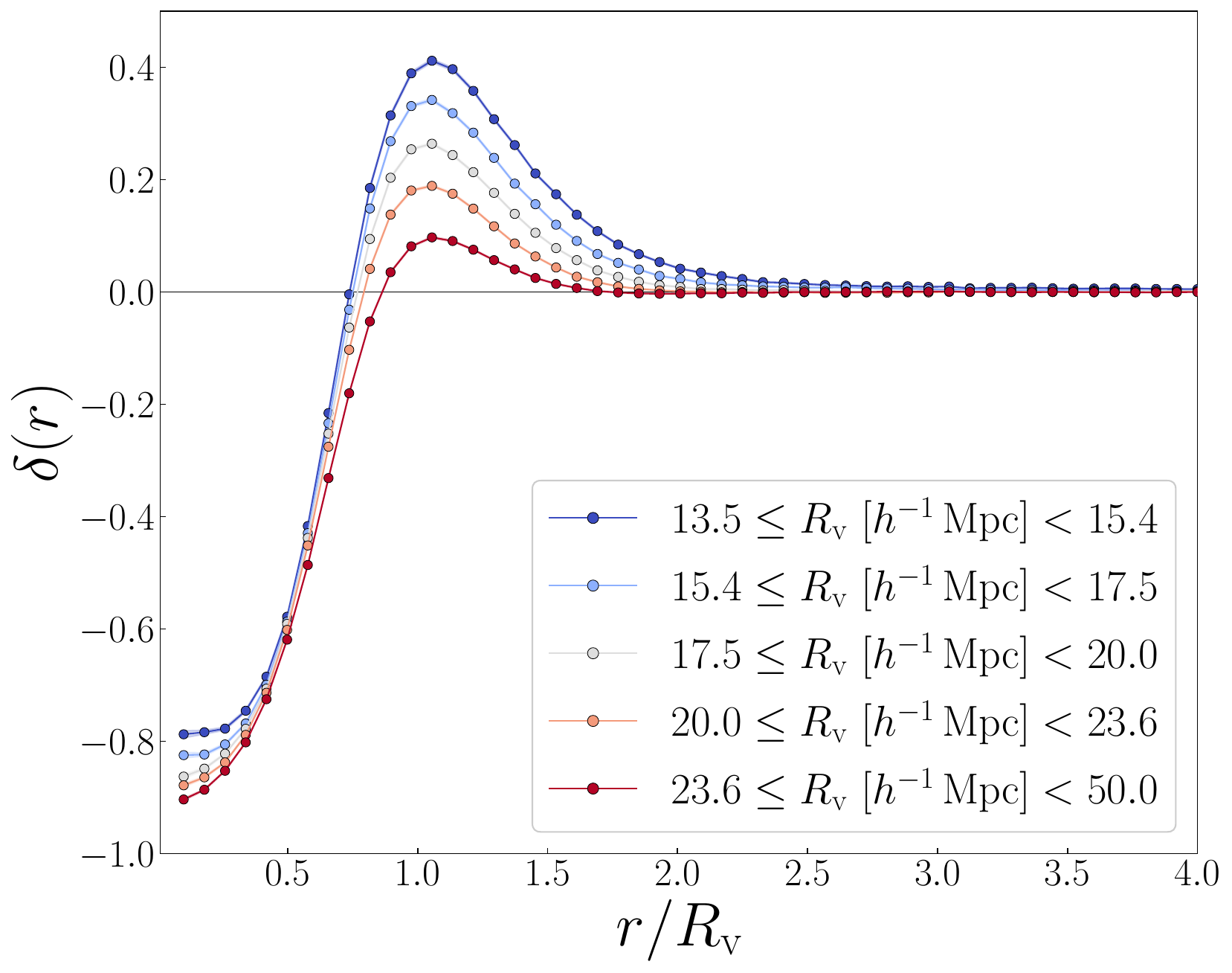}
    \includegraphics[height=5.4cm]{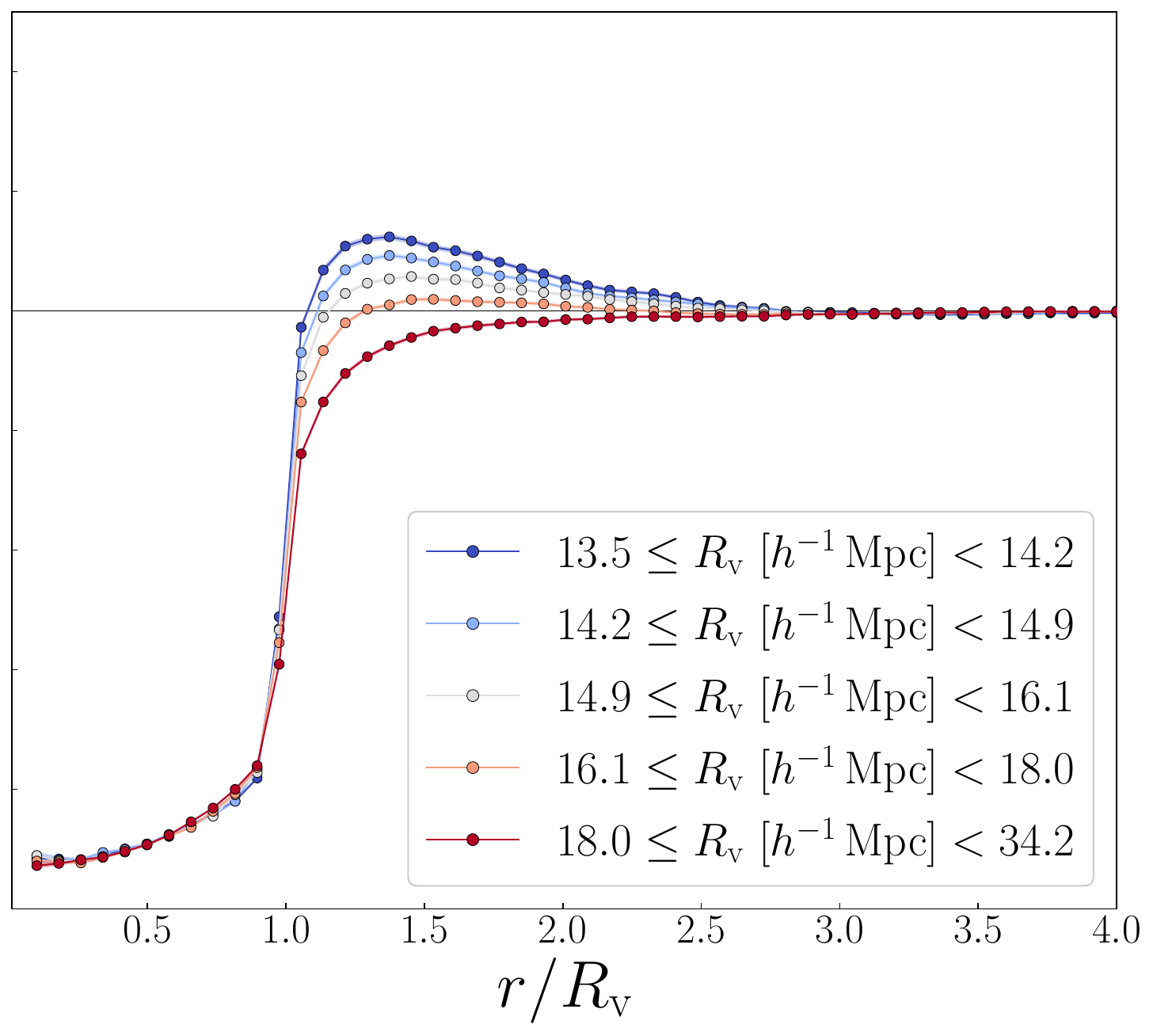}
    \caption{Comparison between the differential density contrast profiles measured using cosmic voids identified with \vide (left) and \spark (right) in the \textsc{AbacusSummit} simulation suite, using dark halos at $z=0.2$ as tracers. In both panels, the void sample is split in equi-populated bins, starting from 2.5 times the mean separation of the halo catalog and extending up to the largest void radius of each sample. Within each bin, voids with similar effective radii are averaged together, and the resulting density contrast profile is represented with different colors according to the void size.}
    \label{fig:delta_comparison}
\end{figure}

In \Cref{fig:delta_comparison}, we show a comparison of the differential density contrast profiles (i.e., computed in concentric spherical shells) of voids identified using \vide and \spark, applied to the dark matter halos of the \textsc{AbacusSummit} simulation suite at $z = 0.2$. Voids of similar size are stacked by dividing the sample into equally populated bins, containing approximately 28,000 voids for \vide and 10,000 for \spark.
Following a number of works \citep[e.g.][]{Pisani_2015a,Ronconi_2019,Contarini_2019,Radinovic_2024}, the minimum void size is set to 2.5 times the mean tracer separation, which in this case corresponds to about $13.5 \ h^{-1} \, \mathrm{Mpc}$. This cut reduces the number of voids to about $82\%$ of the original catalog for \vide, and to about $22\%$ for \textsc{Sparkling}.
We note that, although \vide typically exhibits higher compensation walls and smoother interiors, the two sets of profiles display the same overall shape and radial dependence---again an indication of coherent physical behavior across different void definitions. Alongside the VGCF, the same statistical information can be measured in Fourier space, as the void-galaxy cross-spectrum \citep{Hamaus_2014c,Chan_2014}.

\subsubsection{Void velocity profiles}
The velocity profiles around cosmic voids are computed by measuring the radial velocity of mass tracers in concentric shells around void centers; the distance from the void center is often normalized by the void radius in order to stack different profiles together. It is important to note that the velocity field becomes biased when estimated with the velocities of individual tracers \citep[see e.g.][]{Bernardeau_1996,Schaap_2000}. Specifically, in the mass-weighted approach, the field is sampled by tracers (e.g. dark matter particles) that predominantly reside in massive structures (e.g. clusters), whose peculiar velocities tend to dominate the resulting average velocity field.
As a consequence, the motion of matter in underdense regions, such as cosmic voids, is significantly underrepresented. An unbiased estimation of the velocity field requires a volume-weighted approach, which samples space uniformly, regardless of the local density \citep[see e.g.][for descriptions of various volume-weighting techniques]{Bernardeau_1996,Cautun_2011,Bel_2019,Esposito_2024}. 

We also note that, when computing stacked velocity profiles (i.e. averaging the radial velocity of voids with similar sizes), empty shells should not be assigned a zero radial velocity. Assuming a null velocity field in regions without tracer particles would indeed be incorrect. Nevertheless, to prevent biased results, it is sufficient to exclude empty radial bins from the averaging procedure \citep[see][for alternative approaches]{Schuster_2023,Schuster_2024}.

Generally, tracers exhibit a coherent motion from the center of voids to their outskirts, with velocities tightly dependent on the depth of the void density profile, retaining the typical void shape. Indeed, following the basic principle of the mass conservation \citep[see e.g.][]{Peebles_1980}, the negative value of the integrated density profile of voids results in a positive radial flow (i.e. directed outward). Outside voids, the velocity profile typically trends toward zero, but may become negative near the compensation wall, where tracers tend to flow toward the overdensity characterizing the void boundary.
The theoretical modeling of velocity profiles is crucial for understanding the effect of redshift-space distortions (RSD) in voids, which manifest as an apparent elongation of voids along the line of sight. The effect of RSD must be incorporated, for example, into the VSF and the VGCF models to ensure accurate theoretical predictions (see \Cref{sec: VSF,Subsection: VGCF Theory}).

\begin{figure}[t]
    \centering
    \includegraphics[height=5.4cm]{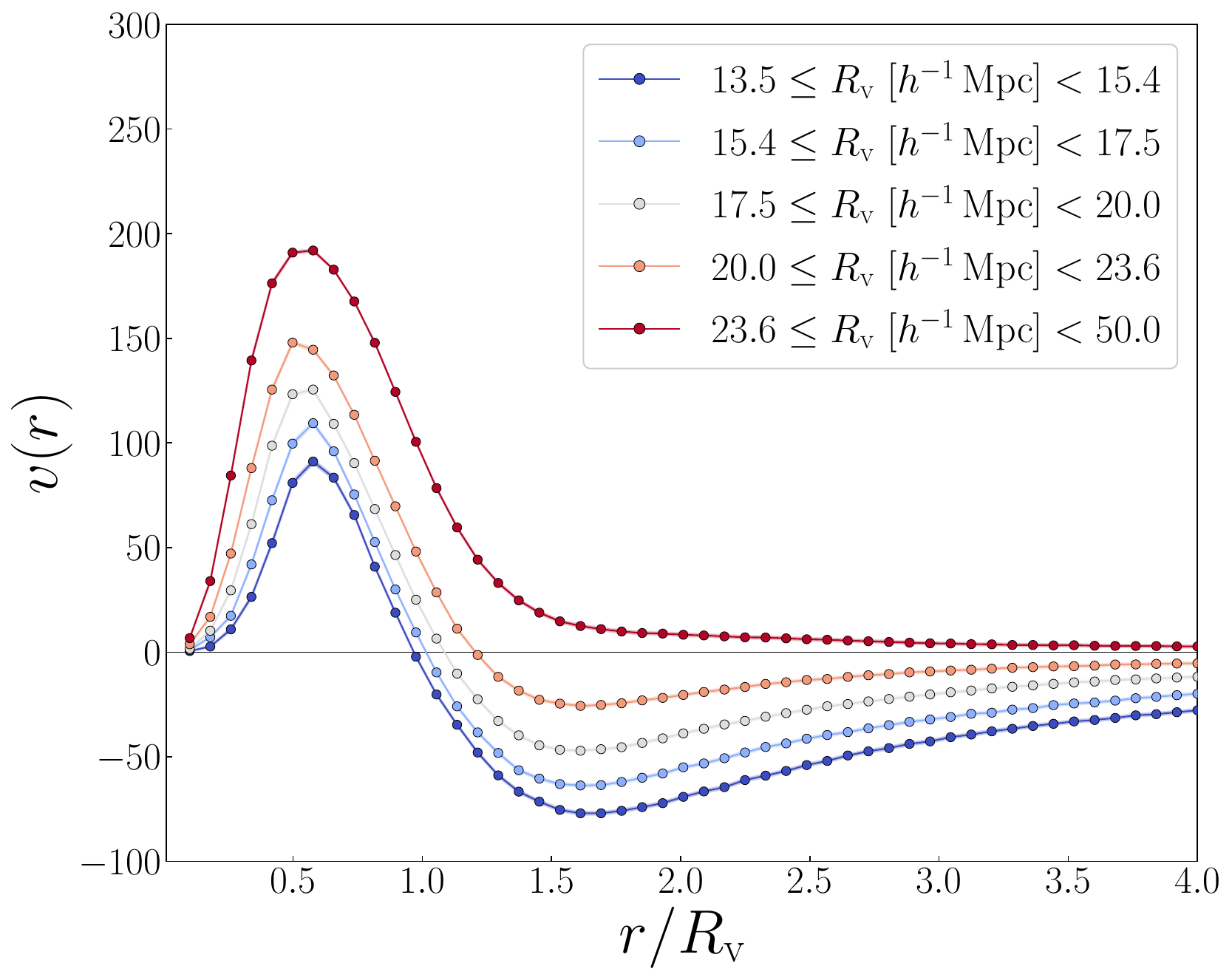}
    \includegraphics[height=5.4cm]{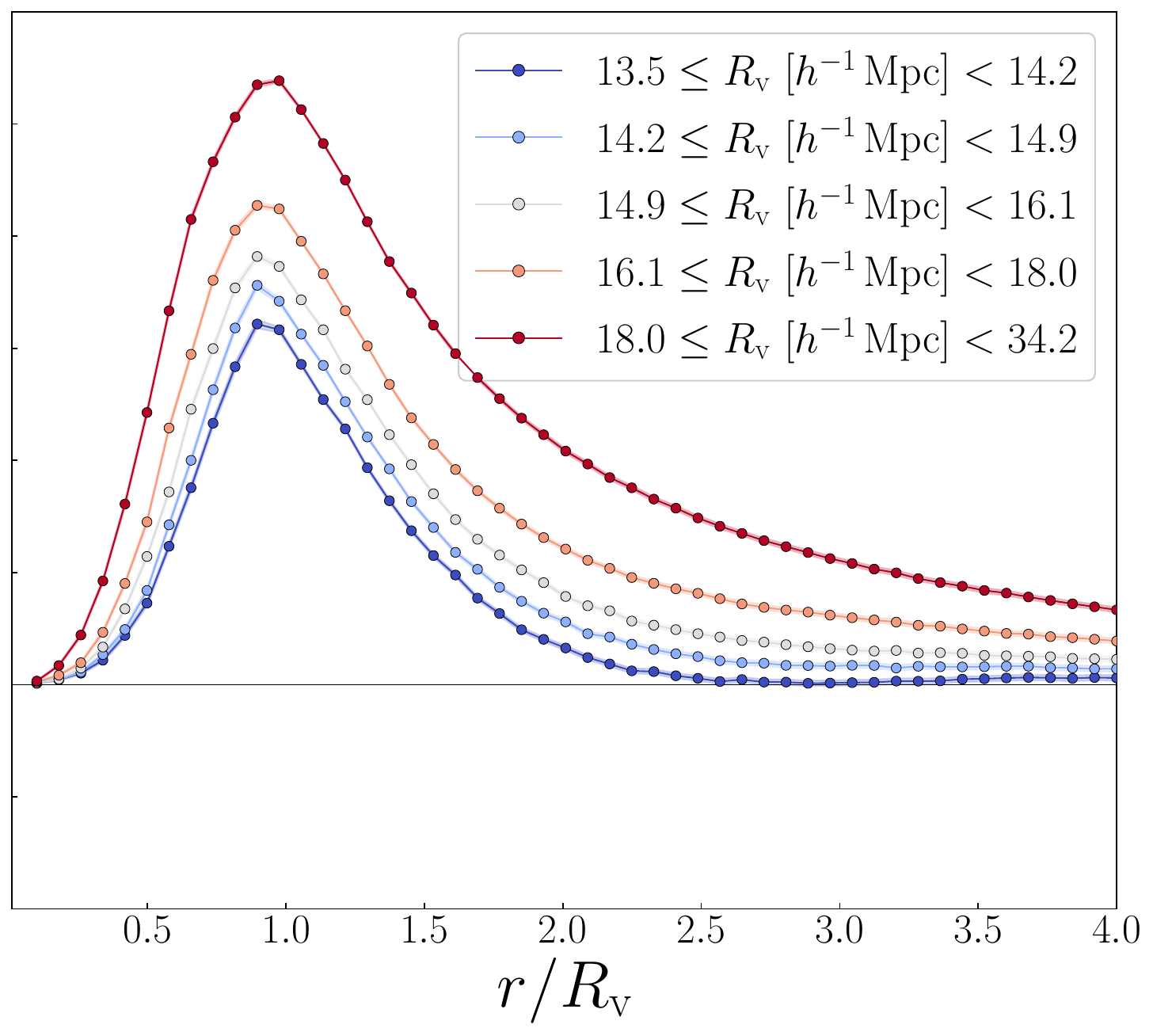}
    \caption{Same as \Cref{fig:delta_comparison}, but for void radial velocity profiles.}
    \label{fig:vr_comparison}
\end{figure}

In \Cref{fig:vr_comparison}, we present the radial velocity profiles extracted from the same void samples used for \Cref{fig:delta_comparison}.
Depending on the void finder used---\vide or \spark---the velocity profiles exhibit slightly different behaviors, although both are characterized by a coherent outflow in their inner regions.
For small voids, the \vide profiles are characterized by negative velocities in the outer regions, whereas the \spark profiles tend to remain positive throughout. This difference is expected and can be attributed to the more pronounced compensation walls observed in the \vide void density profiles (\Cref{fig:delta_comparison}). High compensation walls indicate the systematic presence of overdensities at a given distance from the void center, which induce a matter inflow. In contrast, smoother density profiles---such as those from \textsc{Sparkling}---that do not exhibit positive values when integrated over the void radius, result in an entirely positive radial velocity profile.

\subsubsection{Void auto-correlation function}
\begin{figure}[ht]
    \centering
    \includegraphics[height=5.3cm]{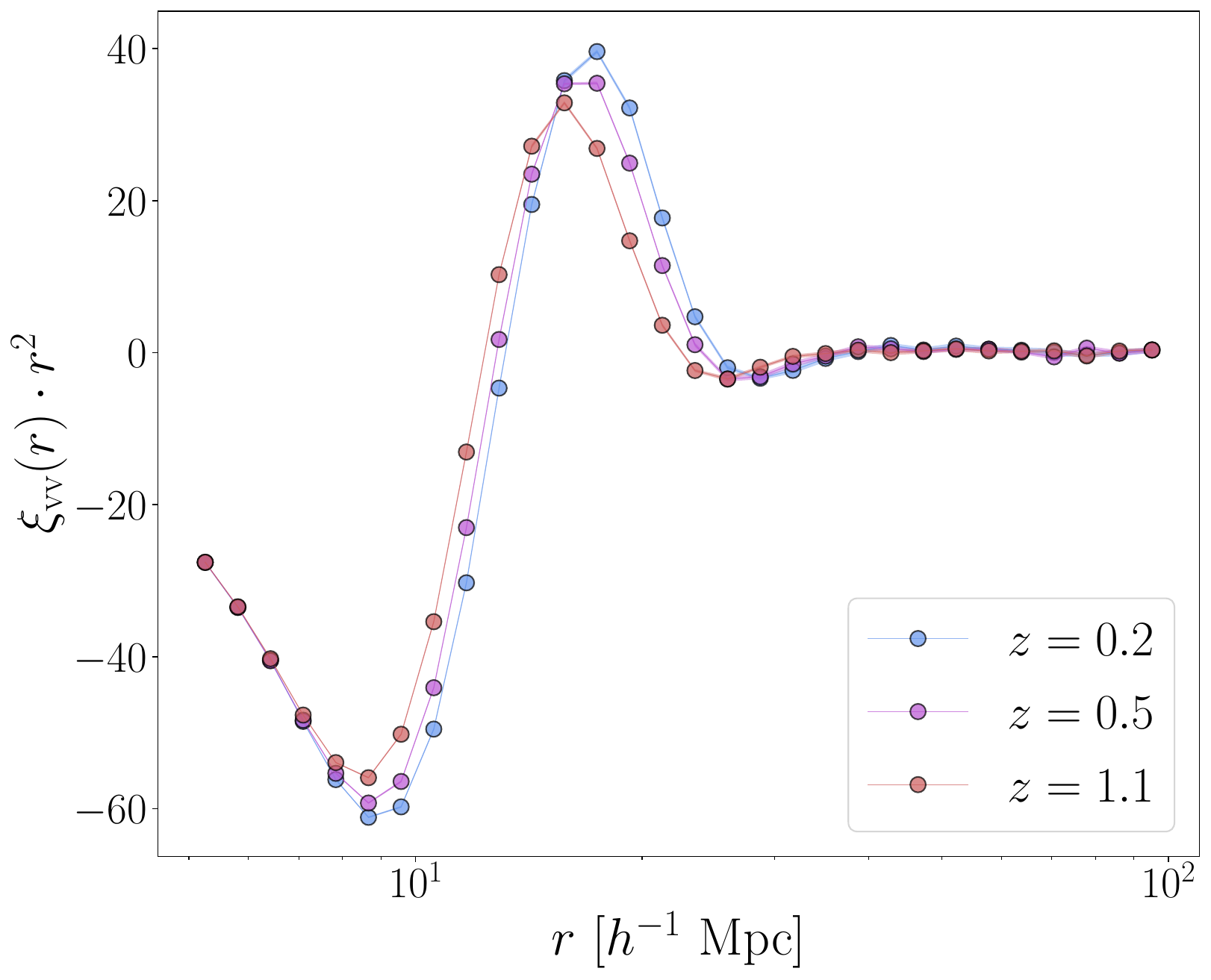}
    \includegraphics[height=5.3cm]{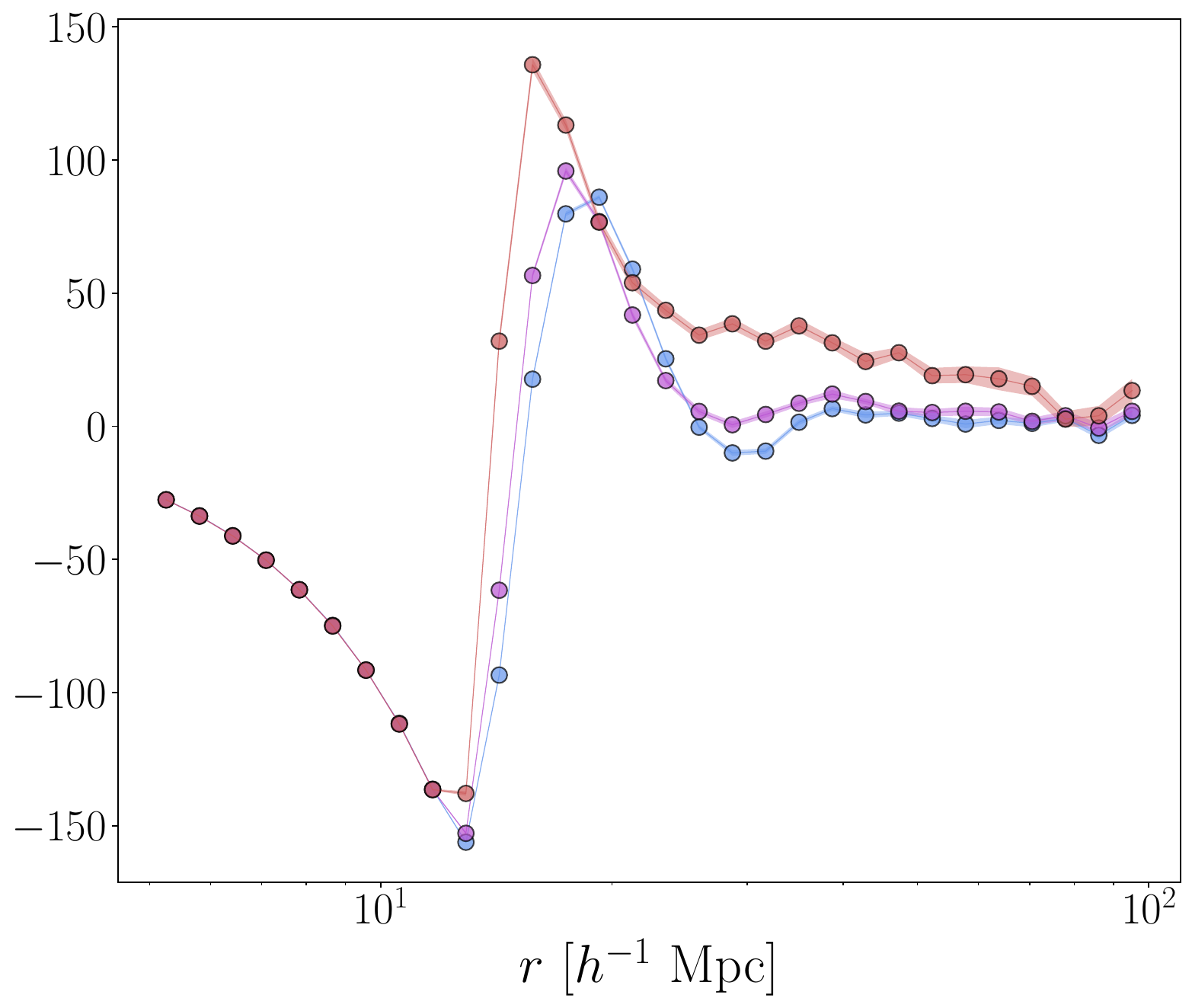}
    \caption{Comparison between the auto-correlation function extracted from voids identified with \vide (left) and \spark (right), using the same simulations introduced for \Cref{fig:VSF_comparison} but selecting voids with radius greater 2.5 times the mean separation of the dark matter particles. Here both measures are multiplied by the squared distances between the pairs of void centers in order to enhance the signal.}
    \label{fig:void-auto_comparison}
\end{figure}
The void auto-correlation function ($\xi_\mathrm{vv}$), is the correlation between different void centers. Despite its cosmological content and the fact that it can be theoretically described \citep{Hamaus_2014c,Chan_2014,Voivodic_2017}, this function has rarely been measured in galaxy surveys, due to its currently low statistical signal, a fact expected to change in coming years with the increase of volume mapped by upcoming surveys (see \Cref{sec:modern_surveys}). As of now, the typical total number and number density of voids detected in galaxy survey is usually too low to perform cosmological analyses on the void auto-correlation function. As is the case for the VGCF, the auto-correlation of voids can be measured in Fourier space, as the void auto-power spectrum \citep{Hamaus_2014c,Chan_2014}.

In \Cref{fig:void-auto_comparison} we present two examples of void auto-correlation functions measured using voids identified in the dark matter distribution of the {\tt AbacusSummit} simulation suite, at redshifts $z = 0.2,\ 0.5,\ 1.1$. In these panels, the auto-correlation functions derived from void catalogs created with \vide and \spark are shown, each multiplied by the squared distance between void centers (which also corresponds to the x-axis of the plot). These results are derived by considering, for both the finding algorithms, voids with radius greater than $2.5$ times the mean separation of the dark matter particles, i.e. $6.33 \ h^{-1} \, \mathrm{Mpc}$. At $z=0.2$, this purity cut reduces the number of voids with respect to the original catalog down to $80\%$ for \vide and $23\%$ for \spark. At other epochs, the reduction in number counts is different as the slope of the VSF varies with the redshift: at $z=0.5$ it is $66\%$ [$18\%$] while at $z=1.1$ it is $57\%$ [$9\%$] for \vide [\textsc{Sparkling}]. We compute the auto-correlation function errors with
the Bootstrap method, using for each sample a random catalog $20$ times denser than the reference void catalog, composed of the 3D positions of void centers; we divide the void catalog into $27$ sub-catalogs and construct $200$ realizations by resampling from the sub-catalogs, with replacement.

Comparing the two sets of results, we observe that both void catalogs produce functions with similar shapes and redshift evolution. The negative part of the function reflects the mutual exclusion of voids: the probability of finding a pair of void centers at separations smaller than the typical void size is essentially null ($\xi_\mathrm{vv}=-1$). Void exclusion is a prominent feature of the void auto-correlation function \citep{Hamaus_2014c,Chan_2014}. Beyond this scale, the probability increases, peaking at the distance where the occurrence of another void center is most likely. The amplitude of the void auto-correlation function obtained with \vide is lower than that of \spark, as the former includes a wider range of void radii, effectively smoothing the signal by averaging on different scales.

\subsubsection{Void ellipticity}
Void shapes carry cosmological information, therefore a void statistic that could provide sensible constraints is the ellipticity of voids.
We note that this is not the ellipticity (or change in shape) caused by geometrical or dynamical distortions (that is RSD or Alcock--Paczy{\'n}ski distortions, see \Cref{sec:VGCF_in_surveys}), but the \textit{intrinsic} ellipticity of voids. The topic has received little attention from the community, in part due to the difficulty of finding a robust correspondence between observed voids and theoretical models, and in part due to the fact that dense (and large!) surveys have not been available so far. Void ellipticity is indeed a statistic that would be particularly boosted by robust measurements of void profiles from dense surveys.

The ellipticity provides information about the anisotropic matter distribution and shape of voids, and the underlying tidal field. There are two main methods to measure void ellipticity. The first consists of measuring void ellipticity through the spatial distribution of tracers belonging to voids. This quantity, usually called Eulerian ellipticity, can be inferred via the ratio of eigenvectors of the corresponding inertia tensor \citep{Lavaux_Wandelt_2010,Sutter_2015}. The other method measures the ellipticity as the ratio of the eigenvalues of the displacement field, and it is called Lagrangian ellipticity \citep{Park_Lee_2007,Lee_Park_2009,Lavaux_Wandelt_2010}. While the Eulerian ellipticity can be directly measured for voids detected with any void finder, the Lagrangian one requires the knowledge of the Lagrangian trajectories of tracers belonging to voids, therefore its measurement relies on reconstruction techniques \citep{Lavaux_Wandelt_2010}. The ellipticity distribution of cosmic voids depends on the chosen void sample (e.g., the applied radius selection), but generally peaks around $0.15$, indicating that most voids are only mildly aspherical \citep{Schuster_2023,Schuster_2024}. The distribution exhibits a tail toward higher ellipticity values, while the majority of voids presents moderate deviations from sphericity. Moreover, smaller voids tend to have higher ellipticity \citep{Leclercq_2015,Wang_2023}, a scenario coherent with voids expanding and increasing their sphericity in the process \citep{Icke_1984}.

\subsubsection{Void CMBX}
Secondary CMB anisotropies are the consequence of the interaction of CMB photons with the diffuse medium (e.g. inter-galactic and inter-cluster medium) and cosmic structures characterized by a gravitational potential. Cosmic voids leave their specific imprint on CMB as well: 
their gravitational potential is the source of void-CMB lensing \citep{Raghunathan_2020,Vielzeuf_2021,Hang_2021,Kovacs_2022c,Vielzeuf_2023}, while its time evolution causes the integrated Sachs-Wolfe \citep[ISW,][]{Sachs_Wolfe_1967} and Rees-Sciama \citep{Rees_Sciama_1968} effects \citep{Granett_2008,Nadathur_Sarkar_2011,Nadathur_2012,Nadathur_2014,Kovacs_Granett_2015,Nadathur_2016,Kovacs_2017,Kovacs_2018,Kovacs_2019,Kovacs_2022b}.
In contrast to standard methodologies developed for the galaxy-CMB cross-correlation, the void-CMB cross correlation is usually measured using a 2D stacking technique, as the imprint of voids on the CMB is an integrated effect. This consists of stacking patches of CMB-lensing convergence map for lensing and temperature for ISW---located in the same direction of voids. In this way, primary CMB fluctuations are averaged away since they do not correlate with voids, while enhancing the signal-to-noise ratio of the signal sourced by voids. Similarly to the VGCF case, in the stacking procedure the patches of CMB maps can be rescaled according to the angular void size.

\subsubsection{Void-galaxy lensing}
The effect of weak lensing from cosmic voids refers to the slight distortion of background galaxy images caused by the deflection of light rays when they pass through underdense regions of the Universe. This phenomenon is specifically referred to as anti-lensing \citep{Bolejko_2013,Chen_2015} because it causes a light distortion opposite to that induced by overdensities. Light rays tend to bend outward from voids, and the background images are de-magnified and radially distorted, rather than magnified and bent inward to form arcs around the lens, as is the case for example for galaxy clusters.
To measure this effect, a statistical approach is generally adopted, stacking the lensing signal generated by voids of similar size. These voids can be identified in 2D or 3D depending on the availability of redshift information for tracers (e.g. photometric or spectroscopic) and thus on the possibility of conducting a tomographic study. One approach involves identifying voids in signal-to-noise weak lensing maps and measuring the tangential shear profile or convergence\footnote{See \Cref{sec:theory:lensing} for details on shear and convergence.} \citep{Davies_2018,Davies_2019,Davies_2021,Maggiore_2025}, which characterizes the distortion of background galaxies. This method maximizes the signal generated by large underdense regions along the observer’s line of sight \citep{Higuchi_2013,Gruen_2016,Davies_2021,Shimasue_2024}.
Alternatively, voids can be identified in thin redshift bins using the 3D distribution of tracers or the projected mass density. The tangential shear profile is often measured, which allows us to model the 3D density profile of voids \citep{Sanchez_2017,Fang_2019,Boschetti_2024}.
Concerning its use as a cosmological probe, weak lensing from voids shows a main advantage: it is sensitive to the full gravitational potential generated by underdense regions, rather than to the tracer density contrast, so its use avoids all the complications related to the estimate of tracer bias.

\section{Theoretical modeling of void statistics}\label{sec: Theoretical models}
After introducing the various void statistics, we delve into the theoretical framework available for each statistic. This section will therefore introduce the reader to the currently available models, with details on their historical development. 

\subsection{Theory for the void size function}\label{sec: VSF}
The theoretical model for the VSF is formally similar to the halo mass function modeling, as both have been developed within the same hierarchical structure formation framework \citep{Sheth_van_de_Weygaert_2004}. To analytically model this kind of statistic, a well-consolidated methodology consists of dividing the description of the statistics in Lagrangian and Eulerian space. In this framework, the Lagrangian space is the initial density field, linearly evolved down to the epoch of interest; where initial means at high enough redshift to be fully described by linear theory. The Eulerian space is instead the fully nonlinear evolved density field. 

The core idea of this framework consists of modeling the statistic we are interested in---the VSF---in Lagrangian space, to then map the statistics in Eulerian space. The advantage of this scheme relies on the fact that the Lagrangian statistics is computed for a linear and Gaussian density field\footnote{This framework, however, can be extended to account for primordial non-Gaussianity \citep{Matarrese_2000,Matarrese_2008,Carbone_2008,Maggiore_Riotto_2010,Achitouv_Corasaniti_2012,Musso_Sheth_2014a}.}, therefore avoiding high-order correlation terms. Then, the mapping is modeled assuming that the number of halos and voids is conserved (see however \Cref{sec:theo:VSF:VSF}) and considering how individual voids (or halos) evolve from Lagrangian to Eulerian space. This approach is much simpler than considering the full non-linear computation. 

\subsubsection{Lagrangian and Eulerian voids}
As for halos, Lagrangian voids, i.e. the objects in Lagrangian space that evolve to form voids in Eulerian space, are commonly defined by thresholding: Lagrangian voids are those fluctuations that reach a specific value in their Lagrangian density contrast, called the void formation threshold. The main difference from dark-matter halos consists of the choice of the threshold value. 
For halo formation, this quantity is the linear density contrast corresponding to the full collapse in the Eulerian space\footnote{This picture can be extended considering other operators of the density field, it has however been demonstrated that the effect of these operators can be approximated with a scale dependent barrier in density only \citep{Sheth_2001,Chiueh_2001,Ohta_2004,Sheth_2013,Lazeyras_2016,Castorina_2013,Musso_Sheth_2014b}.} \citep{Bond_1991}. However, cosmic voids do not experience any event similar to the full collapse of halos in their evolution: they start expanding faster with respect to the background and continue their expansion. As long as the size and density of Eulerian voids can be mapped into the corresponding Lagrangian quantities, a one-to-one mapping between the linear and nonlinear density contrast threshold exists. As a consequence, the linear void formation threshold can be any negative value, which can be mapped into the corresponding Eulerian quantity. Different thresholds describe the distribution of voids characterized by different depths.

Throughout the review we denote Lagrangian (or linear) quantities with the subscript L. To avoid confusion, in this section only we use the subscript E to indicate Eulerian quantities. In the rest of the text no subscript is specified, since in cosmology all measured quantities are defined in Eulerian space.

\subsubsection{The excursion-set formalism}\label{subsec:excursion-set}
Focusing on the VSF in Lagrangian space, the main approaches explored in the literature are the excursion-set formalism \citep{Sheth_van_de_Weygaert_2004,Paranjape_2012a,Jennings_2013}, peak theory \citep{Sheth_van_de_Weygaert_2004}, and their combination \citep{Verza_2024a}. In these frameworks, the fundamental quantity is the smoothed, linearly evolved initial density contrast field at the Lagrangian position ${\bf q}$:
\begin{equation}\label{eq:delta definition}
\Delta({\bf q},R_{\rm L})=\int {\rm d}^3x W(|{\bf x}|,R_{\rm L}) \, \delta({\bf q} + {\bf x}) = \int \frac{{\rm d}^3k}{(2 \pi)^3} W(kR_{\rm L}) \, 
\delta({\bf k}) e^{-i {\bf k} \cdot {\bf q}}\;,
\end{equation}
where $W(kR_{\rm L})$ is the smoothing kernel in Fourier space (see details below) and $R_{\rm L}$ the smoothing length, i.e. the Lagrangian radius. 

The excursion-set framework \citep[][for a review see \citealt{Zentner_2007,Desjacques_2016}]{Bond_1991,Peacock_Heavens_1990} focuses on the evolution of the filtered density contrast field with respect to the smoothing length. The specific position ${\bf q}$ of the field is not considered, and is assumed to be random. A void with Lagrangian radius $R_{\rm L}$ is considered formed if $R_{\rm L}$ is the largest smoothing scale at which the Lagrangian density contrast field reaches the void formation threshold, without having crossed the halo formation threshold on a larger scale \citep{Sheth_van_de_Weygaert_2004}. 
Since the fluid elements inside an overdense region that will collapse to form a halo in Eulerian space are involved in nonlinear collapse, it is therefore required that the halo formation threshold on a larger scale is not crossed.
Hence, it is not relevant whether an underdense region is embedded in it at a smaller scale, it will collapse with the forming halo. This type of Lagrangian underdensity is known as ``void-in-cloud'' \citep{Sheth_van_de_Weygaert_2004}. 
The requirement to first cross the void formation threshold, i.e. the condition that $R_{\rm L}$ is the largest smoothing scale at which the field reaches the threshold, is to avoid double counting sub-voids. In the excursion-set, when a void is formed, the internal structure is not important; it may happen that at smaller scale the density contrast field crosses the void formation threshold several times, a set-up called ``void-in-void'', but only the fluctuation (the void) at larger scale is considered. It can also happen that on a smaller scale the collapse threshold is reached, i.e. the ``cloud-in-void'' case, but this does not impact the formation and evolution of a void at larger scale. In fact, halos and galaxies (that is what is usually referred to as ``clouds'') exist within cosmic voids (since these are underdense---not empty---regions).

\begin{figure}[ht]
    \centering
    \includegraphics[width=1\linewidth]{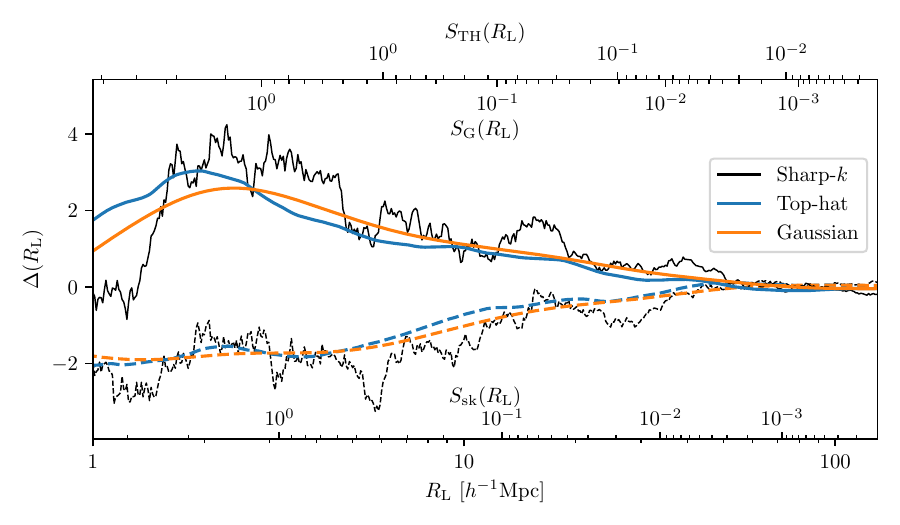}
    \caption{Two different stochastic realizations (solid, dashed) of the random walk (see Eq.~\ref{eq:langevin}), considering the sharp-k (black), top-hat (blue) and Gaussian (orange) filter. The secondary axes show the variance $S(R_{\rm L})$ evaluated with each of the filter function considered: sharp-k (sk), top-hat (TH), and Gaussian (G).}
    \label{fig: random walks}
\end{figure}

To find the first crossing distribution, one has to solve the first crossing problem of the stochastic evolution of the field given in Eq.~\eqref{eq:delta definition} with respect to the smoothing radius $R$, which is described by the Langevin equations \citep{Porciani_1998}:
\begin{equation}
\begin{cases}\label{eq:langevin}
\dfrac{\partial \Delta(R_{\rm L})}{\partial R_{\rm L}} = \displaystyle\int \dfrac{\mathrm{d}^3 k}{(2 \pi)^3} \dfrac{\partial W(k R_{\rm L})}{\partial R_{\rm L}} \delta(\mathbf{k}) e^{-i\mathbf{k \cdot q}} = {\cal Q}(R_{\rm L})\\[2ex]
\langle {\cal Q}(R_1) {\cal Q}(R_2) \rangle = \displaystyle\int_0^\infty \dfrac{\mathrm{d}k \, k^2}{2 \pi^2} P(k) \dfrac{\partial W(k R_1)}{\partial R_1} \dfrac{\partial W^*(k R_2)}{\partial R_2}\;,
\end{cases}
\end{equation}
where $P(k)$ is the linear power spectrum of matter fluctuations and ${\cal Q}$ represents the stochastic force, $R_1$ and $R_2$ are two different smoothing lengths. The filter function $W(k R_{\rm L})$ plays an important role: it is statistically related to the definition of voids (and halos) in Lagrangian space through Eq.~\eqref{eq:delta definition}, and drives the correlation between different smoothing lengths through the second equation in Eq.~\eqref{eq:langevin} \citep{Porciani_1998,Verza_2024a}. \cref{fig: random walks} shows an illustrative example of different filters (i.e. sharp-$k$, top-hat and Gaussian) applied to smooth the matter field.
If a sharp-$k$ filter is chosen---i.e. a top-hat in Fourier space, $\Theta_H(1-kR_{\rm L})$, where $\Theta_H$ is the Heaviside step function---the first crossing problem of Eq.~\eqref{eq:langevin} can be analytically solved, as the correlation among different smoothing lengths vanishes, reducing the problem to a Markovian process. An analytical solution exists for both the single \citep{Chandrasekhar_1943,Bond_1991} and double threshold \citep{Sheth_van_de_Weygaert_2004} with constant height. 

\subsubsection{The Sheth and Van de Weygaert model}
According to the void formation model described above, the Lagrangian void formation distribution is formally described by the double barrier first crossing distribution: the first barrier representing the void formation threshold, $\delta_{\rm v}$, and the second the halo formation one, $\delta_{\rm c}$. We note that, although in this review we adopt the convention of denoting the integrated density contrast with $\Delta$, we report these thresholds using the symbol $\delta$ to remain consistent with the literature. Common values that can be used for these thresholds are $\delta_{\rm v}=-2.71$, corresponding to the shell-crossing\footnote{In the spherical model of an initial underdensity, represented as concentric shells, the inner shells expand faster than the outer ones. As a result, they eventually overtake the surrounding shells, a process known as shell-crossing. While shell-crossing was historically considered the standard threshold for void formation \citep{Sheth_van_de_Weygaert_2004,Jennings_2013}, alternative choices can be employed to characterize void populations of different depths. See \citet{Moretti_2025} for a thorough discussion, as well as \citet{Massara_Sheth_2018,Contarini_2022}.} for voids, and $\delta_{\rm c}=1.69$, the critical value for the collapse of overdensities, both derived in an Einstein–de Sitter universe.

\citet{Sheth_van_de_Weygaert_2004} solved the corresponding first crossing distribution, also known as multiplicity function, by considering a sharp-$k$ filter in Eq.~\eqref{eq:langevin}. This allows the derivation of an analytical expression for the multiplicity function:
\begin{equation}\label{eq:SVdW}
f(S) \, {\rm d} S = \frac{1}{S}\sum_{i=1}^\infty e^{-\frac{(i \pi x)^2}{2}} i \pi x ^2\sin (i \pi {\cal D})  \, {\rm d} S\;,
\end{equation}
where ${\cal D} = |\delta_{\rm v}| / (\delta_{\rm c}-|\delta_{\rm v}|)$, $x = \sigma {\cal D} / |\delta_{\rm v}|$ and
\begin{equation}\label{eq:sigma2}
S(R_{\rm L}) = \sigma^2(R_{\rm L}) = \langle \Delta^2(R_{\rm L}) \rangle = \int \frac{{\rm d} k \, k^2}{2 \pi^2} P(k) W(kR_{\rm L})^2
\end{equation}
is the variance of the linear density contrast filtered on a scale $R_{\rm L}$. The series appearing in Eq.~\eqref{eq:SVdW} is rapidly converging, therefore, for practical applications a number of iterations between one and four are enough, depending on the threshold value and on the considered scale \citep{Jennings_2013}. The void-in-cloud effect, modeled by the presence of the collapsing barrier, impacts the population of small voids, while at large scale it converges to the well-known multiplicity function with a single barrier \citep{Bond_1991}. 

\subsubsection{The void size function}\label{sec:theo:VSF:VSF}
The multiplicity function describes the fraction of fluctuations with scale $R_{\rm L}$ that become voids in Lagrangian space. Therefore, to obtain the corresponding Lagrangian VSF we need to multiply the multiplicity function by the number density of Lagrangian fluctuations at each scale $R_{\rm L}$,
\begin{equation}\label{eq:VSF def}
\frac{{\rm d} n(R_{\rm L})}{{\rm d} R_{\rm L}} = \frac{3}{4 \pi R_{\rm L}^3} f(S) \frac{{\rm d} S}{{\rm d} R_{\rm L}}\;,
\end{equation}
where we use the relation $f(S) {\rm d} S = f(R_{\rm L}) {\rm d} R_{\rm L}$ to change variables, as the multiplicity function is a differential quantity. We note that the $3/(4 \pi R_{\rm L}^3)$ term appearing in Eq.~\eqref{eq:VSF def} is the inverse of the volume of the sphere of radius $R_{\rm L}$, i.e. the volume of the top-hat filter. In principle, this quantity should be the inverse of the volume enclosed in the smoothing filter $W(kR_{\rm L})$ considered. In the literature, however, the volume of the sphere is always used (together with other approximated methods) to map the specific filter to top-hat voids, as discussed in the following. By inserting Eq.~\eqref{eq:SVdW} in Eq.~\eqref{eq:VSF def}, we obtain the \citet{Sheth_van_de_Weygaert_2004} Lagrangian VSF. 

The final step consists of obtaining the Eulerian multiplicity function from the Lagrangian formulation (Eq.~\ref{eq:VSF def}). This is performed by considering how voids evolve. Assuming mass conservation, the Eulerian void radius\footnote{In the general case this relation is true for the effective radius, as the quantity which is conserved is the mass, not the shape. \citep{Sheth_van_de_Weygaert_2004}} is $R_{\rm E} = (1+\delta_{\rm v}^{\rm E})^{-1/3} R_{\rm L}$ (see Eq. (11) in \citealp[][]{Jennings_2013}). Assuming that the number of voids is conserved from Lagrangian to Eulerian space, we obtain:
\begin{equation}\label{eq:eulerian_VSF}
\frac{{\rm d} n(R_{\rm E})}{{\rm d}R_{\rm E}} = \frac{3 f_{\rm E}(R_{\rm E}) }{4 \pi R_{\rm L}^3(R_{\rm E})}\;.
\end{equation}
where $f_{\rm E}(R_{\rm E}) = f[R_{\rm L}(R_{\rm E})] \, {\rm d} R_{\rm L} / {\rm d} R_{\rm E} = (1+\delta_{\rm v}^{\rm E})^{1/3}f[R_{\rm L}(R_{\rm E})]$ \citep{Verza_2024a}. 

The standard VSF model presented here provides a first approximation in modeling the distribution of void sizes, however, it presents some theoretical issues. The first concerns how voids are defined in Lagrangian space, through Eq.~\eqref{eq:delta definition}. The \citet{Sheth_van_de_Weygaert_2004} model relies on the sharp-$k$ filter for voids definition, for which, however, the physical interpretation is not clear. In fact, in configuration space, the sharp-$k$ filter is an oscillating function that takes both positive and negative values. As a consequence, the amount of mass enclosed in the filter is ill-defined, as well as the interpretation of the corresponding void radius. To overcome this issue, it is common to use the multiplicity function of Eq.~\eqref{eq:SVdW}, but with the variance $\sigma^2$ (Eq.~\ref{eq:sigma2}) computed using the top-hat filter instead of the sharp-$k$ one. In this way, the void radius is implicitly defined as $R_{\rm L}$ such that $\sigma_\mathrm{TH}^2(R_{\rm L}) = \sigma_{\rm sk}^2(k)$, where the subscript TH denotes the top-hat filter while sk the sharp-$k$ one \citep{Porciani_1998}. 

However, \citet{Jennings_2013} noted that spherically mapping the \citet{Sheth_van_de_Weygaert_2004} model in Eulerian space (that is Eq.~\eqref{eq:eulerian_VSF} with Eq.~\eqref{eq:SVdW}) leads to problems in the normalization of the VSF, resulting in the total volume fraction of Eulerian voids being greater than 1.
In other words, it predicts that the sum of all void volumes is larger than the volume of the entire universe. To solve this problem, \citet{Jennings_2013} propose the so-called ``volume-conserving model'' (Vdn hereafter). Instead of assuming that voids are conserved from Lagrangian to Eulerian space, this model imposes that the volume fraction of voids is conserved from the Lagrangian to Eulerian space. This solves the normalization problems from a mathematical perspective, but at the expenses of the one-to-one mapping of Lagrangian to Eulerian voids, on which the excursion-set framework is based. However, it has been shown that the normalization problem does not appear when a physically motivated filter function is consistently considered, together with a better definition of the multiplicity function of voids in Lagrangian space \citep{Verza_2024a}, as described in the next paragraph.

\subsubsection{Beyond the excursion-set: peak theory and the effective barrier}
The excursion-set model is not the only framework in which it is possible to theoretically model the VSF. \citet{Sheth_van_de_Weygaert_2004} also proposed a Lagrangian VSF in the context of peak theory \citep[][see \citealt{Desjacques_2016} for a review]{Bardeen_1986}. As in the case of the excursion-set, the fundamental quantity is the filtered density contrast field (Eq.~\ref{eq:delta definition}), but instead of considering the evolution of the field with respect to the smoothing length it focuses on the spatial distribution. 

The basic idea consists of studying the distribution of the local extrema of the field that reaches the void formation threshold, by considering the value of the field, its corresponding gradient and Hessian. From this it is possible to obtain the number density of the extrema at a given smoothing length $R_{\rm L}$. However, converting this quantity into a multiplicity function is far from being straightforward, due to the spatial dependency of extrema when varying the smoothing scale, void-in-void problems, and other mathematical difficulties \citep{Paranjape_2012b}. In fact, even in the excursion-set formalism the physical interpretation of the multiplicity function requires a careful treatment. It is obtained by solving the first crossing problem over random field positions ${\bf q}$, and it is therefore weighted over the entire Lagrangian space. As a consequence, the corresponding multiplicity function describes the fraction of volume where the density contrast field first reaches the threshold value. Relating this quantity to a number density of objects such as voids and halos is not straightforward.

To solve these problems, it has been proposed to merge the excursion-set with peak theory, by performing the excursion-set over the subset of Lagrangian extrema, and this approach is valid to describe the multiplicity function of both halos and voids \citep{Verza_2024a}. The precise definition of a Lagrangian void follows accordingly. A Lagrangian void with radius $R_{\rm L}$ corresponds to the filtered density contrast $\Delta({\bf q},R_{\rm L})$ satisfying the following conditions: \textit{i}) the Lagrangian position ${\bf q}(R_{\rm L})$ is not contained in any larger void; \textit{ii}) the Lagrangian position ${\bf q}(R_{\rm L})$ is a minimum of the density field filtered on the scale $R_{\rm L}$; \textit{iii}) $R_{\rm L}$ is the largest scale at which $\Delta({\bf q},R_{\rm L})$ crosses the void formation threshold.
Both the void-in-void and peak-in-peak (or trough-in-trough) conditions are now satisfied. 

Given the high dimensionality of the problem, deriving the corresponding multiplicity function by directly solving the corresponding Langevin equations can be extremely challenging. It has been shown, however, that the excursion-peak multiplicity function can be well approximated in the standard excursion-set framework by using an effective moving barrier. This effective barrier does not correspond to the physical void formation threshold; however, it is a function of it, and it contains the statistical information related to the fact that we are now solving the first crossing problem over the subset of extrema. The moving barrier is a scale-dependent threshold, as the one developed in the halo mass function context \citep{Sheth_2001,Sheth_Tormen_2002}. This can be evaluated by numerically solving the first crossing problem directly using a Gaussian realization of the density contrast field, representing the Lagrangian space \citep{Verza_2024a}. The exact multiplicity function, properly accounting for the filter function, can be computed by numerically solving Eq.~\eqref{eq:langevin}, or by using approximated analytical formulations, such as:
\begin{align}
f(S) &= \frac{e^{-B^2_S / 2S}}{\sqrt{2 \pi S}}  \left\llbracket  \sqrt{\frac{\Gamma_{\delta \delta}}{2 \pi S}} \exp \left[ -\frac{S}{2\Gamma_{\delta \delta}} \left(\frac{B_S}{2S} - B'_S \right)^2\right]\right. +  \\
& \left. \frac{1}{2} \left(\frac{B_S}{2S} - B'_S \right) \left\{{\rm erf}\left[\sqrt{\frac{S}{2\Gamma_{\delta \delta}}} \left(\frac{B_S}{2S} - B'_S \right)\right]+1\right\} \right\rrbracket , \nonumber
\end{align}
where $\Gamma_{\delta \delta} = SD_S - 1/4$, $D_S=\langle ({\rm d} \Delta(S) / {\rm d}S)^2 \rangle$, while $B_S=B(S,\delta_{\rm v})$ is the moving barrier and $B'_S = {\rm d} B(S) / {\rm d} S$.
The corresponding Lagrangian VSF follows from Eq.~\eqref{eq:VSF def}. This model, together with solving the problems of the \citet{Sheth_van_de_Weygaert_2004} and \citet{Jennings_2013} models, shows that the spherical approximation is already sufficient to map voids of intermediate and large size at $z=0$ in Eulerian space (see Eq.~\ref{eq:eulerian_VSF}). The higher the redshift, the more reliable the spherical approximation is, while for small voids tidal effects are not negligible, requiring to go beyond the spherical mapping for an accurate modeling \citep{Verza_2024a}.

\subsubsection{Final remarks}

Although the theoretical VSF is uniquely defined and effectively captures the number density of voids as a function of their size in both simulations and observations, there are some caveats to be considered. Firstly, each void-identification algorithm relies on a particular definition of voids; the adopted definition, which varies from one void finder to another, leads therefore to a different VSF. For this reason, either the measured VSF or the corresponding theoretical modeling may need to be adjusted for the specific case (see \Cref{sec:obs_VSF}). This issue is well known in the context of dark matter halos as well as for each specific halo finder and definition the theoretical model needs to be slightly modified from the pure theoretical excursion-set model~\citep{Sheth_Tormen_1999,Tinker_2008,Watson_2013,Bocquet_2016,Despali_2016,Comparat_2017,Diemer_2020,Seppi_2021}. Secondly, the theoretical VSF describes voids in the matter distribution. In the real universe, however, we observe voids in the distribution of biased tracers (e.g. galaxies, clusters). 
The effect of bias has to be taken into account to recover the observed VSF \citep{Furlanetto_Piran_2006,Pollina_2017,Contarini_2019}. Together with tracer bias, other systematic effects impact the observed VSF. Examples include the effect of RSD, that will modify voids in observed space and therefore affect the VSF, or the impact of tracer sparseness.
We discuss in \Cref{sec:obs_VSF} how to deal with such issues, even if alternative approaches relying on emulators have recently been explored \citep[see e.g.][]{Lehman_2025,Salcedo_2025}, as properly accounting for all effects may prove challenging. As a final remark, we highlight that the tools publicly available for computing the theoretical VSF and for constructing catalogs consistent with the corresponding void definitions are summarized in Appendix~\ref{app:A}.

\subsection{Theory for the void auto- and cross-correlation function}\label{Subsection: VGCF Theory}
Void clustering refers to the correlations of void centers either among themselves or with other tracers, studied in configuration or Fourier space. The main quantities explored in the void literature so far are the void auto-correlation function and the void-tracer correlation function (corresponding to the so-called VGCF when the mass tracers are galaxies). The methodology developed to model void clustering relies on the existing framework for galaxy clustering, with some peculiarities. Exactly as for the galaxy clustering case, a fully predictive void clustering model does not exist to date. Nevertheless, theoretical models involving free parameters to be fitted against simulations or real data, or accurate semi-analytic models, have been developed in the last decade; we describe these in the following sections. 

As previously mentioned, the VGCF represents the void density profile in the galaxy field. While some work aims at modeling the VGCF from first principles, it is hard to obtain a match between theoretical predictions and observed void density profiles, due to the rich number of effects that impact void profiles in observations. We note, indeed, that the \textit{observed} VGCF is affected by RSD and other geometrical effects, introducing anisotropies along the line of sight. For this reason, the observed VGCF is often decomposed into multipoles (see below, Eq.~\ref{eq: multipoles}), which is particularly useful for the cosmological interpretation of such effects. In this section, we do not consider observational effects, focusing instead on the modeling in real-space. 
Later in this review (see \Cref{sec:VGCF_in_surveys}), we will discuss how observational distortions---arising from tracer peculiar velocities (RSD, i.e. dynamical distortions) and from the use of a fiducial cosmology to convert redshifts into distances (geometrical distortions)---affect the VGCF. As we will discuss in \Cref{sec:VGCF_in_surveys}, while these effects must be accurately modeled, they also represent an additional source of cosmological information.

\subsubsection{The void model}
The void model \citep{Hamaus_2014b} was developed in analogy to the halo model \citep{Cooray_Sheth_2002}, and was later extended to the halo-void and halo-void-dust models \citep{Voivodic_2020}. The key idea consists of considering the entire matter (galaxies) in the universe as belonging to voids. This allows us to split the correlation function of matter (galaxies) in terms that include the correlation of matter (galaxies) inside the same void, or belonging to different voids. Let us consider the void-galaxy power spectrum ($P_\mathrm{v g}$) as an example. In this framework, it can be divided into two parts, the one- and the two-void terms:
\begin{equation}\label{eq:Pvg}
P_\mathrm{v g} = P_\mathrm{v g}^{(1 {\cal V})} + P_\mathrm{v g}^{(2 {\cal V})}\;.
\end{equation}
The one-void term reads:
\begin{equation}
P_\mathrm{v g}^{(1 {\cal V})} = \frac{1}{\bar{n}_{\rm v}\bar{n}_{\rm g}} \int \frac{{\rm d} n_{\rm v}(r_{\rm v})}{{\rm d} r_{\rm v}} N_{\rm g}(r_{\rm v}) u_{\rm v}(k|r_{\rm v}) {\rm d}r_{\rm v}
\end{equation}
and describes the correlation among the $N_{\rm g}$ galaxies inside any void with radius $r_{\rm v}$ and the void center; $u_{\rm v}(k|r_{\rm v})$ is the normalized density profile for voids of radius $r_{\rm v}$ in Fourier space, satisfying  $u_{\rm v}(k|r_{\rm v})\rightarrow 1$ for $k \rightarrow 0$. The two-void term describes the correlation of void centers and galaxies (matter) belonging to distinct voids:
\begin{equation}\label{eq:void_model_vg}
P_\mathrm{v g}^{(2 {\cal V})} = \frac{1}{\bar{n}_{\rm v}\bar{n}_{\rm g}} \int \frac{{\rm d} n_{\rm v}(r_{\rm v})}{{\rm d} r_{\rm v}} \frac{{\rm d} n_{\rm g}(m_{\rm g})}{{\rm d} m_{\rm g}} b_{\rm v}(r_{\rm v}) b_{\rm g}(m_{\rm g}) u_{\rm v}(k|r_{\rm v}) P_{\rm mm}(k) {\rm d} r_{\rm v} {\rm d} m_{\rm g}\;,
\end{equation}
where $n_{\rm v}$ and $n_{\rm g}$ are the number density of voids of radius $r_{\rm v}$ and galaxies of mass $m_{\rm g}$, $b_{\rm v}$ and $b_{\rm g}$ are the void and galaxy radius- and mass-dependent bias respectively\footnote{These bias terms correspond to the linear large-scale structure terms, therefore they do not depend on the scale $k$, but only on the void radius or galaxy mass (and other properties). The scale dependence in this formulation is encapsulated in the function $u_v(k)$ of Eq.~\eqref{eq: bias1}.}.
Considering a narrow range in the radius and mass distribution, Eq.~\eqref{eq:Pvg} becomes:
\begin{equation}
P_\mathrm{v g} \simeq b_{\rm v} b_{\rm g} u_{\rm v}(k) P_{\rm mm}(k) + \bar{n}_{\rm v}^{-1} u_{\rm v}(k)\;,\label{eq: bias1}
\end{equation}
where the density profile $u_{\rm v}(k)$ shapes the cross-spectrum at scales of the order of the void size. The same arguments can be applied for the auto-power spectra of both voids and galaxies, obtaining: 
\begin{equation}
\begin{split}
P_{\rm vv} &\simeq b_{\rm v}^2 u_{\rm v}^2(k) P_{\rm mm}(k) + \bar{n}_{\rm v}^{-1} \;, \\
P_{\rm gg} &\simeq b_{\rm g}^2 P_{\rm mm}(k) + \bar{n}_{\rm g}^{-1} \;,  
\end{split}
\end{equation}
where $\bar{n}_{\rm v}^{-1}$ and $\bar{n}_{\rm g}^{-1}$ correspond to the shot noise terms for voids and galaxies, respectively. In the void-void power spectrum, the density profile acts as an exclusion effect at low $k$, while at large scale, where $u_{\rm v}(k) \rightarrow 1$, we recover the standard auto-correlation function $b_{\rm v}^2 P_{\rm mm}(k)+ \bar{n}_{\rm v}^{-1}$.
This framework has been further extended by explicitly dividing the distribution of matter (galaxies) as belonging to either voids or halos, and also in dust, i.e. not belonging neither to halos or voids \citep{Voivodic_2020}.
One crucial quantity in the void model is the void matter-density profile, either in configuration or in Fourier space. A predictive theoretical model that matches measured voids would be extremely helpful but unfortunately it does not currently exist. In the absence of such a model, we will describe existing attempts to model profiles by fitting against simulations, or directly inferring them from data \citep{Hamaus_2014b,Chan_2014,Voivodic_2020}.
A different option to model the density profile originates from the fact that it can be normalized in such a way that:
\begin{equation}
\frac{\delta \rho_{\rm m}}{\bar{\rho}_{\rm m}} u_{\rm v}(k) = b_{\rm v}(k)\;.
\end{equation}
The void bias $b_{\rm v}$ can be interpreted as a scale-dependent void bias $b_{\rm v}(k) = b_{\rm v} u_{\rm v}(k) $, with $b_{\rm v}$ the large-scale void bias, and the function $u_{\rm v}(k)$ encapsulating the scale dependence. 
Assuming that the galaxy bias weakly depends on the void environment, and given that $\rho_{\rm m} \propto n_{\rm g}b_{\rm g}$, the above equation can be rewritten as:
\begin{equation}
\frac{\delta n_{\rm g}}{\bar{n}_{\rm g}} u_{\rm v}(k) \simeq \frac{b_{\rm v}(k)}{b_{\rm g}}\;,
\end{equation}
entailing that the scale-dependent void bias is observable up to the normalization constant given by the galaxy bias.

\subsubsection{Renormalized bias expansion}
Another possibility to model void clustering consists of bias expanding the loop-expanded matter power spectrum. Here the bias coefficients are free parameters of the model to fit against data. This methodology has been explored considering contributions up to the third order in the local bias expansion and 1-loop correction of the matter power spectrum \citep{Chan_2014}. In this approach, void exclusion plays a crucial role in modeling the void auto-power spectrum, and can be considered as an extra-term in the expansion, $P_{\rm excl}$.  
From these arguments, the void auto-power spectrum can be expanded as: 
\begin{equation}
P_{\rm vv} = P_{1,1} + P_{2,11}, P_{11,11} + P_{\rm excl} \;,
\end{equation}
where $P_{1,1}=b_1^2 P_L$, $P_L$ is the linear matter power spectrum, and the 1-loop corrections are:
\begin{equation}
\begin{split}
P_{2,11} &= 2 b_1 b_2 \int {\rm d}^3q F_2({\bf q},{\bf k-q}) P_L(|{\bf k-q}|)\;, \\ 
P_{11,11} &= \frac{b_2^2}{2}\int {\rm d}^3q P_L(q) P_L(|{\bf k-q}|)\;,
\end{split}
\end{equation}
where $F_2({\bf p},{\bf q})$ is the mode coupling kernel \citep{Chan_Scoccimarro_2012}. The exclusion term has been modeled using various ansatzes for the transition from the exclusion zone to the outer region \citep{Chan_2014}.

\subsubsection{Peak theory}
The theory of Lagrangian density peaks \citep{Bardeen_1986} provides an additional framework for theoretically modeling the void density profile. Following the original work of \citet{Bardeen_1986}, the mean density profile around extrema, i.e. the peak-matter correlation function, can be modeled as \citep{Nadathur_2012}: 
\begin{equation}\label{eq:xi_vm}
\xi_{\rm vm}(r) = \frac{1}{\sigma (R_{\rm L})} \int \frac{k^2}{2 \pi^2} \frac{\sin (kr)}{kr} W^2(kR_{\rm L}) P_L(k) \left[ \frac{\nu-\nu \gamma^2 - \gamma \Theta}{1-\gamma^2} + \frac{\Theta R_*^2 k^2}{3\gamma(1-\gamma^2)} \right] {\rm d} k\;.
\end{equation}
In the above equation, $\nu=\delta_{\rm v}/\sigma(R_{\rm L})$ where $\delta_{\rm v}$ is the density contrast of the minimum at the smoothing scale $R_{\rm L}$, $W(kR_{\rm L})$ is the smoothing kernel as defined in \Cref{sec: VSF}, ${\gamma=\sigma_1^2/\sigma_2\sigma_0}$ and $R_*=\sqrt{3}\sigma_1/\sigma_2$ where $\sigma_n^2=\int {\rm d}^3kW(kR_{\rm L})P_L(k)k^{2n}$ is the $n-$spectral moment. By construction, this formula directly applies to voids defined as minima at a specific smoothing length $R_{\rm L}$, with density contrast in the minimum equal to $\delta_{\rm v}$ when smoothed on the scale $R_{\rm L}$. It is important to note, however, that the peak theory framework in which Eq.~\ref{eq:xi_vm} has been derived is in Lagrangian space; therefore to apply this model to data or simulations it is necessary to consider a mapping similar to the one discussed in \Cref{sec:theo:VSF:VSF}. 
To conclude, this model is an approximation in the proximity of the minima and breaks down in the outer part of void regions. For this reason it has been explored in the context of ISW in voids, as the ISW signal peaks around minima \citep{Nadathur_2012}.

Beyond the explicit profile formulation, peak theory suggests a specific void profile shape, i.e. a corresponding void scale-dependent bias, which behaves as $k^2$. This idea has been exploited to implement a parametric function for the scale dependent bias in the void model, providing good fit to the simulated void auto- and cross-matter power spectra \citep{Chan_2014}.

\subsubsection{Excursion-set and excursion-peak}
The excursion-set approach provides another framework to derive the theoretical void matter-density profile. As discussed in \Cref{sec: VSF}, voids in the excursion-set formalism are modeled as fluctuations in the Lagrangian matter density field, first crossing the void formation threshold. Therefore, the stochastic trajectory of the filtered density contrast field forming a void represents a stochastic realization of a void profile \citep{Achitouv_2016,Verza_2024a}. As a consequence, having the stochastic differential equation describing the trajectory evolution of fluctuations forming voids can provide a theoretical model for the void matter-density profile. In the standard excursion-set model however, the information on the position of the field is not available, and therefore a direct application of the Langevin equations (Eq.~\ref{eq:langevin}) is not meaningful. The excursion-peak theory solves this problem, and the effective barrier approach provides a possible method to perform computations. The effective moving barrier provides the statistical information required for the multiplicity function, and can be extended to model the density profile. The basic idea consists of solving the first crossing problem with the Langevin equations and the effective moving barrier with Monte-Carlo methods, to obtain the distribution of trajectories forming voids. Since the effective barrier is not physical, the derived density profile should be rescaled by the barrier values. By construction, this method is exact for regions within the void radius, and approximate in the outer part, providing a good match with simulations, for both the mean value and the distribution around it \citep{Verza_2024a}. The profile obtained in this way is for Lagrangian voids, and has to be mapped in Eulerian space to compare it with observations.

\subsubsection{Void density profile evolution}
The evolution of the void density profile has been explored in several works, as it connects the void matter-density profile in Lagrangian space (i.e. the linear profile) with voids at any redshift and cosmology \citep{Blumenthal_1992,Sheth_van_de_Weygaert_2004,Jennings_2013,Demchenko_2016,Massara_Sheth_2018,Stopyra_2021,Moretti_2025,Schuster_2025}.
\citet{Demchenko_2016} tested the spherical evolution model using spherical voids identified at ${z=0}$ in the halo field, comparing their matter density profiles with those measured at different redshifts while keeping the same void centers. Around the void radius, the agreement between model and simulations is usually good, however, the overall profile slope is not perfectly reproduced. 
\citet{Massara_Sheth_2018} considered the evolution of  matter-density profiles for topological voids. They extended the spherical model by considering that void centers move during they evolution, and also improved the void profile modeling, based on the excursion peak. By considering the profile evolution around a moving center, the agreement between theory and simulations increases.
To go beyond the spherical evolution, \citet{Stopyra_2021} additionally explored the accuracy of the Zel'dovich approximation \citep{Zel'dovich_1970}, i.e. the first-order Lagrangian perturbation theory, finding that it can provide an accurate prediction for the matter-density profile of voids. We finally emphasize that, when comparing the theoretical evolution of density profiles with results from numerical simulations, spatial resolution effects play an important role, particularly in the innermost regions of voids. Moreover, the evolution of voids identified in the distribution of biased tracers must account for the impact of tracer bias and its time evolution, as it will be discussed in the following.

\subsubsection{Biasing the void density profile}
A standard way to connect matter- and galaxy-density profiles of voids consists of modeling the galaxy (halo) bias in void profiles. This is useful to connect the void-matter cross-correlation function with the corresponding one in the tracer distribution, and {\it vice versa}. Many studies focused on the bias of halos or galaxies in voids \citep{Furlanetto_Piran_2006,Neyrinck_2014,Pollina_2017,Pollina_2019,Contarini_2019,Massara_2022,Verza_2022,Montero-Dorta_2025}, however in this section we focus on the void-tracers correlation bias. A simple and robust model consists of using a linear relation between the void-tracer and void-matter correlation function \citep{Pollina_2017,Pollina_2019}. A semi-analytical bias expression can be obtained by using the theoretical conditional mass function \citep{Furlanetto_Piran_2006,Neyrinck_2014}. This can be obtained by measuring the halo mass function along the void density profile \citep{Verza_2022}. Interestingly, comparisons with simulations show that the halo mass function in voids is less evolved with respect to the mean halo distribution in the universe, leading to a scale-dependent bias in the VGCF \citep{Furlanetto_Piran_2006,Contarini_2019,Schuster_2019,Massara_2022,Verza_2022}. This is in agreement with astronomical observations showing that the galaxy population in voids is less evolved than in filaments and clusters \citep{Bolzonella_2010,Ricciardelli_2014,Liu_2015,Davidzon_2016,Kreckel_2016,Habouzit_2020,Dominguez_2023,Curtis_2024,Rodriguez-Medrano_2024,Argudo-Fernandez_2024,Perez_2025,Ceccarelli_2025}.

\subsubsection{Alternatives to a pure theoretical model}\label{subsec:density_profile_fit}
Due to the difficulty in modeling void profiles from first principles, different solutions have been proposed. To model the redshift-space VGCF and extract cosmological information, a model or prescription for the void profile in real space is necessary, and four options have been used so far.

Firstly, it is possible to use simulations to get the real-space density profile of a void stack. The obvious issue with this method is that a simulation or mock has a well defined cosmology, therefore the void real-space profile obtained with this method would, if used to extract constraints, introduce some cosmology dependence. 

Secondly, one can use a fitting function characterized by coefficients that are left free when fitting the data. A particularly successful recipe has been suggested by \citet{Hamaus_2014b}, commonly referred to as the Hamaus–Sutter–Wandelt (HSW) profile, which reads:
\begin{equation}\label{eq:HSW}
 \frac{\rho(r)}{\bar{\rho}} - 1 = \delta_c\,\frac{1-(r/r_s)^\alpha}{1+(r/r_\mathrm{v} )^\beta}\;. 
\end{equation}
One of the advantages of this prescription is that the parameters have an intuitive meaning: $\delta_c$ corresponds to the central density contrast, $r_s$ to a scale radius at which $\rho_\mathrm{v}=\bar{\rho}$, and $\alpha$ and $\beta$ determine the inner and outer slope of the void’s compensation wall, respectively. Indications of universality\footnote{By universality, \citet{Hamaus_2014b} refers to the capability to describe different kind of voids (including, but not limited to, voids found in different tracer fields such as halos or galaxies). A further universality would include the capability of this function to fit voids in other cosmologies, this has been explored in e.g. \citet{Zivick_2015}.} are present for this recipe, since some parameters can be expressed as a function of the others.
This recipe has inspired the use of analogous functions to parametrize the profile \citep{Barreira_2015, Nadathur_2015c}. 

\citet{Paz_2013} also proposed an analytical function for the void density profile based on the error function, which has the advantage of being a simple differential form, useful for many applications. Its most simple form reads:
\begin{equation}
    \Delta(r)=\frac{1}{2} \bigg[ {\rm erf}\bigg(S \, {\rm log(r/R_{\rm v})}\bigg) -1 \bigg]\;,
\end{equation}
where $R_{\rm v}$ is the void radius and $S$ is called \textit{steepness} coefficient. This functional form can be extended to model integrated density profiles of voids with pronounced compensation walls by adding an exponential term, which introduces an additional coefficient controlling the peak of the profile.

Beyond the \citet{Hamaus_2014b} and \citet{Paz_2013} formulas, \citet{Voivodic_2020} proposed a fitting formula with a different functional shape, typically used for spherical void finders:
\begin{equation}\label{eq:voivodic}
\frac{\rho(r|R_{\rm v})}{\bar{\rho}} = \frac{1}{2}\left[1+  \tanh \left( \frac{y-y_0}{s(R_{\rm v})}\right) \right]\;,
\end{equation}
where $y=\ln (r/R_{\rm v})$ and $y_0=\ln (r_0/r_{\rm v})$.
This formula involves three parameters: $r_{\rm v}$, $r_0$, and $s$. In the original work, $r_{\rm v}$ is the void radius, and $r_0$ is fixed by requiring that the profile integral up to $r_{\rm v}$ is equal to the threshold value used to identify voids. The only remaining free parameter is therefore $s$, which depends on the void radius, $r_{\rm v}$. This parameterization considers that voids are empty in the center, $\rho(0|r_{\rm v})=0$. Despite their simplicity, fitting functions have the drawback of introducing extra parameters that need to be marginalized over, which in turn degrades the constraining power.

Thirdly, it is possible to obtain the real-space density profile of voids directly from data with a de-projection technique \citep{Pisani_2014, Hawken_2017}. 
This technique has also been used to obtain the profile of matter in voids from lensing measurements \citep{Sanchez_2017,Fang_2019,Hunter_2025}, it has the considerable advantage that it does not rely on simulations. The method exploits the idea that the Alcock-Paczy{\'n}ski test and RSD will distort voids along the line of sight (see \Cref{sec:VGCF_in_surveys}). Therefore the projection of the void stack along the line of sight does not depend on such distortions. It is possible to de-project the projected profile to obtain the real-space spherically symmetric density profile, relying on the Abel inverse transform. The problem is, however, ill-conditioned, and therefore needs large numbers, otherwise it can become noise dominated. Interest has rapidly grown for the latter methodology, given that it performs well when enough voids are available, as is the case for current survey data (see \Cref{sec:VGCF_in_surveys}).

Finally, a fourth possibility consists of modeling the void density profile using emulators based on cosmological simulations, as recently explored by \citep{Fraser_2024,Lehman_2025}. This methodology allows us to model void profiles with high accuracy, without free parameters. The disadvantage is that emulators depend on a large number of expensive simulations and they are not as flexible as the analytical fitting functions presented above. They have a strong dependence on the cosmological models, on the parameters characterizing the simulations used for calibration, and on the method and parameters used to model the galaxy population.

\subsubsection{Final remarks}
To conclude this section, we note that the VGCF is already used successfully in survey analyses, combining the prescriptions introduced in \Cref{subsec:density_profile_fit} with models that account for dynamical and geometrical distortions around voids. Its application to galaxy surveys will be discussed in detail later in this review (\Cref{sec:VGCF_in_surveys}). Nonetheless, it is worth highlighting that some caveats remain. In particular, the theoretical modeling of void auto- and cross-correlation functions is still being tested against numerical simulations. This is because a fully theoretical framework that incorporates observational effects and achieves unbiased constraints on all relevant cosmological parameters has yet to be achieved. Moreover, as mentioned earlier in this review, no cosmological constraints currently exist for the void auto-correlation function, as it has rarely been measured in galaxy surveys due to its limited statistical significance. Its use will likely increase in the coming years thanks to data from modern surveys that cover larger volumes. Finally, we refer the interested reader to Appendix~\ref{app:A}, which provides a comprehensive list of publicly available tools for measuring and analyzing void auto- and cross-correlation functions.

\subsection{Theory for void velocity profiles}\label{sec:theory_velocity} 
The radial velocity field around cosmic voids can be modeled by imposing mass conservation through spherical shells centered on the void. To linear order in the density contrast, this can be written as:
\begin{equation}
\nabla \cdot \mathbf{v} (\mathbf{x}, z) = - f(z) \frac{H(z)}{1+z} \delta(\mathbf{x}, z)\;,
\end{equation}
where $\mathbf{v}$ is the peculiar velocity field of a pressure-less fluid, $H(z)$ is the Hubble parameter, and $f(z)$ is the linear growth rate, defined as the logarithmic derivative of the linear growth factor $D(z)$, which we define below \citep[see e.g.][]{Dodelson_2003}:
\begin{equation}\label{eq:hubble_rate}
H(z) = H_0 \left[ \Omega_{\rm m} (1+z)^3 + \Omega_\Lambda \right]\;,
\end{equation}
\begin{equation}\label{eq:growth_rate_factor}
f(z) = -\frac{\mathrm{d}\ln D(z)}{\mathrm{d}\ln(1+z)} \quad \text{and} \quad D(z) = \frac{5 \Omega_{\rm m} H(z)}{2} H_0^2 \int^\infty_z \frac{1+z'}{H^3(z')} {\rm d}z'\;.
\end{equation}
In the expression for the Hubble parameter, we assume a standard $\Lambda$CDM model and neglect the contributions from curvature and radiation.

Integrating over the volume of a sphere with radius $r$ and applying the divergence theorem to convert the volume integral to a surface one, we obtain a relation between the radial velocity field and the average density contrast within the sphere \citep{Peebles_1980}:
\begin{equation}\label{eq:radial_velocity}
    v(r,z) = - \frac{1}{3} \, f(z) \, \frac{H(z)}{1+z} \, r \, \Delta(r,z) \quad \text{where} \quad \Delta(r,z) \equiv \frac{3}{4 \pi r^3} \int_{V_\mathrm{s}} \delta(\mathbf{x},t) \, \mathrm{d}^3x\;.
\end{equation}
As already mentioned, this prescription is used to model the RSD around cosmic voids in the context of different void summary statistics, such as the VSF and the VGCF.
The analysis of velocity profiles around voids provides a powerful cosmological probe also on his own, but its practical implementation is hindered by challenges in accurately measuring tracer velocities \citep[see, however,][for pipelines for velocity reconstruction]{Sarpa_2019,Nadathur_2019c,Ata_2021,Ganeshaiah_Veena_2023}. Despite these difficulties, recent pioneering studies with low-redshift observational data \citep{Courtois_2012,Paz_2013,Li_2020,Courtois_2023} confirmed simulation-based results \citep{Massara_Sheth_2018,Massara_2022,Schuster_2023,Schuster_2024}. 

Various analyses \citep{Massara_Sheth_2018,Massara_2022,Schuster_2023,Schuster_2024} show that Eq.~\eqref{eq:radial_velocity} works remarkably well in simulations, especially in the outer parts of voids and when the velocity field is sampled with a volume-weighted approach. In contrast, within the inner parts of voids, the model is expected to remain valid down to scales where the linear regime holds ($|\delta|\ll1$), with the exact limiting scale depending on the adopted void definition. It is finally important to highlight that a correct treatment of the velocity profiles of voids identified in realistic density fields must include the modeling of tracer bias, which depends on the environment and physical scale \citep{Pollina_2017,Pollina_2019,Contarini_2019,Contarini_2021,Verza_2022}.

\subsection{Theory for void ellipticity} 
 
An analytical model for the ellipticity of voids based on the Zel'dovich approximation has been developed by \citet{Park_Lee_2007} and later improved by \citet{Lavaux_Wandelt_2010}, by considering that voids correspond to maxima of the source of displacement. These analytical formulae describe the ellipticity of the displacement field and are therefore valid in Lagrangian space. This work has investigated the precision of the model with $N$-body simulations, achieving a precision of $\sim0.1 \% $ for the mean ellipticity for voids larger than $4 \ h^{-1} \, \mathrm{Mpc}$. Given the considered range of void scales it is easy to see how this model is relevant to be tested with upcoming surveys, while demanding further tests on mocks accounting for galaxy properties. 

As the intrinsic Eulerian ellipticity is linked to tidal fields \citep{Park_Lee_2007, Lavaux_Wandelt_2010}, we first obtain the the eigenvalues of the inertial mass tensor \citep{Shandarin_2006}, which is defined as:
\begin{equation}
  M_{xx} = \sum_{i=1}^{N_p} m_i (y_i^2 + z_i^2),\hspace{.3cm} M_{xy} =
  - \sum_{i=1}^{N_p} m_i x_i y_i\;,
\end{equation}
where $x_i$, $y_i$, $z_i$ are the coordinates and $m_i$ is the mass
of the $i$-th particle of the void (with respect to its center of
mass). Through cyclic permutation of $x$, $y$ and $z$ it is possible to obtain the other matrix elements. 
These matrix elements can be expressed in terms on the Lagrangian displacement field, linking the analytical description of Lagrangian statistics to the Eulerian ones \citep{Lavaux_Wandelt_2010}.
The (Eulerian) ellipticity is then defined as:
\begin{equation}
\mathcal{E}=
  1-\left(\frac{I_2+I_3-I_1}{I_2+I_1-I_3}\right)^{1/4}\;, \label{eq:epsilon}
\end{equation}
where $I_{j}$ are the $j$-th eigenvalues of the tensor $M$, with $I_{1}
\leq I_{2} \leq I_{3}$. 

It is also possible to define the ellipticity distribution of voids (see \citealt{Lavaux_Wandelt_2010} for more details). Considering $N$-body simulations, the mean ellipticity has been shown to be particularly sensitive to the level of clustering and the properties of dark energy \citep{Platen_2007, Lee_Park_2009}.
Besides theoretical modeling, other works relying on machine learning techniques have explored the link of ellipticity with cosmology (using voids found in halo catalogs). In particular \citet{Wang_2023} used the set of void catalogs \textsc{GIGANTES} \citep[][see also Appendix~\ref{app:A}]{Kreisch_2022} at $z=0$ to constrain $\Omega_\mathrm{m}$, $\sigma_8$ and $n_s$ with mean relative errors of $10\%$, $4\%$ and $3\%$, respectively.

\subsection{Theory for CMBX voids}\label{sec:theory:CMBX}

CMB photons traveling from the last-scattering surface to the observer are impacted by the gravitational potential of the cosmic structures they cross through their journey. In particular, the underdense and expanding nature of cosmic voids induces characteristic secondary anisotropies, most notably through gravitational lensing and the ISW, which we will analyze in the following from a theoretical point of view.

\subsubsection{CMB lensing}
The deflection due to weak gravitational lensing can be computed by solving the null geodesic equation in the weak gravitational limit of General Relativity. Given a gravitational potential $\Phi$ and denoting $\beta$ the true angular position, the local deflection angle for a photon traveling a small comoving distance ${\rm d} \chi$ is ${\rm d} \beta = -2 {\rm d} \chi \nabla_\perp \Phi$. The deflection angle of CMB photons moving from the last scattering surface at comoving distance $\chi_*$ is obtained by integrating along the line of sight\footnote{This equation is for a flat universe, see \citet{Lewis_Challinor_2006} for the general case.}:
\begin{equation}
\alpha= -2 \int_0^{\chi_*} {\rm d} \chi \frac{\chi_* - \chi}{\chi_*} \nabla_\perp \Phi (\chi \hat{n},t)\;.
\end{equation}
The observed CMB temperature in the direction $\hat{n}$ is $T_{\rm o}(\hat{n})=T(\hat{n}+\alpha)$. The lensing signal can be described by the magnification matrix:
\begin{equation}\label{eq:lensing_matrix}
A_{ij} = \delta_{ij} + \frac{\partial \alpha_j}{\partial \theta_i} = \delta_{ij} + \begin{pmatrix} \kappa - \gamma_1 & & -\gamma_2 + \omega \\
-\gamma_2 - \omega & & \kappa + \gamma_1
\end{pmatrix}\;,
\end{equation}
where $\theta$ is the observed angular position and $\delta_{ij}$ is the Kronecker delta. The magnification is the inverse of the determinant, which at lowest order depends only on the convergence $\kappa$: $\det A^{-1} \simeq 1 + 2\kappa$, where $\kappa=-\partial_i \alpha_i/2$. The shear $\gamma = \gamma_1 + i \gamma_2$ is the distortion, while $\omega$ is the rotation, which vanishes at lowest order due to its antisymmetry \citep[see][for reviews and details on computations]{Bartelmann_Schneider_2001,Lewis_Challinor_2006}.

The dominant part of the lensing signal in the CMB temperature sourced by voids is the convergence $\kappa$, which can be written as: 
\begin{equation}\label{eq:convergence}
\kappa (\theta) = \frac{1}{2 c^2} \int \frac{\chi (\chi_* - \chi)}{\chi_*} \nabla^2 \Phi(\chi,\theta) {\rm d} \chi = \frac{3 \Omega_{\rm m} H_0^2}{2 c^2} \int \frac{\chi (\chi_* - \chi)}{\chi_*} \frac{\delta(\theta,\chi)}{a(\chi)} {\rm d} \chi\;.
\end{equation}
From the above equation it can be shown that, unlike clusters and filaments, cosmic voids induce a de-magnification on the CMB, due to the negative density contrast $\delta(\theta,\chi)$, representing the matter density contrast field of the photon path crossing voids\footnote{In principle, the quantity $\delta(\theta,\chi)$ can be modeled using the theory presented in \Cref{Subsection: VGCF Theory} but, for an effective usage, its derivation from simulation-based measurements is currently preferred (see \Cref{subsec:obs_CMB_lensing}).}. This means that voids act as divergent lenses. The CMB lensing signal from an individual void is dwarfed by the amplitude of noise and the primary temperature fluctuation in the CMB. The typical method to detect this signal consists of stacking the CMB map in the direction of voids. In this way, noise and the primary CMB fluctuation are averaged out, while the lensing signal is summed \citep{Nadathur_2017,Raghunathan_2020,Vielzeuf_2021,Hang_2021,Kovacs_2022c,Vielzeuf_2023,Camacho-Ciurana_2024,Demirbozan_2024,Sartori_2025}. In addition, from Eq.~\eqref{eq:convergence}, it is easy to prove that the peak of the convergence signal comes from voids detected around $z\simeq 1$ \citep{Lewis_Challinor_2006,Sartori_2025}.

\subsubsection{Integrated Sachs--Wolfe}

The integrated Sachs--Wolfe effect \citep{Sachs_Wolfe_1967}, together with its nonlinear extension \citep{Rees_Sciama_1968}, describes the impact of the time variation of the gravitational potential of structures to photons crossing them, and in particular to CMB photons.
The total amount of change in the photon's energy is given by the time integral of the potential variation along the line of sight. Changing coordinates from time to redshift, this reads:
\begin{equation}
\frac{\Delta T}{T}(\hat{\bf n}) = \frac{2}{c^2}\int_{t_*}^{t_0} \dot{\Phi}\big[\hat{\bf n},t\big] {\rm d} t = 2\int_0^{z_*} \frac{a}{H(z)} \dot{\Phi}\big[\hat{\bf n},\chi(z)\big] {\rm d} z\;.
\end{equation}
Then, by combining this equation with the linear Poisson equation for the gravitational potential $\Phi$, we obtain:
\begin{equation}
\frac{\Delta T}{T}(\hat{\bf n}) = -2\int_0^{z_*} a[1 - f(z)] \Phi\big[\hat{\bf n},\chi(z)\big] {\rm d} z\;,
\end{equation}
where $f(z)$ is the linear growth rate introduced in Eq.~\eqref{eq:radial_velocity}. The measurement of the ISW signal is a direct detection of the accelerated expansion of the Universe. As an example, in a matter dominated universe $f=1$, corresponding to a null ISW signal.

As for CMB lensing in voids, the ISW signal in voids is measured by stacking the CMB patches in the direction of cosmic voids detected in the galaxy distribution. This method allows us to increase the signal-to-noise ratio with respect to the standard cross-correlation with galaxies \citep{Granett_2008,Nadathur_2012,Nadathur_2014,Cai_2014,Kovacs_Granett_2015,Nadathur_Crittenden_2016,Kovacs_2017,Kovacs_2018,Kovacs_2019,Kovacs_2020,Kovacs_2022b,Kovacs_2022a}. By comparing the ISW signal obtained by stacking large voids with cosmological simulations, current analyses show an excess of ISW signal with respect of $\Lambda$CDM expected one to the one expected for a standard $\Lambda$CDM cosmology. \citep{Kovacs_2017,Kovacs_2018,Kovacs_2019,Kovacs_2020,Kovacs_2022b,Kovacs_2022a}. For further details on the observational results related to CMBX voids see \Cref{sec: CMBX_obs}.

\subsection{Theory for void-galaxy lensing}\label{sec:theory:lensing}

With the terminology ``void-galaxy lensing'' we usually refer to voids acting as lenses of the background distribution of galaxies (see \citealt{Bartelmann_Schneider_2001} for a review on weak lensing). Even if the theory is the same as that already explored for the CMB lensing, void-galaxy lensing differs from the former in at least two features. First, the comoving distance of the lensed CMB last-scattering surface is fixed, while the comoving distance of lensed galaxies changes. Second, instead of the convergence, void-galaxy lensing focuses on the shear field, inferred from galaxy shape measurements. This is for a twofold reason: first, the shear of background galaxies is a more straightforward measurement than the magnification; second, the shear is expected to produce a much higher signal-to-noise ratio in voids \citep{Amendola_1999,Krause_2013}. 

The shear is a spin-2 field (see Eq.~\ref{eq:lensing_matrix}), and can therefore be decomposed into two components: the tangential shear, $\gamma_+$, and the cross shear, $\gamma_\times$, oriented at $45^\circ$ with respect to each other. Let us consider the stacked signal around voids, which, due to isotropy, is expected to exhibit rotational symmetry along the line of sight. In the axially symmetric case, only the tangential component survives, while the cross component, oriented at $45^\circ$, vanishes. Therefore $|\gamma_+|=|\gamma|$ \citep{Gruen_2016,Amendola_1999}. In that case, the (tangential) shear can be expressed as \citep{Bartelmann_Schneider_2001,Amendola_1999,Krause_2013,Higuchi_2013,Barreira_2015,Baker_2018,Paillas_2019,Davies_2019,Bonici_2023,Boschetti_2024,Maggiore_2025,Su_2025}:
\begin{equation}\label{eq:shear_esmd}
\langle \gamma \rangle (\Theta) = \frac{\Delta \Sigma (\Theta)}{\Sigma_{\rm cr}} = \frac{\bar{\Sigma} (\Theta) - \Sigma (\Theta)}{\Sigma_{\rm cr}}\;,
\end{equation}
where $\Delta \Sigma (\theta)$ is the excess surface mass density, and $\Sigma (\theta)$ in the surface mass density, which for a single void at redshift $z_{\rm v}$ is given by: 
\begin{equation}
\Sigma(\theta) = \int d D_A(z) \delta \big(D_A(z_{\rm v})\theta,D_A(z) \big)\;.
\end{equation}
Here $D_A(z)$ is the comoving angular-diameter distance, which for a flat universe ($\Omega_{\rm k}=0$) is defined as:
\begin{equation}\label{eq:DA}
D_A(z)=\int_0^z \frac{c}{H(z')} {\rm d}z'\;.
\end{equation}
The quantity $\bar{\Sigma} (\theta)$ denotes the mean surface mass density within an angle $\theta$:
\begin{equation}
\bar{\Sigma}(\theta) = \frac{2}{\theta}\int_0^\theta {\rm d} y \, y \Sigma(y)
\end{equation}
and $\Sigma_{\rm cr}$ is the critical mass density weighted by the source galaxy distribution for lenses at redshift $z_{\rm v}$:
\begin{equation}
\Sigma_{\rm cr} = \frac{c^2}{4 \pi G}\int_{z_{\rm v}}^\infty {\rm d} z \frac{{\rm d} n(z)}{{\rm d} z} \frac{D_A(z)}{D_A(z_{\rm v})D_A(z,z_{\rm v})} \left[ \int_{z_{\rm v}}^\infty {\rm d} z \frac{{\rm d} n(z)}{{\rm d} z} \right]^{-1}\;.
\end{equation}
In the inner region of voids, the shear is negative, meaning that the weak lensing distortion is aligned in the radial direction. For a positive density contrast, as around galaxy clusters or even in proximity of the compensation wall of voids, the shear is positive and the weak lensing distortion is aligned tangentially \citep{Amendola_1999}. 

After describing the theory for the various statistics, we will address the sensitivity of voids to cosmology in the next section.

\section{Constraints from surveys}\label{sec:constraints_from_surveys}

At the beginning of the 21st century, observations of the large-scale structure of the Universe were still insufficient to provide a statistically significant sample of voids that could be used for deriving constraints on the cosmological model. Despite this, the scientific community's interest in cosmic voids had already emerged through early theoretical studies focused on the modeling of the statistics presented in \Cref{sec: Void statistics}. Alongside this, increasingly large and high-resolution cosmological simulations were produced testing such theoretical models and calibrating their free parameters.

The combination of theoretical studies and simulation development enabled a progressively deeper understanding of the fundamental properties of voids and allowed to test their sensitivity to cosmological parameters. For example, modern cosmological simulations allowed the community to study the behavior of cosmic voids in mock universes characterized by different cosmological models.

In this section, we first focus on the sensitivity of void-based statistics to cosmology, highlighting the property of orthogonality with respect to overdensity-based probes. We then summarize the available results obtained by applying void statistics to survey data, including a detailed description of the steps to obtain such constraints. These analyses rely on various galaxy survey data and cross-correlations with CMB data.

\subsection{Sensitivity to fundamental physics}\label{sec: sensitivity}
Over the years, various analyses have shown that voids are particularly useful for cosmology, both as a stand-alone probe and in joint analyses with overdensity-based statistics. Their underdense nature and unparalleled spatial extent make them strongly sensitive to cosmological parameters that are typically challenging to constrain, such as those governing the growth of large-scale structures, the equation of state of dark energy, gravity beyond General Relativity, and the total neutrino mass
\citep{Hamaus_2015,Hamaus_2016,Pisani_2015b,Massara_2015,Cai_2015,Verza_2019,Verza_2023,Contarini_2021,Contarini_2022,Radinovic_2023}.

\subsubsection{Dark energy}\label{sec:darkenergy}
Cosmic voids have emerged as highly sensitive probes of dark energy, as their evolution is directly governed by the interplay between gravitational collapse and cosmic expansion. Unlike overdense regions, where the influence of dark energy quickly becomes negligible due to collapse, in voids its contribution remains relevant throughout their entire evolution \citep{Goldberg_Vogeley_2004,Bromley_Geller_2025,Moretti_2025}. This makes voids a particularly suitable environment for studying dark energy. Another consequence of their underdense nature is that voids expand faster than the background Universe, rendering their density profiles, shapes, and sizes particularly sensitive to the dark energy equation of state, $w(z)$ \citep{Park_Lee_2007,Lee_Park_2009,Biswas_2010,Lavaux_Wandelt_2010,Pisani_2015b,Pollina_2016,Verza_2019,Sahlen_2019,Contarini_2022,Bonici_2023,Verza_2023,Verza_2024b,Moretti_2025}.
For example, when the growth of structures is enhanced, void density profiles become deeper and develop more pronounced compensation walls \citep{Lehman_2025}. In such scenarios, matter flows more rapidly from the void interior toward its boundary, directly impacting void sizes and abundances.

This dynamical sensitivity also leaves geometric imprints on void shapes: measuring the distribution of their ellipticities can, in principle, constrain the properties of dark energy \citep{Park_Lee_2007,Lee_Park_2009,Lavaux_Wandelt_2010}. Moreover, dark energy---together with the total matter density and spatial curvature---governs the expansion history of the Universe, which can be probed using voids as standard spheres. The average shape of stacked voids is spherical by construction, enabling the application of the Alcock–Paczy{\'n}ski test \citep[][see \Cref{subsec:vsf_distortions,subsec:vgcf_distortions} for a description of this effect]{AlcockPaczynski_1979}. This method yields competitive constraints on the evolution of $w(z)$ and complements traditional probes such as baryon acoustic oscillations (BAO) and weak lensing \citep{Lavaux_Wandelt_2012,Sutter_2012,Hamaus_2016,Hamaus_2022,Radinovic_2023,Nadathur_2020b,Woodfinden_2022,Woodfinden_2023,Hamaus_2014c,Mao_2017,Aubert_2022,Hamaus_2020}.

\subsubsection{Modified gravity}\label{sec:modgravity}
Modified gravity theories produce characteristic imprints on void profiles and affect the growth of cosmic structures by altering the gravitational interaction. To remain consistent with General Relativity in high-density regions---and in particular with the stringent constraints from Solar System dynamics---these models typically invoke screening mechanisms that suppress deviations around matter overdensities \citep{Clifton_2012,Saridakis_2021}. Cosmic voids, being underdense, are mostly unaffected by such mechanisms and therefore provide a natural environment where the signatures of modified gravity are amplified. This leads to the enhanced formation of underdensities and an overall acceleration of void evolution, with the exact impact depending on the strength of the modification to gravity \citep{Li_Efstathiou_2012,Clampitt_2013,Cai_2015,Zivick_2015,Barreira_2015,Voivodic_2017,Baker_2018,Davies_2019,Paillas_2019,Perico_2019,Contarini_2021,Wilson_Bean_2023,Mauland_2023,Maggiore_2025}. Therefore, voids offer a particularly suitable environment for studying modified gravity models, also because baryonic processes have a much weaker impact in underdense regions than in overdense ones \citep{Schuster_2024}. It is finally worth noting that other relativistic effects are subdominant in void evolution~\citep{Williams_2025}.

\subsubsection{Neutrinos and intrinsic degeneracies}\label{sec:neutrinos}
Massive neutrinos, by contrast, suppress the clustering of matter on scales smaller than their free-streaming length due to their large thermal velocities. Voids are especially sensitive to this effect for two main reasons. First, the typical neutrino free-streaming scale is comparable to the characteristic size of voids, making them an ideal environment to probe the scale dependence of neutrino free streaming. Second, because voids contain less matter overall, the relative influence of neutrinos on their structure is enhanced. This manifests as a reduced rate of structure growth, with fewer and shallower voids that are surrounded by less pronounced compensation walls \citep{Massara_2015,Kreisch_2019,Sahlen_2019,Schuster_2019,Contarini_2021,Contarini_2022,Kreisch_2022,Verza_2023,Vielzeuf_2023,Mauland_2023,Maggiore_2025}.

Although they arise from different physical mechanisms, dark energy and modified gravity can leave signatures on large-scale observables that are opposite to those of massive neutrinos \citep{Hamann_2012,Wang_2012,Zablocki_2016,Allison_2015,Wang_2016,Lorenz_2017,Mishra-Sharma_2018,DiazRivero_2019,Zhao_2020}. In some cases, their combined effects may even mimic the predictions of the standard $\Lambda$CDM model \citep[see e.g.][]{Baldi_2014,Peel_2018,Giocoli_2018,Contarini_2021}. However, owing to their distinct sensitivity to these different components, cosmic voids provide a powerful tool to disentangle their contributions and to test extensions of the standard cosmological framework \citep{Bayer_2021,Contarini_2021,Contarini_2022,Bonici_2023,Verza_2023,Mauland_2023,Bayer_2024,Thiele_2024,Verza_2024b}. Moreover, intrinsic degeneracies between cosmological parameters can be further alleviated by combining void statistics with other cosmological probes, as we will discuss in the following.

\subsubsection{Orthogonal constraints}\label{sec: orthogonality}

Recent studies have also highlighted how the cosmological constraints derived from the study of cosmic voids often yield confidence contours orthogonal to those obtained from probes based on overdensities \citep{Sahlen_2016,Sahlen_2019,Bayer_2021,Kreisch_2022,Contarini_2022,Contarini_2023,Bonici_2023,Pelliciari_2023}. This complementarity is of extreme importance in the context of a combined analysis of voids and overdensities, as it can break the natural degeneracies between the parameters of the cosmological model. 

\begin{figure}[ht]
    \centering
    \includegraphics[width=0.47\linewidth]{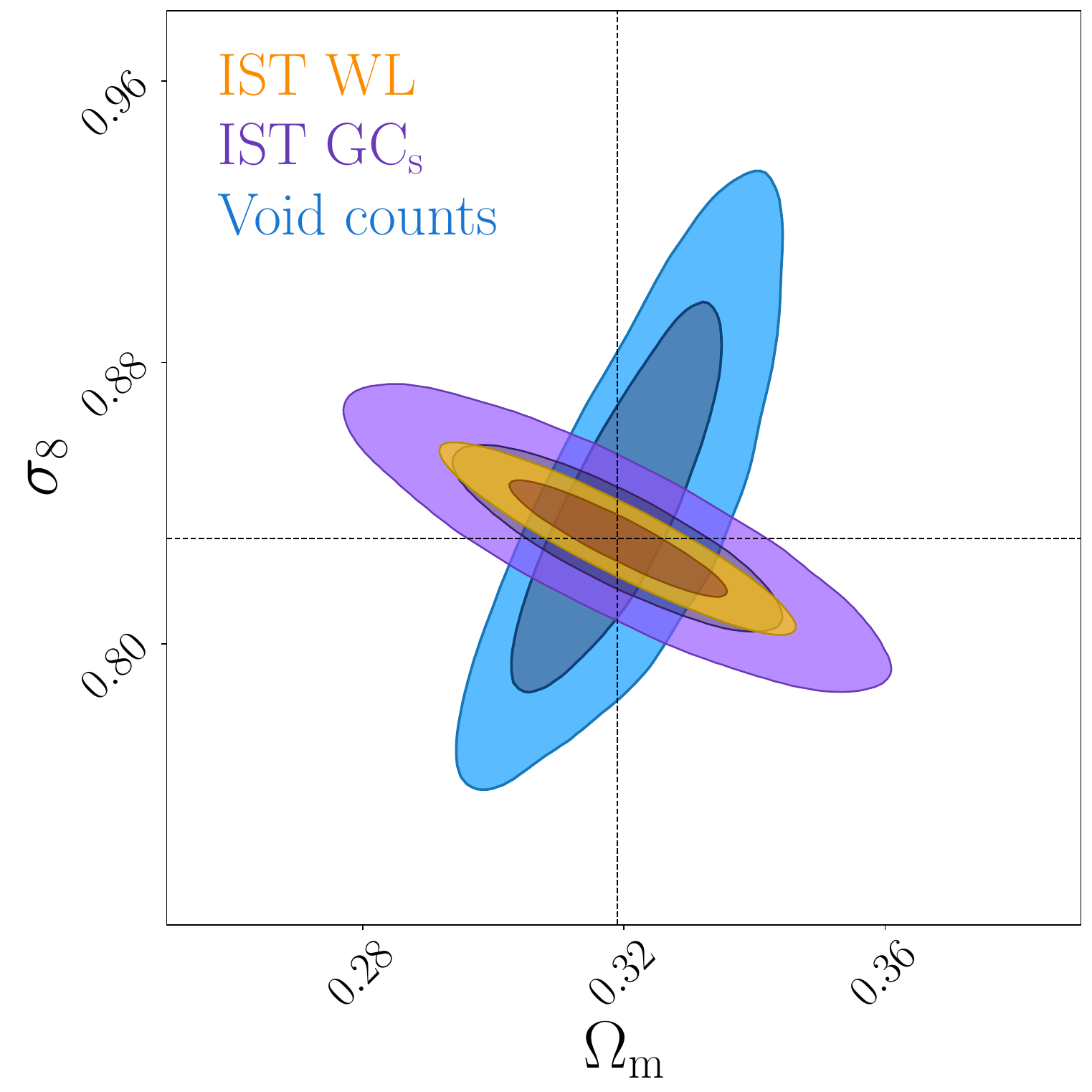}
    \includegraphics[trim=0cm -0.5cm 0cm 0cm, clip, width=0.488\linewidth]{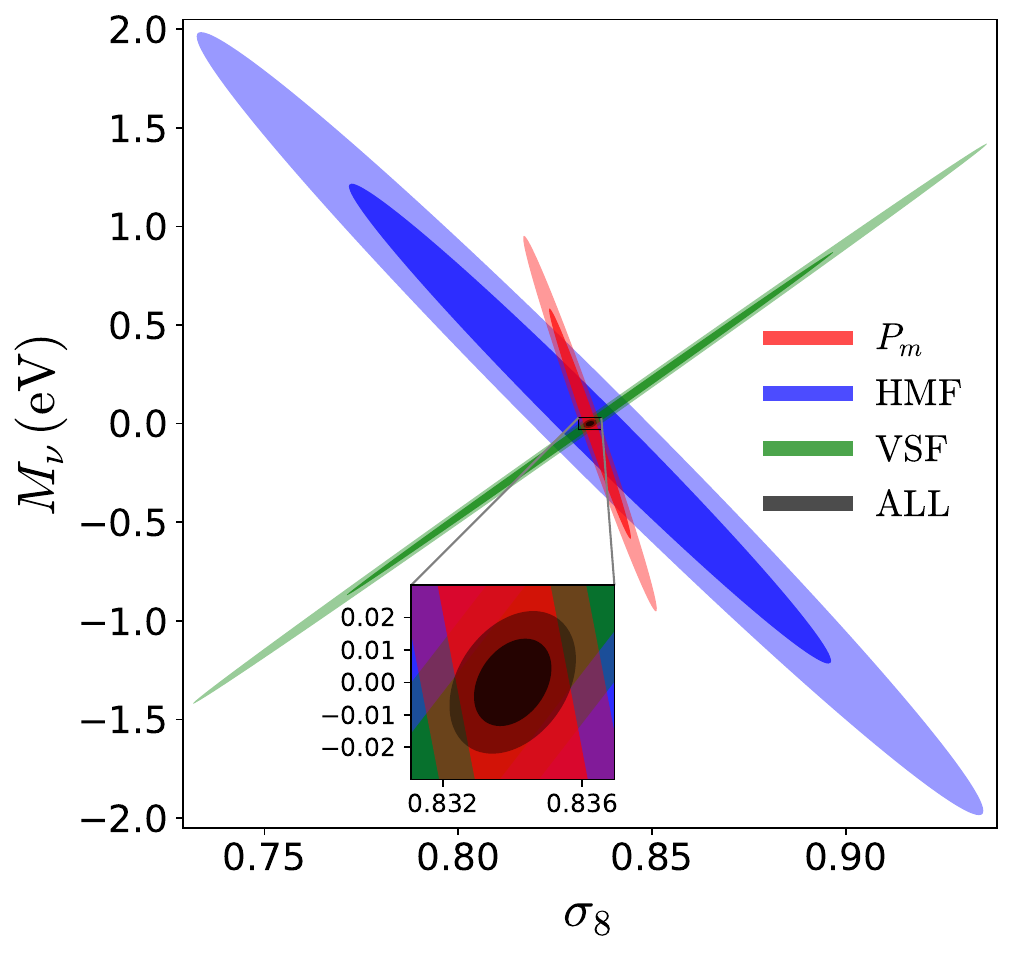}
    \caption{\textit{Left}: Cosmological forecasts for the \textit{Euclid} mission in the $\Omega_\mathrm{m}$–$\sigma_8$ parameter space, obtained using different probes: void counts (blue), spectroscopic galaxy clustering (purple), and weak lensing (orange). \textit{Right}: Cosmological constraints in the $M_{\nu}$–$\sigma_8$ parameter space, derived through a Fisher analysis using the matter power spectrum (red), halo mass function (blue), and VSF (green). For further details, we refer the reader to the original sources \citep{Contarini_2022,Bayer_2021}.}
    \label{fig: VSF orthogonality}
\end{figure}

The complementary constraining power is clearly illustrated in \Cref{fig: VSF orthogonality}, showing how cosmological contours derived from void analyses are oriented almost perpendicularly to those from other probes. In the left panel, the $\Omega_\mathrm{m}$–$\sigma_8$ parameter space showcases how constraints from the VSF are highly complementary to those obtained from galaxy clustering and weak lensing. The right panel instead focuses on the $M_{\nu}$–$\sigma_8$ plane, where the VSF appears nearly orthogonal to both the matter power spectrum and the halo mass function. In this latter case, the predicted combination of the three probes demonstrates the potential for significantly improved cosmological constraints. In fact, the combination between cosmic void and overdensity-based statistics is expected to be particularly effective since their covariance is low \citep{Bayer_2021,Kreisch_2022,Pelliciari_2023}.

The additional information present in cosmic voids represents therefore a decisive contribution to improve the already valuable cosmological constraints extracted using the regions of our Universe with positive matter fluctuations. Thanks to all these characteristics, cosmic voids have acquired increasing prominence in the landscape of cosmological analyses. Nowadays, different void summary statistics have been extracted from modern redshift surveys, yielding results that are competitive with those from standard probes \citep[see][and references within]{Pisani_2019,Moresco_2022,Cai_Neyrinck_2025}.

\subsection{The observed void size function}\label{sec:obs_VSF} 
In recent years, the theoretical model of the VSF finally allowed to extract constraints from data of current redshift surveys. This represents only a first step toward the comprehensive and robust use of this void statistic, but the results obtained so far are highly promising. The main challenge in this context lies in using data that are subject to noise and observational limitations (e.g., redshift errors, sample incompleteness and purity, irregular survey edges).
Additional observational effects arise when dealing with voids from redshift surveys:

\begin{enumerate}[i)]
    \item Tracers of matter such as galaxies and galaxy clusters provide a biased representation of the density field, and the density profile of voids is influenced by this bias.
    \item The peculiar velocities of observed tracers affect their measured positions, since tracer positions are inferred from redshifts. This causes dynamical distortions.
    \item The mapping to convert tracer coordinates in the sky to distances---and hence infer their position in the sky---needs to assume a cosmology. If the cosmology assumed to compute cosmological distances is not the real one, geometrical distortions arise.
\end{enumerate}

\subsubsection{Tracer bias in voids}
The VSF theory presented in \Cref{sec: VSF} describes voids emerging from the total matter density field. When analyzing real survey data, however, 3D voids are usually traced by galaxies or galaxy clusters. 
These astrophysical objects represent biased tracers of the total matter density field and the density contrast threshold used to identify voids, $\Delta_\mathrm{v, tr}$, is affected by their characteristic bias, $b(z)$. To recover the value of the threshold in the unbiased field we can assume the simple relation:
\begin{equation}
    \Delta_\mathrm{v, DM} = \frac{\Delta_\mathrm{v, tr}}{b(z)}\;.
\end{equation}
Various works \citep{Pollina_2017,Pollina_2019,Contarini_2019,Contarini_2021,Verza_2022}, however, have shown that the value of $b(z)$ does not necessarily correspond to the large-scale effective bias of the same tracers \citep[see][for an extensive review]{Desjacques_2016}. The latter can be inferred, in the case of redshift-space galaxies for example, by modeling the two-point correlation function multipoles \citep{Taruya_2010,Beutler_2014}.
In voids, due to their characteristic underdense nature and distinct dynamical evolution, the bias of cosmic tracers can deviate from its linear large-scale value \citep{Pollina_2017,Pollina_2019,Contarini_2019,Contarini_2021}. This deviation can be studied in numerical simulations by measuring the ratio between the void density profiles traced by biased tracers and those traced by dark matter particles, both computed around the same set of voids. The mean value of this ratio tends to be systematically higher in the central parts of voids than on large scales. This behavior can be intuitively understood: the formation of collapsed structures is suppressed in the low-density environment of voids, making massive objects rare. As a result, tracers that reside in these regions are associated with a higher bias.

Accounting for the environmental dependence of tracer bias is non-trivial, and several methodologies have been developed over the years. As already mentioned in \Cref{Subsection: VGCF Theory}, one of the simplest approaches assumes a linear relationship between the void–tracer and void–matter correlation functions \citep{Pollina_2017,Pollina_2019}. 
Alternatively, a semi-analytical expression for the bias can be derived from the theoretical conditional mass function \citep{Furlanetto_Piran_2006,Neyrinck_2014,Verza_2022}.
Other works \citep{Contarini_2019,Contarini_2021} have proposed a semi-analytical framework based on calibrating the relation between the effective large-scale bias and its counterpart inside voids. While these approaches provide effective estimates of the tracer bias in underdense regions, they all rely on cosmological simulations to calibrate their free parameters, and, as such, have hidden dependencies on the simulation features (including, but not limited to, cosmology and modeling of the tracer properties). At the time of writing this review, a fully predictive theoretical approach to model $b(z)$ in voids remains an open challenge.

\subsubsection{Redshift and geometrical distortions}\label{subsec:vsf_distortions}
The matter flowing from the interior to the boundary of voids, caused by the time evolution of underdensities, induces a deformation of the apparent shapes of voids. In fact, beyond the velocity field driven by the Hubble flow, the peculiar velocities of matter tracers introduce a shift in their observed positions, which, in the case of cosmic voids, results in an apparent elongation along the observer’s line of sight. With $r$ and $s$ denoting real- and redshift-space coordinates, respectively, the average size of voids along the line of sight, $s_\parallel$, can be derived from the radial velocity profiles around voids $v(r)$ (see \Cref{sec:theory_velocity}) and therefore parametrized as:
\begin{equation}\label{eq:RSD_elongation}
    s_\parallel = R_\mathrm{v} + v(r) \frac{1+z}{H(z)} \bigg|_{r=R_\mathrm{v}}\;,
\end{equation}
where $R_\mathrm{v}$ is the void radius in real space. Assuming the validity of the linear regime and extending Eq.~\eqref{eq:radial_velocity} to biased tracers using $\beta(z)\equiv f(z)/b(z)$ (see Eq.~\ref{eq:growth_rate_factor}), \citet{Correa_2021} proposed the following form: 
\begin{equation}\label{eq:RSD_radius}
    R^z_\mathrm{v} = \left (1 - \frac{1}{3} \mathrm{f}_\mathrm{RSD} \beta(z) \Delta_\mathrm{v} \right) R_\mathrm{v}\;.
\end{equation}
Here, $\mathrm{f}_\mathrm{RSD}$ represents a fudge factor introduced to account for deviations from linear theory, void identification variations, and other possible systematic effects \citep[see][]{Contarini_2019,Verza_2022}; while the integrated density contrast corresponds, in this case, to the threshold used to define voids ($\Delta_{\rm v}$).

Geometrical distortions are instead due to the (possible) inconsistency between the true cosmology and the assumed one (known as the \textit{fiducial} cosmology). In fact, when dealing with observed astrophysical objects, the 3D position of the mass tracers must be inferred from angular positions and redshifts. The conversion to physical distances passes through the assumption of a cosmology, and if the latter deviates from the true one, the resulting spatial distribution of tracers results distorted, giving rise to the so-called Alcock--Paczy{\'n}ski effect \citep{AlcockPaczynski_1979}. Since it changes the conversion to distances, it also affects the volume of cosmic voids, making their average shape appear as an ellipsoid (in other words, a squeezed or elongated sphere, depending on the fiducial cosmology assumed). The true parallel and perpendicular extensions of voids, $r_\parallel$ and $r_\perp$, are linked to their distorted counterparts measured with a given fiducial cosmology ``fid'', by the following relations:
\begin{equation}\label{eq:AP}
    r_\parallel = \frac{H^\mathrm{fid}(z)}{H(z)} r_\parallel^\mathrm{fid} \equiv q_\parallel r_\parallel^\mathrm{fid} \;, \quad
    r_\perp = \frac{D_\mathrm{A}(z)}{D^\mathrm{fid}_\mathrm{A}(z)} r_\perp^\mathrm{fid} \equiv q_\perp r_\perp^\mathrm{fid}\;,
\end{equation}
where $H(z)$ is the Hubble parameter (Eq.~\ref{eq:hubble_rate}) and $D_\mathrm{A}(z)$ is the comoving angular-diameter distance (Eq.~\ref{eq:DA}).
Now, considering the equivalent sphere with the same
volume of the observed ellipsoid, we can derive the true value of the void radius by knowing its value measured in the fiducial cosmology \citep{Hamaus_2020,Correa_2021}: 
\begin{equation} \label{eq:geom_distortions}
    R_\mathrm{v} = q_\parallel^{1/3} q_\perp^{2/3} R_\mathrm{v}^\mathrm{fid}\;.
\end{equation}
This correction, together with the one due to the RSD, must be applied to the theoretical VSF model to correctly predict the size of voids observed in redshift surveys. While dynamical distortions generally translate into an increase of void sizes, geometrical distortions may result in a shift of the VSF towards smaller or larger radii, depending on the parameter values assumed for the fiducial cosmology.

\subsubsection{Observational results}

So far in this section, we have presented the fundamental observational effects that must be taken into account when modeling void number counts. The impact of other systematic errors, arising purely from survey limitations, is expected to be less significant. This is however still under investigation, and a proper inclusion into the theoretical model has not yet been achieved. Nevertheless, with this initial and basic modeling of the VSF, it has been possible to proceed with the analysis of voids using real data.

In one of the most recent studies published at the time of this review, \citet{Contarini_2023} used cosmic voids identified in the 12th data release (DR12) of the Baryon Oscillation Spectroscopic Survey \citep[BOSS,][]{Dawson_2013}, which is part of the Sloan Digital Sky Survey (SDSS) III project. In this work, the authors modeled void counts through a ``cleaning'' approach aimed at filtering and adjusting the void sample to align it with the properties of voids described by the theoretical model.

This cleaning process arises from the necessity of adapting the output of different void finders to the definition used in the void number counts theory. In fact, to align the sample of voids identified with a watershed-type finder like \vide to the Vdn model, \citet{Jennings_2013} proposed rescaling voids by modifying their radius such that they have, by construction, an integrated density contrast within the sphere equal to the value $\delta_\mathrm{v}^\mathrm{NL}$ defined in the model. In cases of overlap between the rescaled voids, the procedure involves discarding the smaller or shallower voids \citep{Neyrinck_2008,Jennings_2013,Pisani_2015a}, as they are more likely to represent noise-induced density fluctuations. This pipeline was presented in a public code by \citet{Ronconi_2017} as part of the \textsc{CosmoBolognaLib} library \citep[][see also Appendix~\ref{app:A}]{Marulli_2016} and has been used in several papers on the VSF \citep{Ronconi_2019,Verza_2019,Contarini_2019,Contarini_2021,Contarini_2022,Pelliciari_2023,Contarini_2023,Contarini_2024}.

\begin{figure}[ht]
    \centering
    \includegraphics[width=\linewidth]{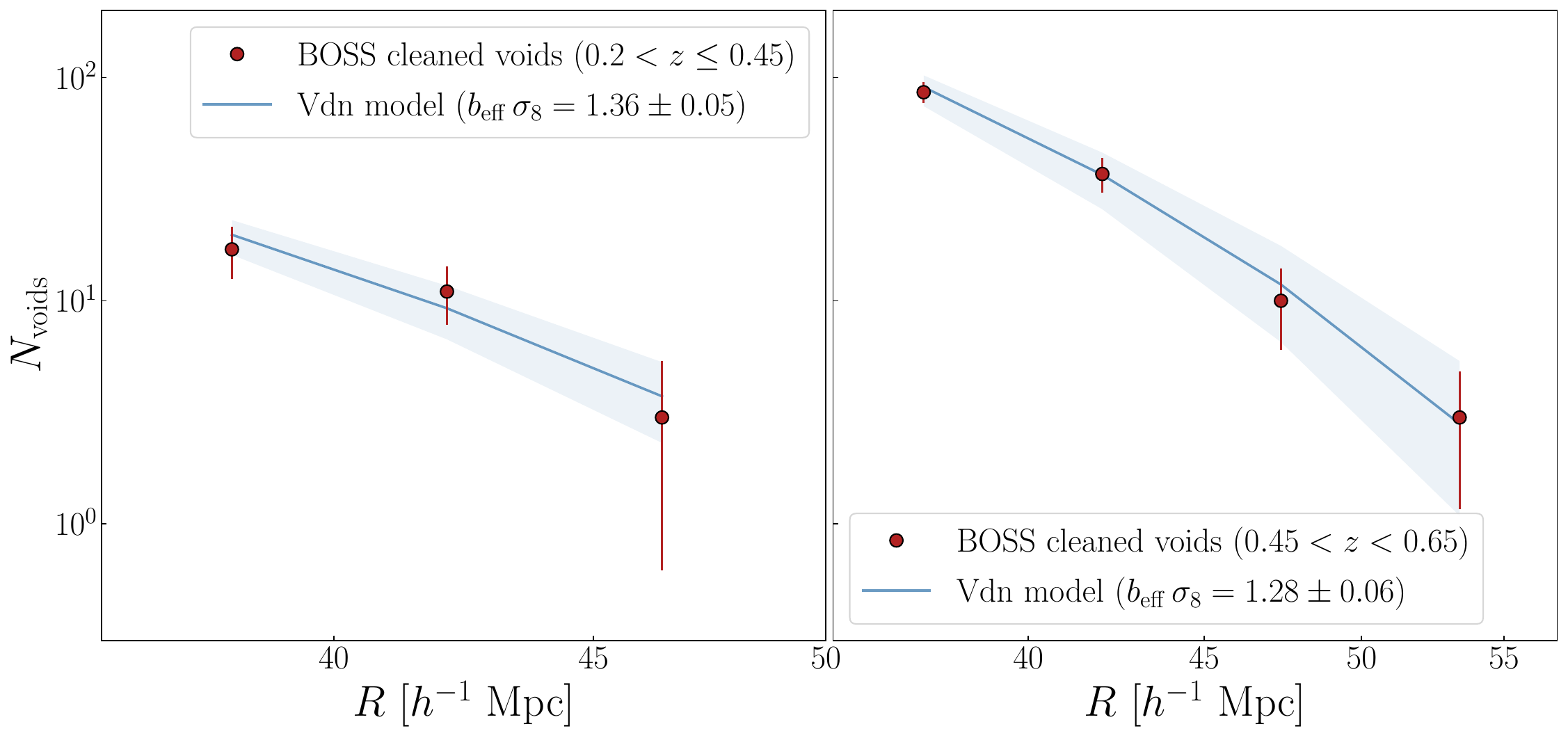}
    \includegraphics[width=0.42\linewidth]{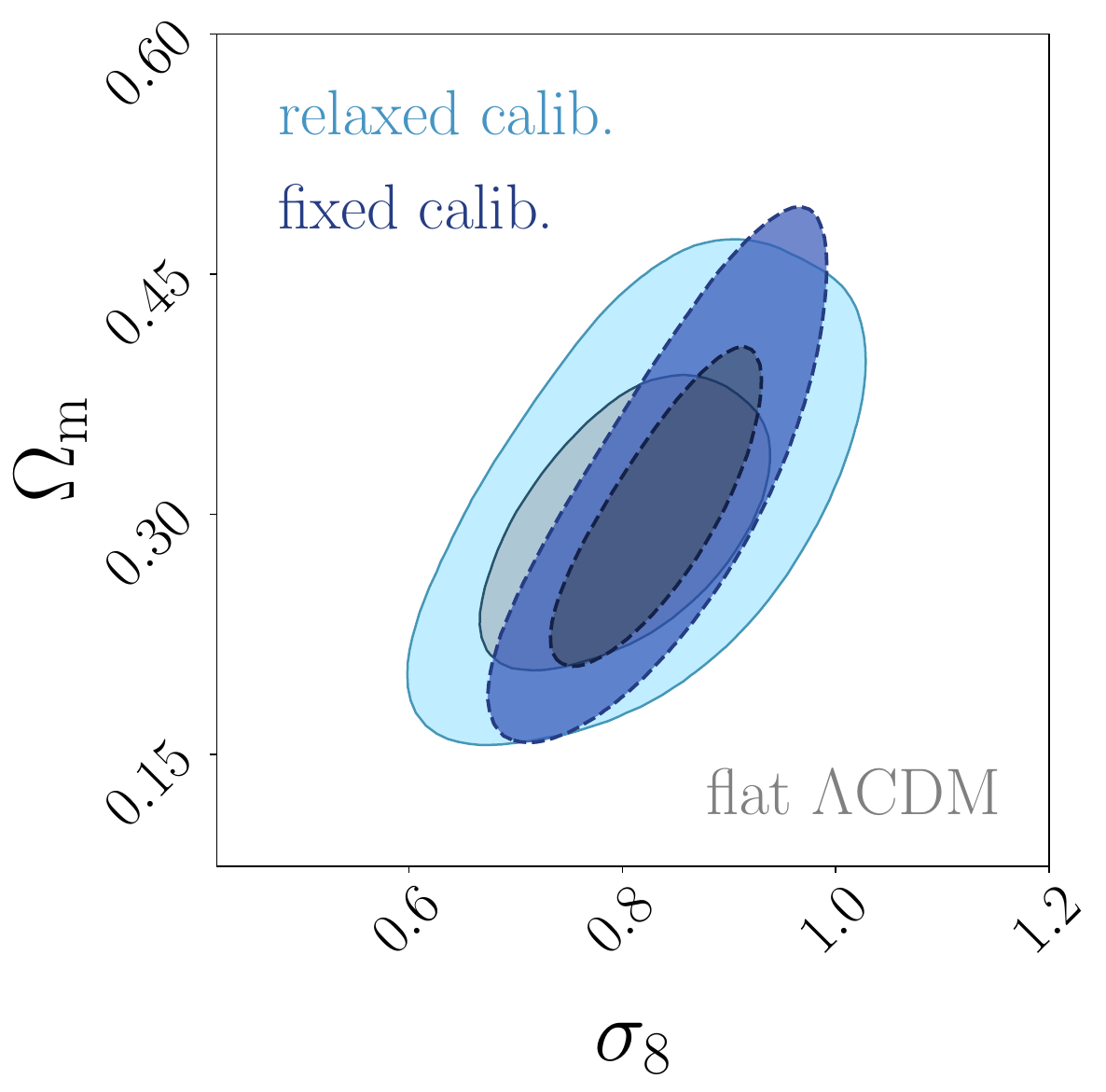}
    \includegraphics[width=0.44\linewidth]{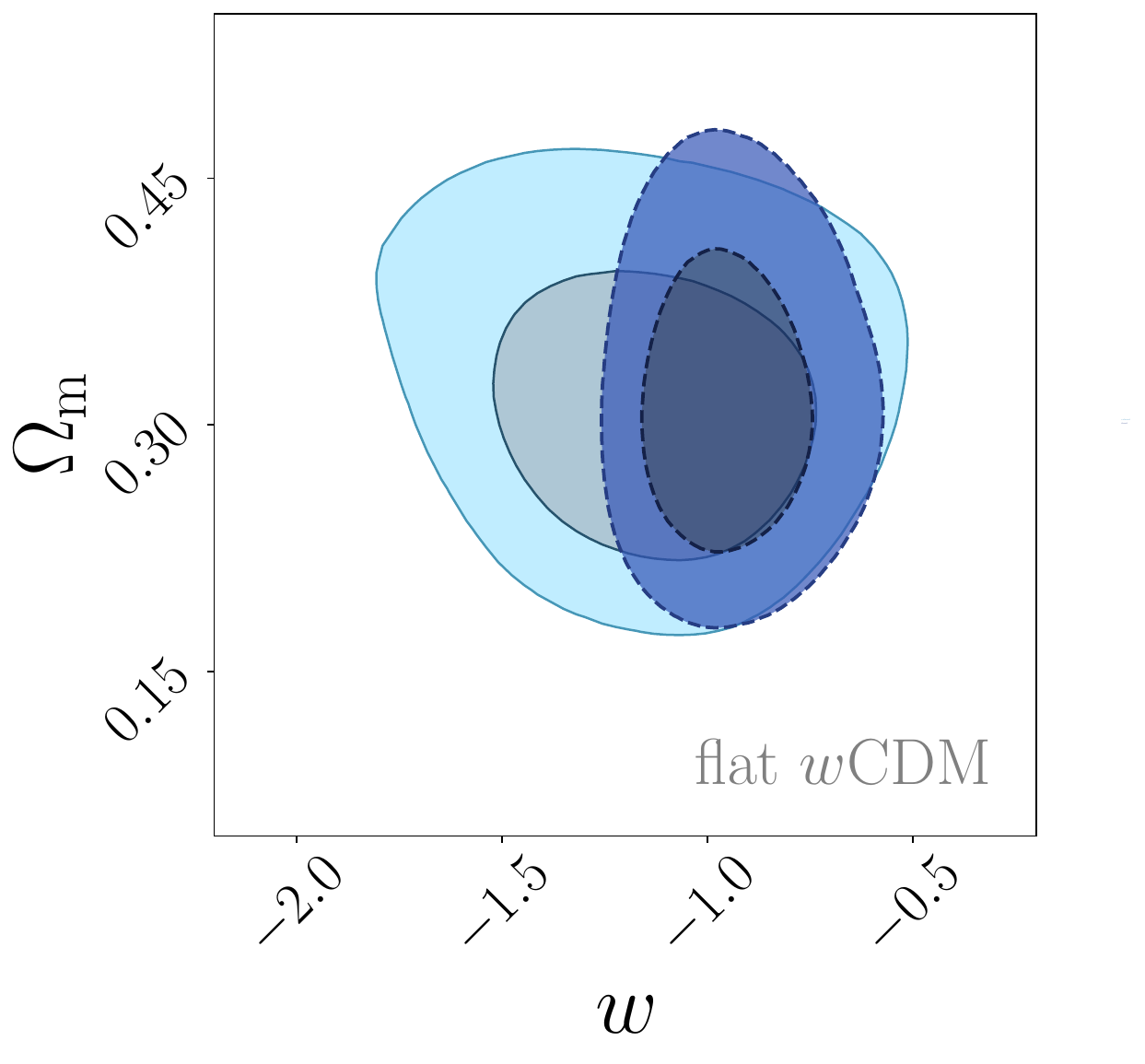}
    \caption{VSF analysis with the BOSS DR12 data set. The upper panels show the measured void number counts in red and the relative best-fit model in blue, for two redshifts intervals: $0.2<z\le0.45$ (left) and $0.45<z\le0.65$ (right). The lower panels report the cosmological constraints derived by jointly modeling both data sets, assuming $\Lambda$CDM (left) and $w$CDM (right) flat cosmologies. Dark and light blue colors here indicate results obtained by fixing or relaxing the prior on the nuisance parameters of the VSF model, respectively. For further details, we refer the reader to the original source \citep{Contarini_2022}.}
    \label{fig: BOSS constraints}
\end{figure}

At this stage, the size function extracted from the sample of cleaned voids has been modeled using the Vdn model (see \Cref{sec: VSF}), extended to account for the observational effects described above. In particular, the increase in void sizes due to RSD and biased tracers was incorporated as a single correction to the void density contrast threshold, $\delta_\mathrm{v}^\mathrm{NL}$, which was calibrated using the official mocks of the BOSS collaboration \citep{Kitaura_2014,Klypin_2016}. In \Cref{fig: BOSS constraints} we present the constraints on $\Omega_\mathrm{m}$ and $\sigma_8$ obtained from modeling the observed VSF with the extended Vdn model. The corresponding relative uncertainties are $\sim20\%$ for $\Omega_\mathrm{m}$ and $\sim10\%$ for $\sigma_8$, while $w$ is constrained up to a precision of $\sim14\%$. The results are consistent with $w=-1$, showing no statistically significant deviation from the standard $\Lambda$CDM prediction.

A similar approach, this time applied to the BOSS DR16 data, was presented by \citet{Song_2025}. In this case, the selection of cosmic voids identified with the \vide void finder was based solely on ellipticity: the authors restricted their modeling to the most spherical voids, without rescaling them to a specific internal density contrast. The counts were then fitted with the Vdn model, leaving the threshold $\delta_\mathrm{v}(z)$ as a free parameter, encapsulating the effect of tracer bias in this parameter. Geometrical distortion corrections were included according to Eq.~\eqref{eq:geom_distortions} and the RSD effects were modeled via Eq.~\eqref{eq:RSD_radius} by treating $\beta(z)$ as a free parameter. This method, previously tested on mock catalogs \citep{Song_2024a}, demonstrated the ability to provide constraints on the standard cosmological model and to offer insights into alternative models, such as $w$CDM.The main parameters investigated in this analysis are $\Omega_{\rm m}$, $\sigma_8$ and $w$, which were constrained with precisions of $\sim19\%$, $20\%$ and $29\%$, respectively.

Another relevant work analyzing the void size function extracted from observational data is presented by \citet{Thiele_2024}. In their study, the authors considered a subsample of the BOSS galaxy catalog and extracted cosmological constraints using a simulation-based approach (namely, implicit likelihood inference). To this end, they generated tailored light-cones with variations in both the cosmological parameters of the $\Lambda$CDM model and the parameters of the halo occupation distribution. By jointly analyzing the void size function, the void-galaxy cross-power spectrum, and the galaxy auto-power spectrum, they were able to place an upper limit of 0.35~eV on the sum of neutrino masses ($95\%$ credible interval).

\subsection{The observed void-galaxy cross-correlation function}\label{sec:VGCF_in_surveys}
The real-space VGCF is expected to be spherically symmetric, given the assumption that our Universe is homogeneous and isotropic on large scales. Nevertheless, voids are observed in redshift space. Therefore so far theoretical models of the VGCF have focused on modeling the redshift-space VGCF from its real-space counterpart. 

Typically, a real-space void stack will appear distorted in redshift space due to both geometrical and dynamical distortions. Indeed, since we observe voids in the RA, DEC, $z$ coordinates and not in physical coordinates, if the correct cosmology is not known when converting to distance, this leads to geometrical distortions. Additionally, dynamical distortions arise due to the presence of peculiar velocities (see \Cref{sec:theory_velocity}). 

While measuring the VGCF in redshift space removes its symmetry, this complication comes with an important gain for cosmology: the measured redshift-space VGCF bears a strong dependence on cosmological parameters. In particular geometrical distortions will depend on the assumed fiducial cosmology and its difference with the true cosmology---therefore embedding information about the Universe's content and evolution; while dynamical distortions will provide information on the structures, therefore directly constraining General Relativity. 

\subsubsection{Redshift and geometrical distortions}\label{subsec:vgcf_distortions}
Modeling the redshift-space VGCF is therefore a necessary challenge, as it will allow us to constrain some among the most relevant unknowns of modern cosmology. Initial work on using the VGCF to extract information considers voids as standard spheres \citep{Ryden_1995,Ryden_Melott_1996, Lavaux_Wandelt_2012}, therefore implicitly relying on the assumption that the Universe is homogeneous and isotropic.

From a practical perspective, to model the observed VGCF---that is the VGCF in redshift space---one has to consider a number of ingredients. First, it is necessary to consider a void profile in real space, and second, to account for the change in coordinates between real and redshift space. Third, one has to consider the impact of peculiar velocities. Starting with a void catalog, if we have computed the void center and we know the positions of galaxies around such void center, then we can measure the VGCF in redshift space, and attempt to model it. This is a vastly explored area of research \citep[see e.g.][]{Hamaus_2015,Paz_2013,Hamaus_2017,Hamaus_2020,Nadathur_Percival_2019}.

In the following, we move to describing the key elements to modeling the VGCF. Following the description presented in \cite{Hamaus_2020}, we use upper-case letters to indicate void center-related quantities, and lower-case letters for galaxy-related quantities. 
We will start with considering the vector $\mathbf{r}$, separating the void center with position $\mathbf{X}$ from the galaxy position $\mathbf{x}$ in real space; and the corresponding vector $\mathbf{s}$ in redshift space. We can express the vector in real space as: $\mathbf{r}\equiv\mathbf{x}-\mathbf{X}$. 

The VGCF is the excess probability of finding a galaxy at a given distance from the void center. It can be written in redshift-space $\xi^s(\mathbf{s})$, or real-space $\xi(r)$\footnote{Due to isotropy, the VGCF in real space depends only on the magnitude of the separation vector $r\equiv|\mathbf{r}|$, and not on its orientation.}. The two are linked by the relation:

\begin{equation}
1+\xi^s(\mathbf{s}) = \left[1+\xi(r)\right]\frac{\mathrm{d}r_\parallel}{\mathrm{d}s_\parallel}\;.
\label{eq:xi^s}
\end{equation}
To model the VGCF in redshift space we need to consider the conversion from $\mathbf{s}$ to $\mathbf{r}$. 
Generically when observing an object in the sky we measure its coordinates (angular sky coordinates and redshift $z$).  Considering the direction of the void center as our line of sight, dynamical distortions for the position of the galaxy are introduced considering that: 
\begin{equation}
\mathbf{x}(z) \simeq \mathbf{x}(z_h) + \frac{1+z_h}{H(z_h)}\mathbf{u}_\parallel\;,
\label{eq:x_rsd}
\end{equation}
where we defined $\mathbf{x}(z_h)$ as the contribute to $z$ from the cosmological Hubble expansion and $\mathbf{v}_\parallel$ as the component of the redshift-space galaxy velocity vector along the line of sight. 

Considering Eq.~\eqref{eq:x_rsd} for the center of the void and a galaxy, this allows us to define our void center-galaxy separation vector in redshift space $\mathbf{s}$ as: 
\begin{equation}
\mathbf{s}\equiv\mathbf{x}(z)-\mathbf{X}(Z) \simeq \mathbf{x}(z_h)-\mathbf{X}(Z_h) + \frac{1+z_h}{H(z_h)}\left(\mathbf{u}_\parallel-\mathbf{V}_\parallel\right) = \mathbf{r} + \frac{1+z_h}{H(z_h)}\mathbf{v}_\parallel\;.
\label{eq:s(r)}
\end{equation}
where we have defined the relative velocity of a galaxy with respect to the void center as $\mathbf{v}$ (and considered its parallel-to-the-line of sight component $\mathbf{v}_\parallel$).
This relationship can be considered along the line of sight, as:
\begin{equation}
s_\parallel = r_\parallel + \frac{1+z_h}{H(z_h)} v_\parallel\;.
\label{eq: s_par(r_par)}
\end{equation}

Taking Eq.~\eqref{eq:xi^s}, which defines the VGCF in redshift space $\xi^s(\mathbf{s})$, we need to evaluate $\frac{\mathrm{d}r_\parallel}{\mathrm{d}s_\parallel}$. From Eq.~\eqref{eq: s_par(r_par)}, we obtain: 
\begin{equation}
\frac{\mathrm{d}r_\parallel}{\mathrm{d}s_\parallel} = \left(1 + \frac{1+z_h}{H(z_h)}\,\frac{\mathrm{d}v_\parallel}{\mathrm{d}r_\parallel}\right)^{-1}\;.
\label{eq: dr_par/ds_par}
\end{equation}
We now need to evaluate $\frac{1+z_h}{H(z_h)}\frac{\mathrm{d}v_\parallel}{\mathrm{d}r_\parallel}$. 

We can assume that the relative velocity field of matter around voids is coupled to the density following the principle of mass conservation at the linear order. This leads to the expression already introduced in Eq.~\eqref{eq:radial_velocity}, but for the specific case $z=z_h$.
Therefore, defining the cosine of the angle between the line of sight $\mathbf{X}/|\mathbf{X}|$ and $\mathbf{r}$ as $\mu_r=r_\parallel/r$, we obtain:
\begin{equation}
\frac{1+z_h}{H(z_h)}\frac{\mathrm{d}v_\parallel}{\mathrm{d}r_\parallel} = -\frac{f(z_h)}{3}\Delta(r)-f(z_h)\mu_r^2\left[\delta(r)-\Delta(r)\right]\;,
\label{eq: du_par/dr_par}
\end{equation} 
which we can finally plug into our definition of the VGCF in redshift space $\xi^s(\mathbf{s})$, to obtain: 
\begin{equation}
1+\xi^s(\mathbf{s}) = \frac{1+\xi(r)}{1-\frac{f}{3}\Delta(r)-f\mu_r^2\left[\delta(r)-\Delta(r)\right]}\;.
\label{eq:xi^s_nonlin}
\end{equation}

This equation shows that we can model the VGCF in redshift space $\xi^s(\mathbf{s})$, provided that we know the VGCF in real space $\xi(r)$ and the void-matter cross-correlation function $\delta(r) $. Since $r$ is needed, we can compute it with $r=(r_\parallel^2+r_\perp^2)^{1/2}$, considering that $r_\perp=s_\perp$, and that $r_\parallel = \frac{s_\parallel}{1 - \frac{f}{3}\Delta(r)}$. We note that $r_\parallel$ can be obtained iteratively (since its derivation requires $r$) by using $\Delta(s)$ as initial step and calculating $r_\parallel$ and $\Delta(r)$ until convergence.

In the above equations, to compute $r_\parallel$ and $r_\perp$ we have used $s_\parallel$ and $s_\perp$. This is, however, a non trivial step, because we need to obtain $s_\parallel$ and $s_\perp$ from the measured redshift and sky coordinates. To do so, a fiducial cosmology needs to be assumed. We can never access distances directly, since we measure redshift and sky coordinates. Therefore there is a cosmology dependence hidden in the conversion from coordinates to distances. While this may seem a curse, it is in fact very empowering: the link between coordinates and distances will provide a further tool to constrain cosmology. 

Considering the comoving distance in redshift space and its link to sky coordinates and redshift, cosmology enters in the conversion, such that: 
\begin{equation}
s_\parallel = \frac{c}{H(z)}\delta z\quad\mathrm{and}\quad s_\perp = D_\mathrm{A}(z)\delta\theta\;, \label{eq:s_comoving}
\end{equation} 
where $H(z)$ is the Hubble parameter (Eq.~\ref{eq:hubble_rate}) and $D_\mathrm{A}(z)$ is the comoving angular-diameter distance (Eq.~\ref{eq:DA}).

If we assume that we know the right cosmology (that is, we know the true values of cosmological parameters entering the definitions of $D_\mathrm{A}(z)$ and $H(z)$), then  all is good. If not, however, by using a \textit{fiducial} cosmology we will introduce a distortion effect on $s_\parallel $ and $s_\perp$. Following the same prescription introduced in Eq.~\eqref{eq:AP} but for $s_\parallel$ and $s_\perp$, we arrive at the expressions:  
\begin{equation}
q_\parallel \equiv \frac{s_\parallel}{s_\parallel^\mathrm{fid}}\quad \mathrm{and} \quad  q_\perp \equiv \frac{s_\perp}{s_\perp^\mathrm{fid}}\;.
\end{equation}
We also define the Alcock--Paczy{\'n}ski parameter as:
\begin{equation}
\varepsilon \equiv \frac{q_\perp}{q_\parallel} =  \frac{D_\mathrm{A}(z)H(z)}{D^\mathrm{fid}_\mathrm{A}(z)H^\mathrm{fid}(z)}\;.
\label{epsilon}
\end{equation}
If we write the cosine of the angle between $\mathbf{s}$ and the line of sight as $\mu_s\equiv \frac{s_\parallel}{s}$, the true $s$ and $\mu_s$ can be expressed as functions of the fiducial $s_\mathrm{fid}$ and $\mu_{s,~\mathrm{fid}}$ with:
\begin{gather}
s = \sqrt{q_\parallel^2 s_{\parallel,\mathrm{fid}}^2+q_\perp^2s_{\perp,\mathrm{fid}}^2} = s_\mathrm{fid} \, \mu_{s,\mathrm{fid}} \, q_\parallel\sqrt{1+\varepsilon^2(\mu_{s,\mathrm{fid}}^{-2}-1)}\;.
\label{s_fid}
\\
\mu_s = \frac{\mathrm{sgn}(\mu_{s,\mathrm{fid}})}{\sqrt{1+\varepsilon^2(\mu_{s,\mathrm{fid}}^{-2}-1)}}\;.
\label{mu_s_fid}
\end{gather}

Bringing everything together, when modeling the VGCF we assume values for $\varepsilon$, $q_\parallel$ and $q_\perp$. Additionally, it is common to assume that the relationship between $\delta(r)$ and $\xi(r)$ is linear, that is $\xi(r) = b \, \delta(r)$, where $b$ can be identified with the large-scale linear bias (but see below the list of assumptions). The same equation can therefore be written using $f/b$ instead of $f$ when we deal with biased tracers. 

Finally, we can define: 
\begin{equation}
\overline{\xi}(r) = \frac{3}{r^3}\int_0^r\xi(r')r'^2\mathrm{d}r'
\label{eq:xibar}
\end{equation}
and re-write Eq.~\eqref{eq:xi^s_nonlin} for the VGCF in redshift space as: 

\begin{equation}
1+\xi^s(\mathbf{s}) = \frac{1+\xi(r)}{1-\frac{1}{3}\frac{f}{b}\overline{\xi}(r)-\frac{f}{b}\mu_r^2\left[\xi(r)-\overline{\xi}(r)\right]}\;.
\label{eq:xi^s_nonlin2}
\end{equation}
This is the nonlinear equation that describes the VGCF in redshift space as a function of the VGCF in real space, and the ratio $f/b$. 

This model takes as input the density profile in real space, a value for the linear growth rate, as well as values for $q_\parallel$ and $q_\perp$ (thereby encoding cosmological parameters through their link to $D_\mathrm{A}(z)$ and $H(z)$). Fitting this model to the measured VGCF therefore allows us to constrain $f/b$, as well as $\Omega_{\rm m}$ (assuming a flat universe) or, for example, the $w_0$ and $w_a$ parameters if considering dynamical dark energy. 

Finally we can perform a further step: since we have considered the mass conservation equation at linear order in $\delta$, we can also expand this equation at linear order, obtaining \citep{Cai_2016,Hamaus_2017}: 
\begin{equation}
\xi^s(\mathbf{s}) \simeq \xi(r) + \frac{1}{3}\frac{f}{b}\overline{\xi}(r) + \frac{f}{b}\mu_r^2\left[\xi(r)-\overline{\xi}(r)\right]\;.
\label{eq: xi^s_lin}
\end{equation}

The above expression is valid under the assumption of linear theory ($|\delta|\ll1$) and for a null dispersion in the peculiar velocity field. It is possible to relax these assumptions by considering the quasi-linear streaming model \citep{Peebles_1980,Paz_2013,Cai_2016,Achitouv_2017,Paillas_2021}, which expresses the redshift-space correlation function as the convolution of its real-space counterpart with the probability distribution function for the peculiar velocities along the line of sight. In particular, it is reasonable to assume a Gaussian form for the pairwise velocity probability distribution function \citep{Fisher_1995}, which gives the approximate form:
\begin{equation}
    1+\xi^s(\mathbf{s}) = \left[1+\xi(r)\right] \int^{+\infty}_{-\infty}\frac{1}{\sqrt{2\pi}\sigma_v}\exp\left(-\frac{[v_\parallel-v(r)\mu_r]^2}{2\sigma_v^2}\right) \mathrm{d}v_\parallel\;,
\end{equation}
where $\sigma_v$ is the velocity dispersion, considered constant or a function of $r$. This model extension can be traced back to the previous expression when $\delta$ is small and $\sigma_v$ can be considered negligible \citep{Cai_2016}.

Until now we have considered the 2D VGCF. Nevertheless, following the strategy adopted by numerous works \citep[see e.g.][]{Cai_2016,Hamaus_2017,Nadathur_2019a,Nadathur_2019b,Hamaus_2020,Hamaus_2022,Aubert_2022,Fraser_2024,Degni_2025} it is possible to decompose it into multipoles with: 
\begin{equation}\label{eq: multipoles}
\xi^s_\ell(s) = \frac{2\ell+1}{2}\int\limits_{-1}^1\xi^s(s,\mu_s)\mathcal{L}_\ell(\mu_s)\mathrm{d}\mu_s\;,
\end{equation}
where $\mathcal{L}_\ell(\mu_s)$ are the Legendre polynomials of order $\ell$. 
In Eq.~\eqref{eq: multipoles}, the only non-vanishing multipoles at linear order in $\xi$ and $\overline{\xi}$ are the monopole ($\ell=0$) and quadrupole ($\ell=2$), giving: 
\begin{equation}\label{eq: mono_quadrupole}
\begin{split}
\xi^s_0(s) &= \left(1+\frac{f/b}{3}\right)\xi(r)\;, \\
\xi^s_2(s) &= \frac{2f/b}{3}\left[\xi(r)-\overline{\xi}(r)\right]\;.
\end{split}
\end{equation}

It is important to note that, while the 2D and multipoles versions of the model are mathematically equivalent, this may not be true from a data analysis perspective. Indeed, the very core of voids is expected to have higher error bars---for the multipoles this will correspond to a higher fraction of the number of bins (since multipoles are one-dimensional vectors) compared to the fraction of the number of bins in the 2D case. This could result in an enhanced signal-to-noise when using the 2D version of the model \citep{Verza_2024b}. 

\subsubsection{Real-space reconstruction and model extensions}
Although modeling the VGCF in redshift space can provide additional cosmological constraints---particularly on the growth of structures via the linear growth rate---it has been shown \citep{Nadathur_2019a, Correa_2022} that performing void identification directly in redshift space introduces challenges for the theoretical framework discussed in the previous section. In particular, this process may lead to a ``selection bias'', whereby voids with their major axis perpendicular to the line of sight in real space are preferentially identified. These voids appear elongated along the line of sight in redshift space due to peculiar velocities, which artificially increases their observed volume compared to voids with different orientations. As a result, these voids are more likely to be detected in redshift space because of their enhanced apparent size. This orientation-dependent selection bias violates the assumption of isotropy---i.e., the average spherical shape of voids in real space---on which the standard RSD modeling is built. It is hard to quantify the exact impact of this effect on constraints, but strategies have been proposed to reduce it. This effect is expected to have a stronger role for future surveys, since error bars will become smaller in light of a higher number of voids. 

With this in mind, a possible solution to the orientation-dependent selection bias is to identify voids in real space using a velocity-field reconstruction technique to recover the real positions of galaxies. Overall, the reconstruction approach is conceptually similar to the reconstruction techniques used in BAO analyses \citep{Eisenstein_2007,Padmanabhan_2012}, and typically relies on an iterative fast Fourier transform (FFT) algorithm \citep{Burder_2015} that solves the Zel'dovich equation in redshift space:
\begin{equation}\label{eq:displacement_rec}
    \nabla \cdot \mathbf{\Psi} + \frac{f}{b} \nabla \cdot [(\mathbf{\Psi \cdot \hat{r}) \, \hat{r}}] = - \frac{\delta_{\rm g}}{b}\;.
\end{equation}
Here, $\Psi$ denotes the Lagrangian displacement field, $f$ is the linear growth rate of structure, $\hat{\mathbf{r}} \equiv \mathbf{r}/r$, $b$ is the galaxy bias, and $\delta_\mathrm{g}$ is the galaxy overdensity field in redshift space. This equation can be solved using tools such as the \textsc{pyrecon} package\footnote{\url{https://github.com/cosmodesi/pyrecon}}, which estimates the density field on a regular grid after Gaussian smoothing, and then computes the displacement field. Since the Zel'dovich approximation is only valid on large scales, the reconstruction outcome depends critically on the choice of the smoothing radius: using too small a radius may over-correct RSD due to non-linear small-scale modes, while too large a radius may fail to remove all RSD contributions. In simulations, the optimal smoothing scale can be determined by minimizing the quadrupole of the reconstructed galaxy power spectrum, which should vanish if the assumed fiducial cosmology is correct.

Once the displacement field is determined, the galaxy positions can be shifted by:
\begin{equation}\label{eq:final_shift}
    \mathbf{\Psi_\mathrm{RSD}} = - f \, (\mathbf{\Psi} \cdot \hat{\mathbf{r}}) \, \hat{\mathbf{r}}\;,
\end{equation}
to approximately recover their real-space configuration, effectively subtracting the RSD contribution. The cosmological dependence of this shift is encapsulated in the parameter $\beta$ (see \Cref{sec:obs_VSF}), which can be marginalized over by performing reconstructions around its fiducial value \citep{Radinovic_2023}. While this method does not fully remove redshift-space effects, it significantly reduces orientation-dependent selection biases, providing a robust starting point for void identification. 

Having introduced reconstruction, it is useful to clarify the different ways in which it can be applied. Early works on reconstruction in VGCF analyses adopted a hybrid strategy: voids were identified in the reconstructed real space, and their centers were subsequently cross-correlated with galaxies in redshift space \citep{Nadathur_2019a,Nadathur_2020a,Nadathur_2020b,Radinovic_2023,Woodfinden_2022}. This \textit{hybrid} approach effectively mitigates void-finder selection effects, but at the cost of a more complex VGCF modeling. In particular, the displacement of void centers between real and redshift space, caused by bulk-flow velocities, introduces non-trivial modifications to the void dynamics in this hybrid setup and generally requires \textit{ad-hoc} mock catalogs for proper calibration (potentially entailing hidden dependencies from the simulation design). More recently, \citet{Degni_2025} proposed an alternative strategy based on cross-correlating both void centers and galaxy positions \textit{both} in reconstructed real space. This method has proven particularly effective, yielding a $23\%$ improvement in Alcock--Paczy{\'n}ski constraints with respect to the analysis fully carried out in redshift space. Despite the lack of rigorous constraints on the parameter $\beta$ (which enters the analysis through the reconstruction process), this approach appears especially promising for accurately testing the cosmological model with current and future galaxy catalogs. 

Finally, besides the methods from \citet{Radinovic_2023} and \citet{Degni_2025}, an alternative strategy to account for selection bias is to incorporate its impact directly into the theoretical modeling of the observed VGCF. A representative example is provided by \citet{Hamaus_2020}, who modified Eq.~\eqref{eq: xi^s_lin} as follows:
\begin{equation}
\xi^s(\mathbf{s}) = \mathcal{M}\left\{\xi(r) + \frac{1}{3}\frac{f}{b}\overline{\xi}(r) + \frac{f}{b}\mathcal{Q}\mu_r^2\left[\xi(r)-\overline{\xi}(r)\right]\right\}\;,
\label{eq: xi^s_lin2}
\end{equation}
where two nuisance parameters, $\mathcal{M}$ and $\mathcal{Q}$, are introduced to correct the VGCF monopole and quadrupole, respectively, for potential systematic effects. These include not only the presence of spurious underdensities, but also orientation-dependent distortions in the redshift-space void sample. The parameters $\mathcal{M}$ and $\mathcal{Q}$ can be calibrated using realistic mock catalogs, or alternatively marginalized over in a more conservative analysis.

The three techniques---hybrid reconstruction, full reconstruction and the inclusion of systematic effects directly in the modeling---have been successfully employed to extract cosmological constraints from spectroscopic survey data or simulations \citep[see e.g.][]{Nadathur_2020b,Hamaus_2020,Woodfinden_2022,Woodfinden_2023,Degni_2025}, as we will discuss in the following.

\subsubsection{Cosmological constraints} 
Over the past decade, a large number of studies have employed the VGCF as a cosmological probe, demonstrating its potential to constrain key parameters of the standard model of cosmology. These works span a wide range of galaxy redshift surveys, and the growing size and quality of the data have progressively enhanced the statistical precision of VGCF measurements. In parallel, the modeling of the signal has become increasingly sophisticated, allowing for a more accurate treatment of statistical uncertainties.

\begin{figure}[ht]
    \centering
    \includegraphics[width=1\linewidth]{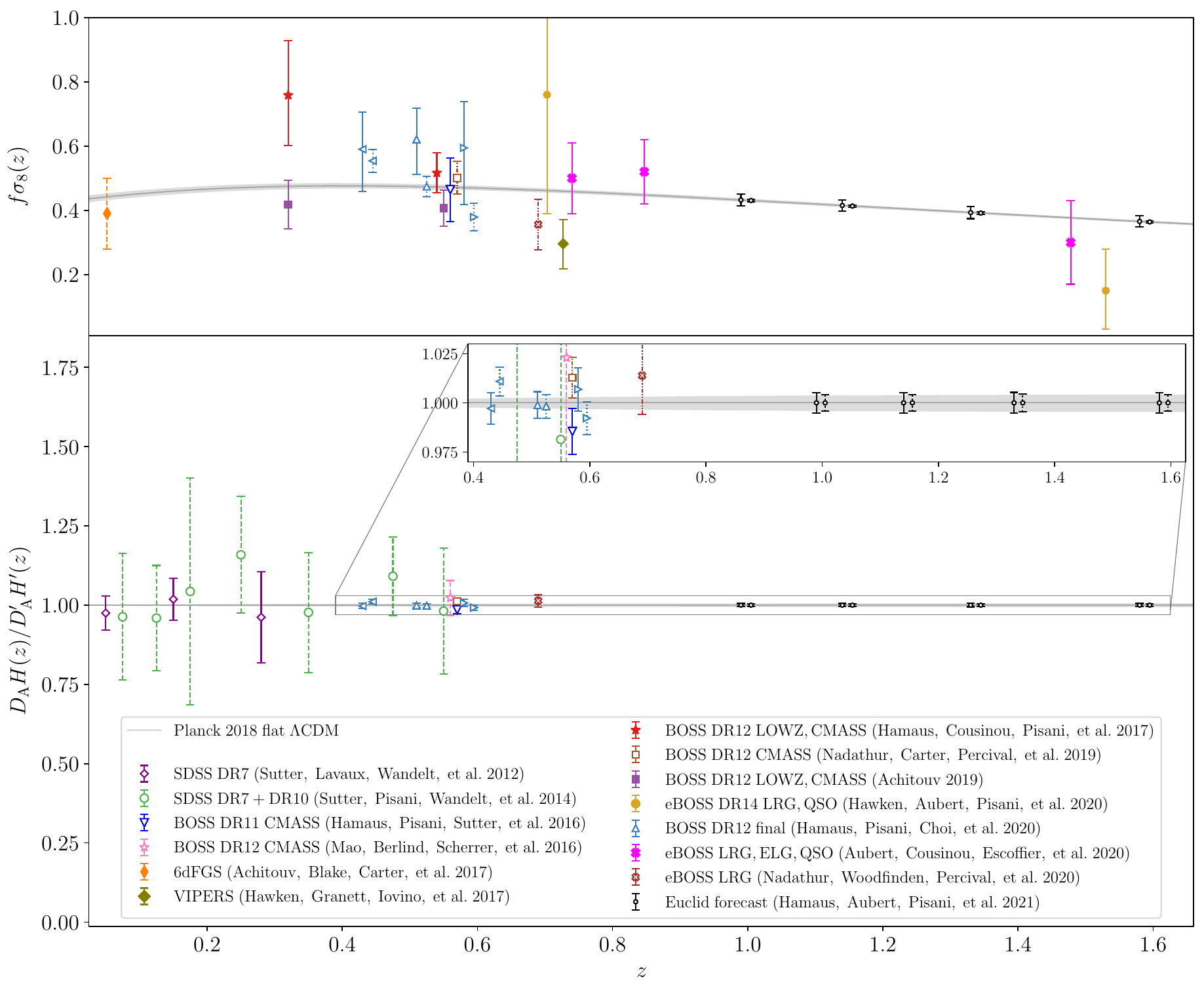}    
    \caption{Constraints on structure growth (top panel) and Universe's geometry (bottom panel) from the VGCF. Filled markers denote growth rate measurements obtained without accounting for the Alcock-Paczy{\'n}ski effect, while open markers include this correction. Different error bar styles reflect varying levels of modeling assumptions: solid lines for model-independent estimates, dashed for those calibrated on simulations, dotted for those based on mock catalogs, and dash-dotted when both simulations and mocks are used for calibration. Gray lines with shaded bands represent the baseline results from \citet{Planck_2020}, shown as a reference and including the corresponding fiducial values of $D'_\mathrm{A}/H'(z)$. For further details, we refer the reader to the original source \citep{Moresco_2022}.
    } \label{fig: AP and RSD Moresco}
\end{figure}

Among the most recent contributions are those by \citet{Hamaus_2016, Hamaus_2017, Hamaus_2020}, \citet{Nadathur_2019b,Nadathur_2020b}, \citet{Aubert_2022} and \citet{Woodfinden_2022,Woodfinden_2023}, each employing different void catalogs, modeling strategies, and data sets. These studies collectively laid the groundwork for applying void-based analyses to current and future galaxy surveys with increased statistical power. A chronological overview of the cosmological constraints obtained from the analysis of the VGCF is presented in \Cref{fig: AP and RSD Moresco}. Such works rely on robust and foundational theoretical work, leveraging on one of the fundamental assets of cosmic voids: their quasi-linear evolution and negligible dependence on baryonic feedback allowing simplistic theoretical models to hold down to small scales. As mentioned above, these methods allowed the community to obtain constraints at the level of traditional methodologies already with the somewhat low statistics from current data. They therefore hold great promise for delivering tighter constraints as we enter the era of big data.

Besides the analyses based on theoretically modeling the VGCF, recent techniques enable the use of machine-learning based approaches. Notably absent from this timeline is the recent analysis by \citet{Fraser_2024}, which presents the first application of a fully field-level inference pipeline to the VGCF. Instead of relying on stacked voids, the authors model the cross-correlation signal in configuration space for individual underdensities identified in the SDSS main galaxy sample. Despite the disadvantage that the method depends on a large number of expensive simulations, this approach enables a refined treatment of selection effects and anisotropies, and allows the authors to marginalize over uncertainties in the void definition itself.

\begin{figure}[ht]
    \centering
    \includegraphics[width=0.46\linewidth]{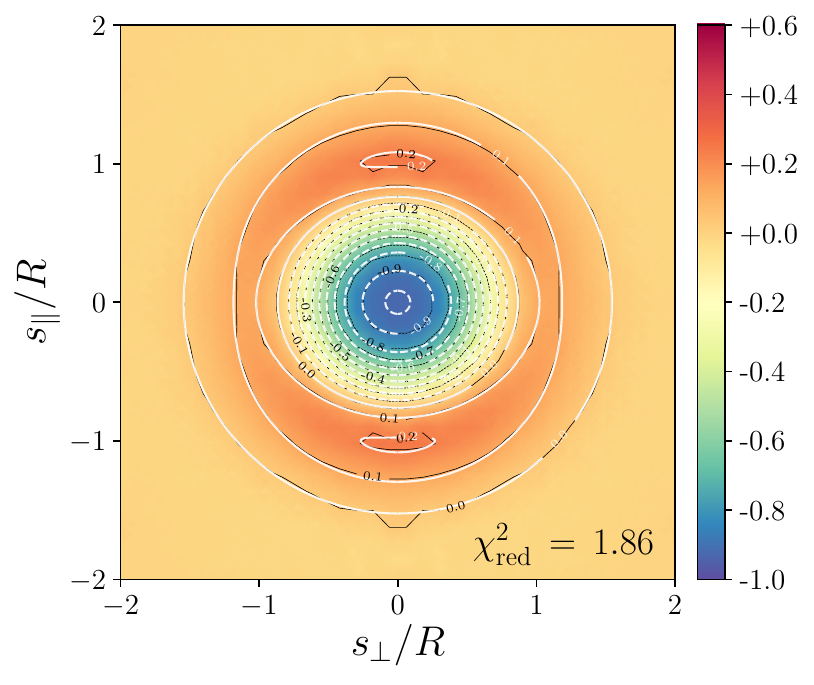}
    \includegraphics[width=0.52\linewidth]{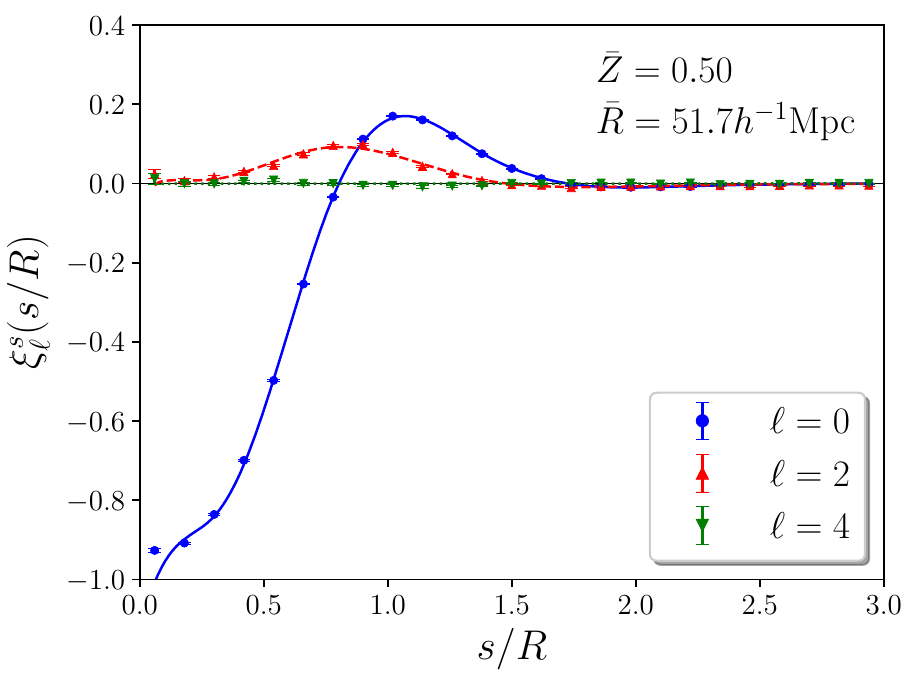}
    \includegraphics[width=0.46\linewidth]{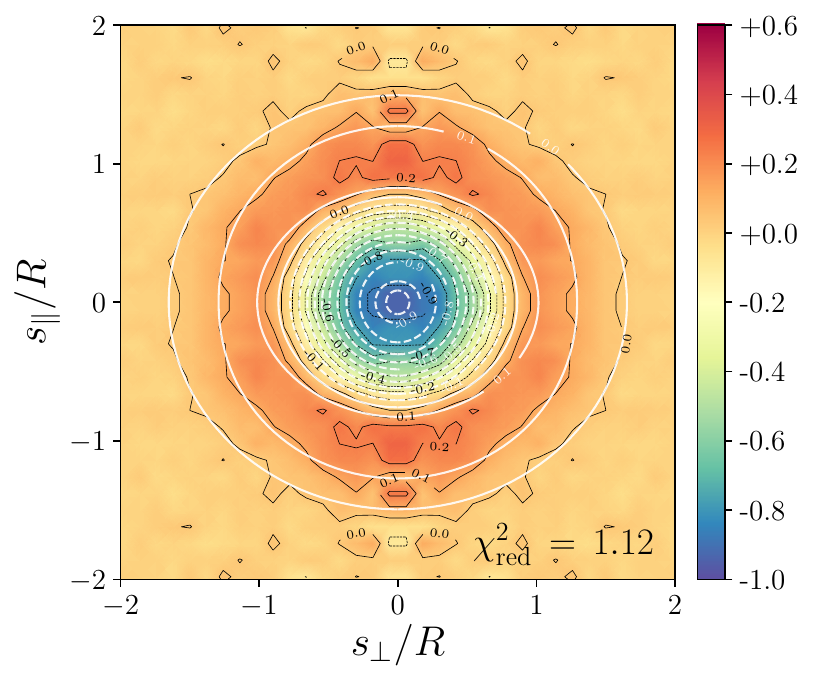}
    \includegraphics[width=0.52\linewidth]{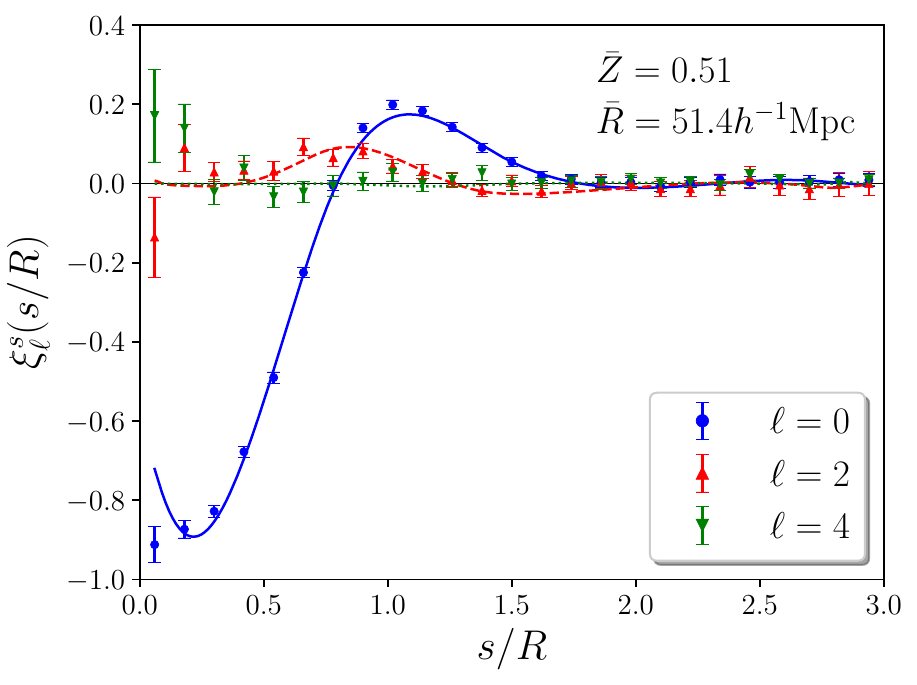}
    \caption{Comparison between the results obtained from the modeling of the VGCF measured in the BOSS collaboration official mocks (\textit{upper panels}) and galaxy catalogs (\textit{lower panels}).
    \textit{Left panels}: 2D redshift-space VGCF, $\xi^s(s_\perp,s_\parallel)$, measured from the two data sets. The mean value of the stacked measures is reported in color scale with black contours, while its best fit obtained with Eq.~\eqref{eq: xi^s_lin} is represented with white contours. \textit{Right panels}: corresponding monopole (blue dots), quadrupole (red triangles), and hexadecapole (green wedges) of the same VGCF, along with their respective model predictions indicated by solid, dashed, and dotted lines. For further details, we refer the reader to the original source \citep{Hamaus_2020}.}
    \label{fig: 2d VGCF BOSS}
\end{figure}

Among the many analyses carried out over the years, here we present the results of \citet{Hamaus_2020} as a representative case study.
Using the model described in Eq.~\eqref{eq: xi^s_lin2} together with data from the BOSS collaboration, these authors found excellent agreement between the VGCF in redshift space---measured both in mock catalogs and observational data---and the theoretical predictions. \Cref{fig: 2d VGCF BOSS} shows this comparison for the BOSS DR12 mocks and galaxy catalogs, presenting the VGCF in two dimensions ($s_\parallel, s_\perp$) as well as its multipole decomposition (monopole, quadrupole, hexadecapole). These results illustrate the success of the theoretical framework and demonstrate its ability to place reliable constraints on the cosmological model.

\begin{figure}[ht]
    \centering
    \includegraphics[width=0.45\linewidth]{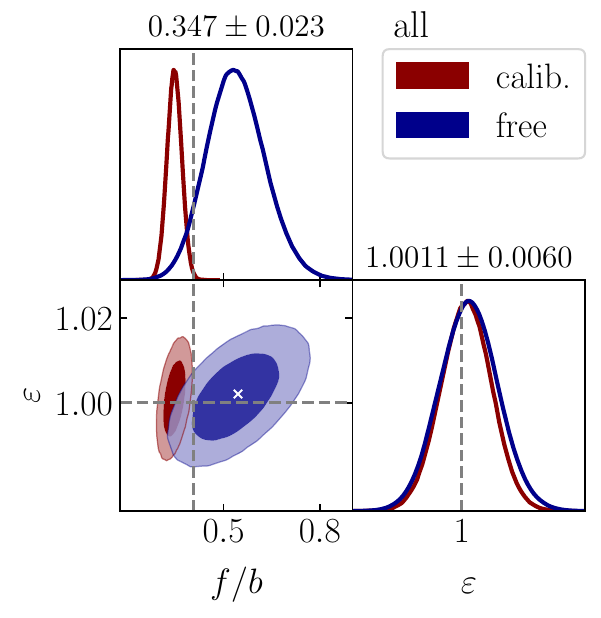}
    \includegraphics[width=0.45\linewidth]{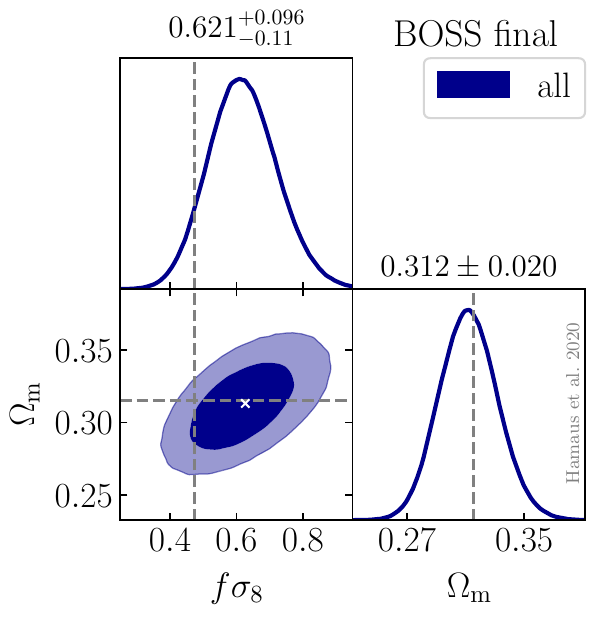}
    \caption{Constraints derived from modeling the VGCF shown in \Cref{fig: 2d VGCF BOSS}. \textit{Right}: confidence contours in the $f/b$–$\varepsilon$ parameter space, represented in blue when obtained by marginalizing over the nuisance parameters $\mathcal{M}$ and $\mathcal{Q}$ (see Eq.~\ref{eq: xi^s_lin}), and in red when calibrating these parameters using mock catalogs. \textit{Left}: Confidence contours in the $f\sigma_8$–$\Omega_{\rm m}$ parameter space, obtained using the calibration-free approach only. For further details, we refer the reader to the original source \citep{Hamaus_2020}.}
    \label{fig: VGCF BOSS constraints}
\end{figure}

\cref{fig: VGCF BOSS constraints} presents the cosmological constraints derived from the full BOSS sample, centered at a mean redshift of $z = 0.51$. From this analysis, \citet{Hamaus_2020} obtained $D_{\rm A} H / c = 0.588 \pm 0.004$ by converting the constraints on $\varepsilon$ (see \Cref{epsilon}); $f\sigma_8 = 0.621 \pm 0.104$ by multiplying $f/b$ by the average galaxy bias $b$ and the value of $\sigma_8$ from \citet{Planck_2020}; and $\Omega_{\rm m} = 0.312 \pm 0.020$ under the assumption of a flat-$\Lambda$CDM cosmology.

These constraints could be further tightened by fixing the nuisance parameters $\mathcal{M}$ and $\mathcal{Q}$ (see Eq.~\ref{eq: xi^s_lin2}) to their best-fit values derived from realistic mock catalogs, as reported in left panel of \Cref{fig: VGCF BOSS constraints}. However, such a calibration relies on simulations and may introduce cosmology-dependent systematic errors. Marginalizing over $\mathcal{M}$ and $\mathcal{Q}$ thus represents a more conservative and model-independent approach providing larger error bars that are, however, more robust. Indeed, even without calibration, the agreement between the VGCF model and the VGCF BOSS observations is outstanding, and the resulting constraints are shown to be both robust and competitive with those obtained from more traditional cosmological probes. Finally, this methodology has also been tested in the context of a parameter-masked mock challenge, further testing its robustness \citep{Beyond2pt_2025}. 

\subsection{Measurements and constraints from CMBX}\label{sec: CMBX_obs}

As presented in \Cref{sec:theory:CMBX}, the imprint of void-CMB lensing and ISW in voids arises from the integrated paths of photons passing through voids. As a consequence, for analysis purposes, the relevant quantity is the projected matter density profile around voids, together with its characteristic properties. Therefore, for CMBX analyses, aside from using spectroscopic 3D voids, it is common to rely on 2D voids, identified in the galaxy distribution projected along the line of sight, in redshift bins. For this reason, photometric galaxy surveys, such as the Dark Energy Survey \citep[DES,][]{DES_2016}, have often been used for CMBX--void analyses.

\subsubsection{CMB lensing}\label{subsec:obs_CMB_lensing}

\begin{figure}[ht]
    \centering
    \includegraphics[width=\linewidth]{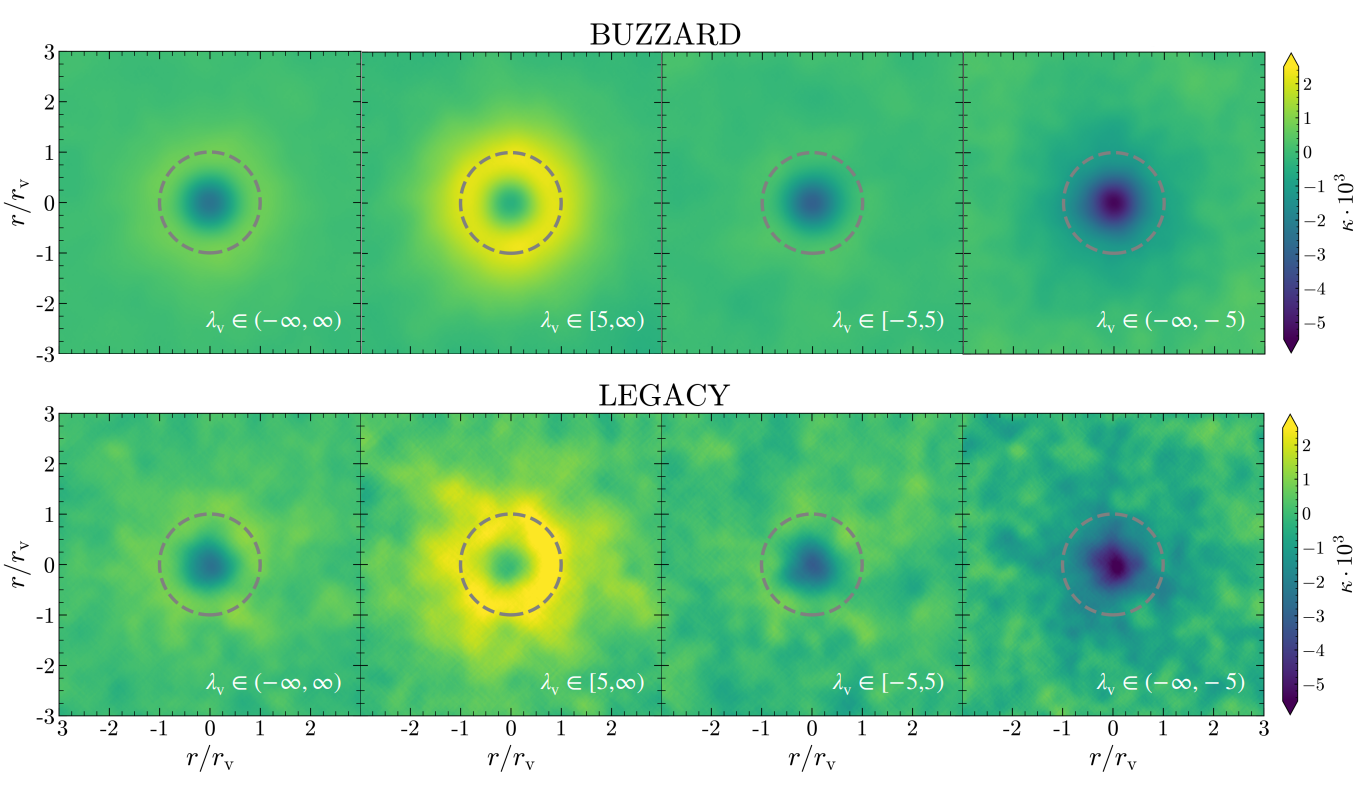}
    \includegraphics[width=\linewidth]{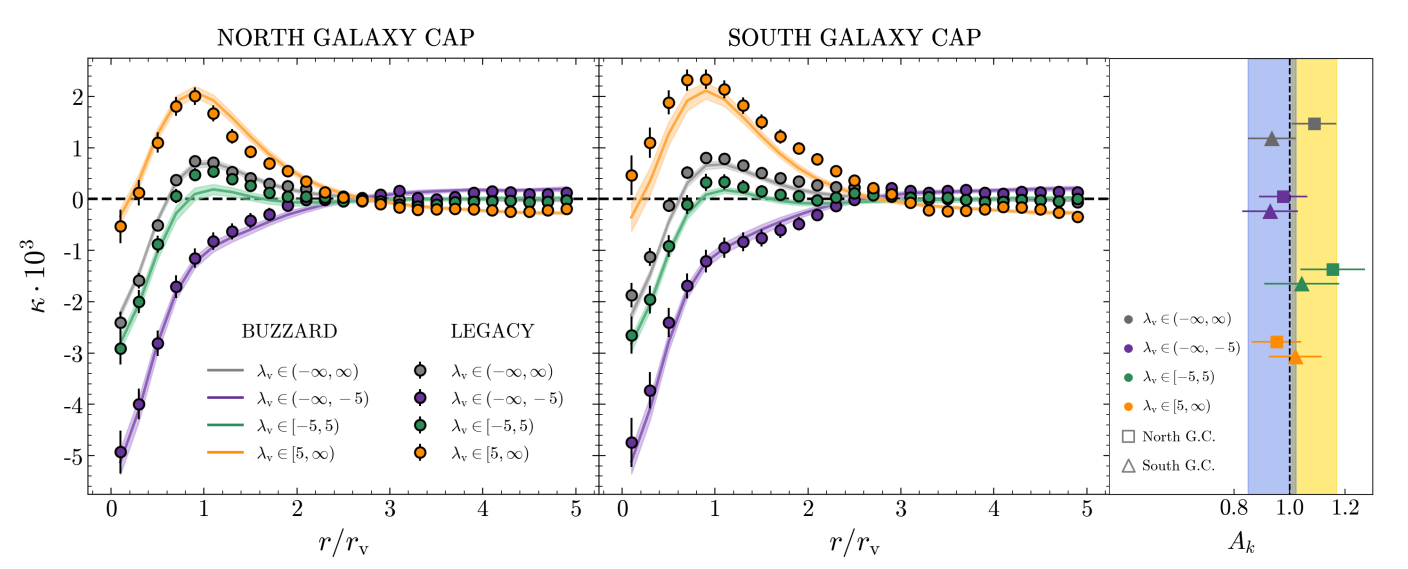}
    \caption{\textit{Top}: comparison among the convergence field $\kappa$ stacking voids from simulations (\textit{upper panels}) and from the Dark Energy Spectroscopic Instrument legacy survey (\textit{lower panels}). \textit{Bottom}: convergence profile in voids, solid lines show the measurements from the simulation, symbols with error bars show the measurements of the stacked $\kappa$ profiles measured from voids in the Dark Energy Spectroscopic Instrument legacy survey. Different colors are used to indicate the signal in voids with different gravitational potential, using $\lambda_{\rm v}$ as proxy. For further details, we refer the reader to the original source \citep{Sartori_2025}.}
    \label{fig:CMBlensing DESI}
\end{figure}

To measure CMB lensing sourced by voids, the standard methodology adopted in the literature is to stack convergence maps of the CMB. The first detection was performed by \citet{Cai_2017} using voids from SDSS and Planck convergence maps, with a significance of $3.2 \sigma$, statistically consistent with $\Lambda$CDM. For CMB lensing, the convergence field $\kappa$ (Eq.~\ref{eq:convergence}) peaks around $z\simeq 1$ \citep{Lewis_Challinor_2006}, therefore high-redshift galaxy surveys are expected to produce a higher significance in the detection. In fact, further analyses measured the lensing signal in both 2D and 3D voids detected in redshift surveys with increasing significance \citep{Raghunathan_2020,Vielzeuf_2021,Hang_2021,Kovacs_2022c,Camacho-Ciurana_2024,Demirbozan_2024,Sartori_2025}. Moreover, for 3D voids, by splitting the void population according to an empirical proxy of their gravitational potential, $\lambda_{\rm v}$ \citep{Nadathur_2017}, the significance of void lensing detection reaches values of 10--15 $\sigma$ \citep{Camacho-Ciurana_2024,Demirbozan_2024,Sartori_2025}.
Conversely to ISW, measurements of CMB lensing in voids have always been consistent with $\Lambda$CDM. \cref{fig:CMBlensing DESI} shows an example of CMB lensing in voids derived in \cite{Sartori_2025} with the data of the Dark Energy Spectroscopic Instrument (DESI) legacy survey \citep{DESI_whitepaper_2016}. 

To conclude this section, it is worth noting that the cosmological inference for the ISW and CMB lensing in voids relies on the density profiles measured around them. These voids are typically traced by luminous tracers such as galaxies, and thus depend on both the evolution of the total matter density field and the tracer bias. Given the complexity of a fully theoretical prediction for these profiles, current analyses usually compare the signals measured in survey data with those extracted from mock catalogs designed to reproduce the main features of the survey. The computational expense of producing large light-cone mocks with the corresponding lensing and ISW maps has so far limited the capability of analyses to verify the consistency of the signal with $\Lambda$CDM. However, it has been shown that the ISW and CMB lensing in voids are powerful probes for inferring cosmological parameters \citep{Chantavat_2016,Chantavat_2017} and measuring physical quantities, such as the total neutrino mass \citep{Vielzeuf_2023}, modified gravity and dark energy models \citep{Kovacs_2022b}.

\subsubsection{Integrated Sachs--Wolfe}

The first measurements of the ISW in voids was performed by \citet{Granett_2008}, studying the population of super-voids in SDSS \citep{Adelman-McCarthy_2008}, i.e. very large and shallow underdense structures in the galaxy distribution. Their study was motivated by the fact that, in the galaxy-CMB angular power spectrum around $z \sim0.5$, the ISW signal peaks at the spherical multipole $l \sim20$. This corresponds to structures with an angular extension of $\sim4^\circ$, or a comoving radius around 100 $h^{-1}$ Mpc. The ISW signal was detected by stacking the CMB temperature map beyond each void. This methodology allows us to increase the signal-to-noise ratio of the ISW in voids, by averaging out primary CMB fluctuations and noise, which do not correlate with voids. This is now considered the standard methodology. Notably, a 4.4$\sigma$ detection of ISW in voids is reached, greatly surpassing the one obtained with CMB-galaxy cross correlation of the same period \citep{Ho_2008,Giannantonio_2008}.

\begin{figure}[ht]
    \centering
    \includegraphics[width=0.95\linewidth]{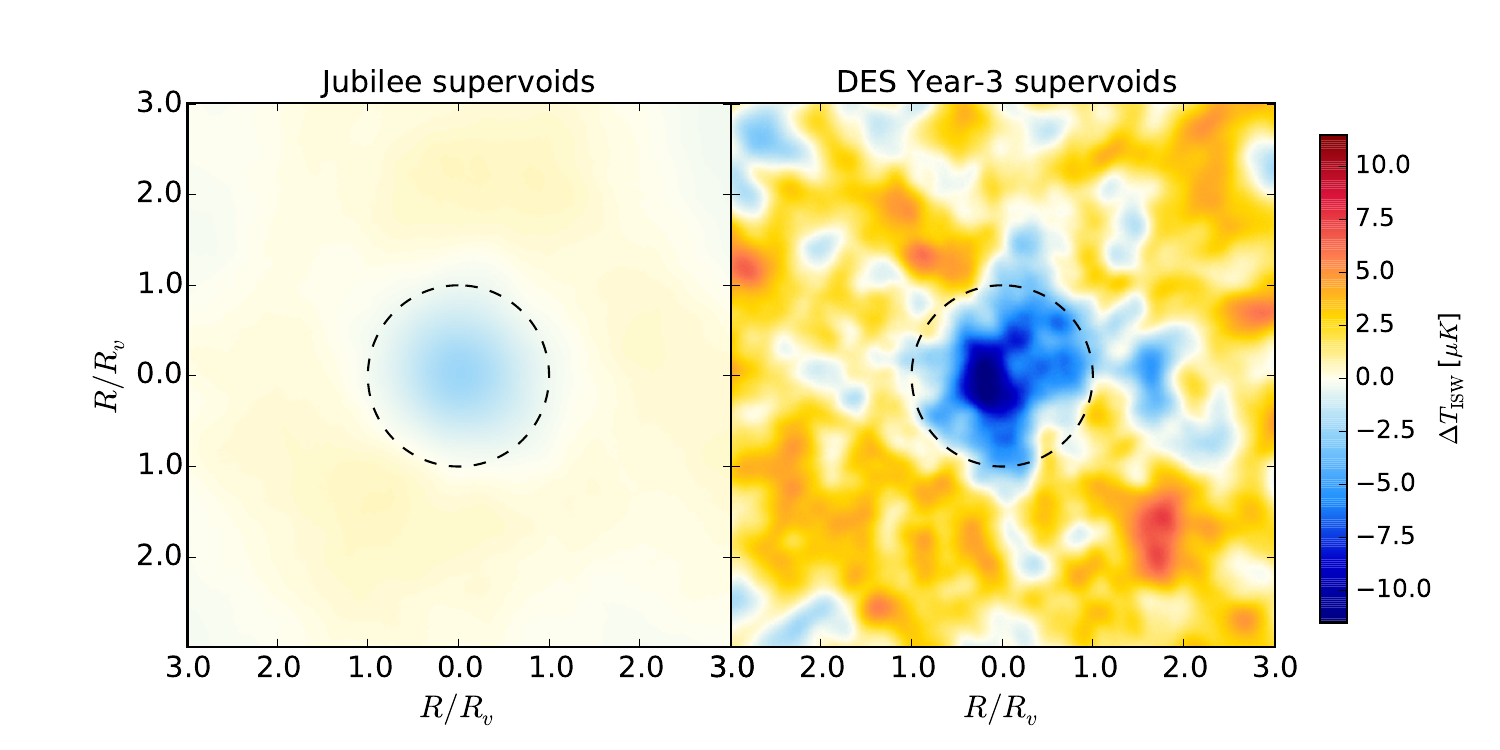}
    \includegraphics[width=\linewidth]{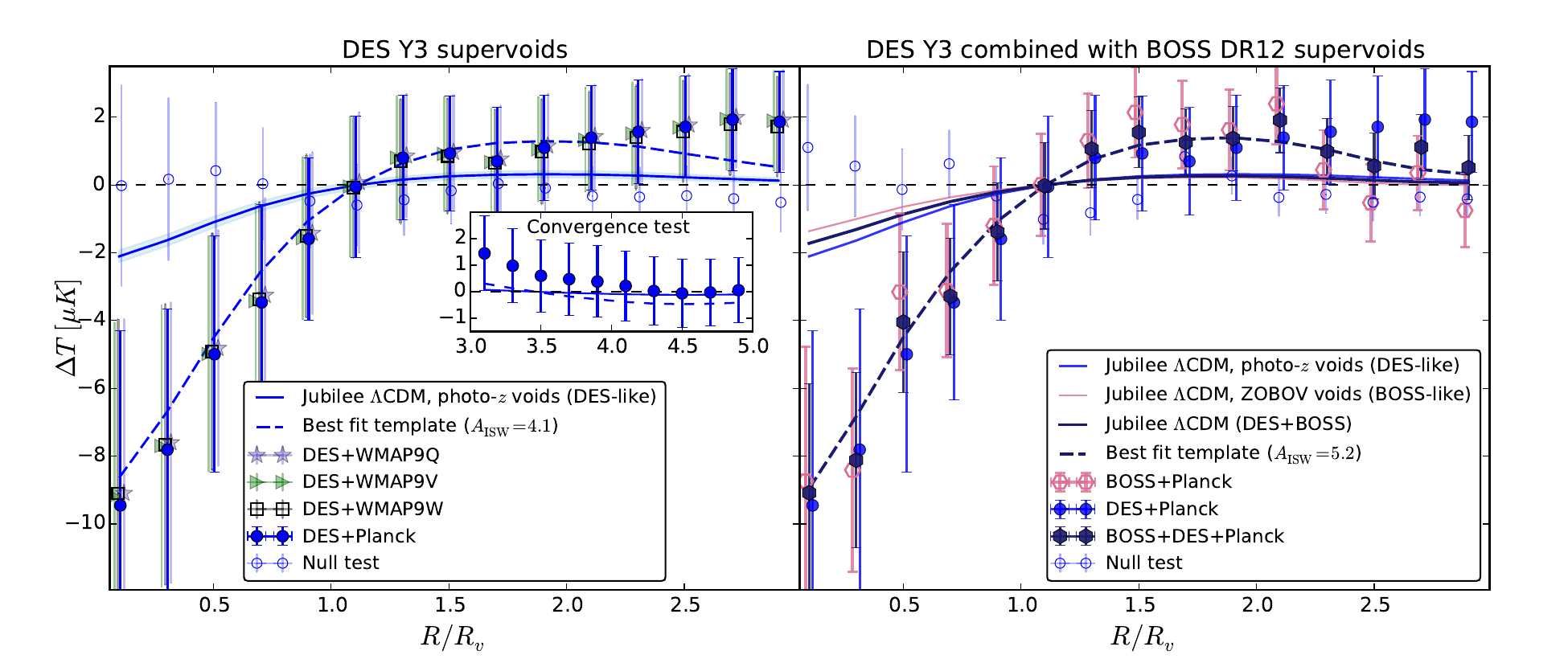}
    \caption{\textit{Top}: comparison among the ISW signal in voids measured in simulations (left) and the signal measured in the DES year-3 voids (right). \textit{Bottom}: temperature profile in voids, solid lines show the measurements from simulations, symbols with error bars show the measurements of stacked CMB temperature in the direction of the DES and the BOSS voids. For further details, we refer the reader to the original source \citep{Kovacs_2019}.}
    \label{fig:ISW_DES}
\end{figure}

After the first detection, several studies investigated the signal with theoretical models and simulations, all showcasing that the detected signal is stronger than the expected one for a $\Lambda$CDM cosmology \citep{Inoue_2010,Cai_2010,Papai_2010,Nadathur_2012,Flender_2013}. These findings opened the questions about a tension in the ISW sourced by voids, which has been thoroughly investigated. In particular, further data analyses and comparison with simulations, show that the signal sourced by small voids is in agreement with $\Lambda$CDM \citep{Nadathur_Crittenden_2016}, while the signal from large voids seems stronger \citep{Cai_2014,Kovacs_Granett_2015,Cai_2016,Kovacs_2017,Kovacs_2018,Kovacs_2019,Kovacs_2022b,Kovacs_2022a}.
Recent measurements still report a tension at $2.6 \sigma$ and a signal 5 times larger than the expected one \citep{Kovacs_2019}. While physics beyond $\Lambda$CDM may be the source of this strong signal, systematic errors may impact measurements as well \citep{Cai_2014,Kovacs_Granett_2015,Cai_2016,Kovacs_2017,Kovacs_2018,Kovacs_2019,Kovacs_2022a,Kovacs_2022b}. \cref{fig:ISW_DES} shows an example of the ISW measured in the DES voids \citep{Kovacs_2019}.
The comparison of the ISW signal with the theoretical expectation is typically done using large simulations that are computationally expensive. This is why until now ISW analyses have not been used to constrain cosmological parameters and models. That said, ISW in voids is recognized as a potentially great probe for dark energy, modified gravity, primordial non-Gaussianity, and non-standard cosmological models \citep{Inoue_2010,Nadathur_2012,Cai_2014,Kovacs_2017,Kovacs_2018,Kovacs_2019,Kovacs_2022b}.

\subsection{Measurements and constraints from void lensing}

\begin{figure}[ht]
    \centering
    \includegraphics[trim=0cm 0cm 2cm 0cm, clip, width=0.49\linewidth]{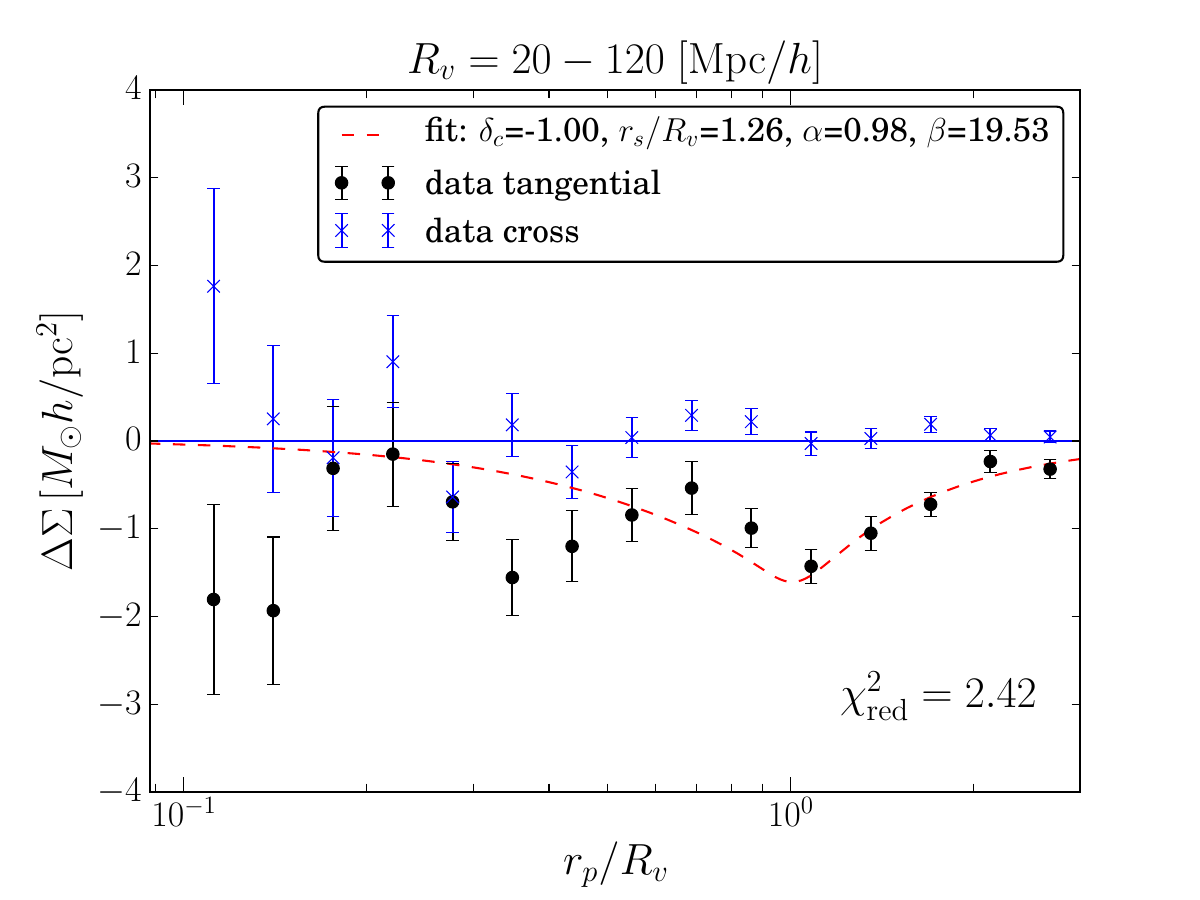}
    \includegraphics[trim=0cm 0cm 2cm 0cm, clip, width=0.49\linewidth]{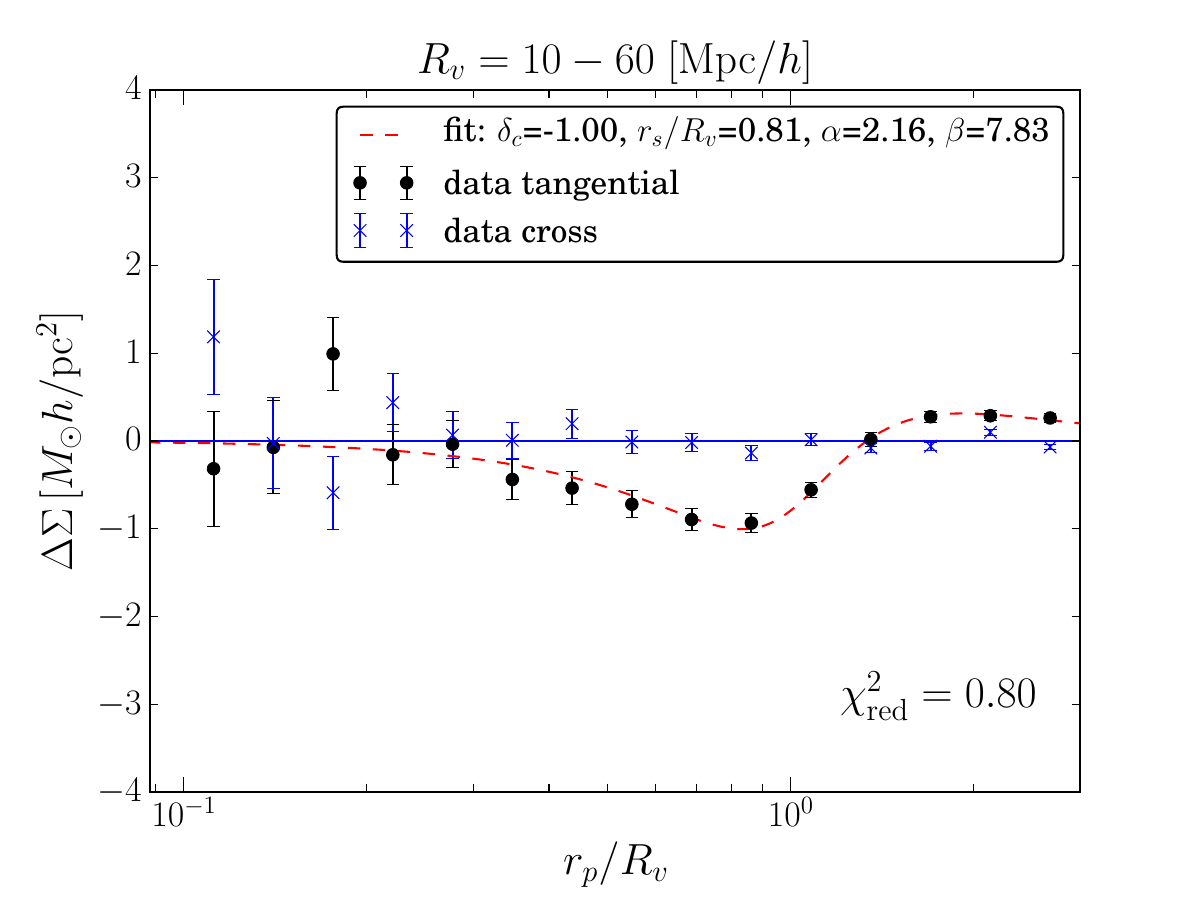}
    \caption{Void-galaxy lensing signal measured in the DES for 2D (left) and 3D (right) voids. The parametric model used for the fit derives from the projection of the void density profile reported in Eq.~\eqref{eq:HSW}. For further details, we refer the reader to the original source \citep{Fang_2019}.}
    \label{fig: Fang2019 lensing}
\end{figure}

In void-galaxy lensing measurements, voids act as lens and background source galaxies are lensed. The shear field (see Eq.~\ref{eq:shear_esmd}) is estimated from the shape of source galaxies, which are distorted by the lenses. To measure the lensing signal in and around voids, the typical method consists of stacking or cross-correlating the shear field with the direction of voids. Historically, there are two possible ways to obtain the lens population, i.e. voids. The first considers void detected in the 3D distribution of galaxies, from either spectroscopic or photometric surveys; the second relies on detecting 2D voids in the projected galaxy density field, usually from photometric galaxy surveys.

The first detection of the void-galaxy lensing signal, by \citet{Gillis_2013}, uses photometric redshifts from the Canada-France-Hawaii Telescope Lensing Survey \citep[CFHTLenS,][]{Erben_2013}, while the first detection from voids detected in a spectroscopic redshift sample is by \citet{Melchior_2014}, using SDSS data \citep{Sutter_2012,Abazajian_2009}. Voids detected in 2D are sometimes called tunnel voids because 2D void finders are particularly adapted to detect underdensities elongated along the line of sight, that is tunnel-shaped underdensities.
It is important to note that this kind of underdensity may break the validity of the thin lens approximation used in the calculations of \Cref{sec:theory:CMBX,sec:theory:lensing}. Even if that simple analytical formulation may break down, the lensing signal can be strong, and it is typically modeled using simulations \citep{Gruen_2016,Davies_2019,Sanchez_2017,Paillas_2019,Fang_2019,Jeffrey_2021,Maggiore_2025}. In the last decade, many other works measured the lensing signal with increasing significance \citep{Clampitt_2015,Gruen_2016,Sanchez_2017,Fang_2019,Jeffrey_2021,Hunter_2025}. \cref{fig: Fang2019 lensing} shows an example of the void-galaxy lensing signal, from the DES survey \citep{Fang_2019}.

As for CMBX, void lensing analyses have until now provided evidence of signal detection and consistency with the $\Lambda$CDM framework, without providing constraints on cosmological parameters. This is due to the difficulty of providing an analytical model for the excess of surface mass density, which strongly depends on the specific void finder considered. Therefore, a direct comparison with simulations is required. To overcome the absence of a full theoretical model, a simulation-base inference approach has recently been explored, but has not yet been applied to real data \citep{Su_2025}. As for future expectations, void-galaxy lensing has been shown to be particularly sensitive to modified gravity and massive neutrinos \citep{Barreira_2015,Baker_2018,Paillas_2019,Davies_2019,Maggiore_2025}.

\section{Voids in modern surveys}\label{sec:modern_surveys}

In this section, we review modern redshift surveys that have been---or are currently expected to be---of fundamental importance for the study of cosmic voids. We report their main characteristics, such as redshift range, sky coverage, and number of observed galaxies, and we also report on the number of voids measured or forecasted.
At this point in the review, the careful reader will have noticed that the optimal definition of cosmic voids can vary significantly depending on the type of void statistic measured and analysis considered. The number counts, size distribution, and required purity cuts of voids are closely linked to the theoretical framework adopted, making both the comparison and prediction of void samples across different surveys inherently challenging.

For this reason, to give an idea of numbers we base our comparison primarily on watershed voids---i.e., those identified by finders such as \textsc{Zobov}, \vide and \textsc{Revolver}---applying purity cuts\footnote{This procedure mainly removes underdensities that fall within the shot noise regime, usually corresponding to scales smaller than at least twice the mean separation of mass tracers. Other quality cuts may involve the rejection of voids intersecting the survey mask or redshift boundaries.}, which we take as our standard reference, given their widespread use in cosmological analyses to date. Watershed voids without purity cuts are instead treated as an upper-limit estimate, since they are generally unsuitable for cosmological analyses (albeit analyses such as the VGCF remain reliable even with very mild cuts).
Conversely, spherical voids---such as those identified with \spark or post-processed via cleaning techniques---are treated as a lower-limit estimate. These typically form a smaller sample composed of voids with very low internal densities. Notably, this latter class of voids has the advantage of being well described by current theoretical VSF models, which can be used to predict their number counts based on the survey volume, tracer number density, and tracer bias. Other types of voids, however, generally require dedicated simulations to estimate their expected abundance, often yielding a stronger constraining power due to the additional information on void shape and orientation \citep{Kreisch_2022}.

\begin{figure}[ht]
    \centering
    \includegraphics[width=0.9\linewidth]{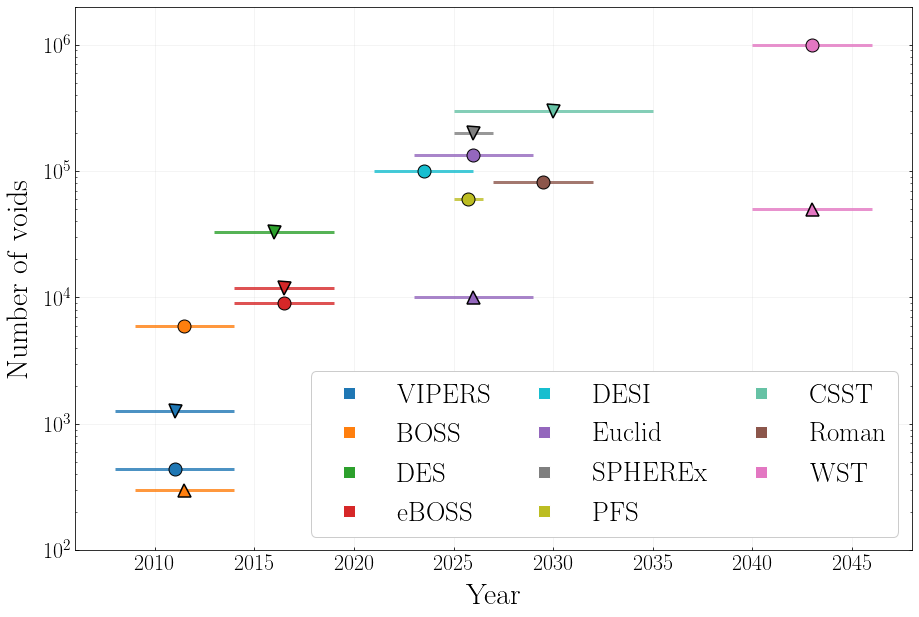}
    \caption{Number of voids detected or expected to be identified in various surveys over the years. Circular markers indicate standard estimates (i.e., watershed voids with applied purity cuts), while downward- and upward-pointing triangles represent potential overestimates and underestimates (i.e. optimistic and pessimistic), respectively. Horizontal bars indicate the planned operational time spans for each survey. The y-axis is shown on a logarithmic scale. The predictions for a number of surveys are notably absent from this representation (see \Cref{sec:other surveys}), as reliable measurements or forecasts of void number counts are not yet available for those.}
    \label{fig:Nvoids-years}
\end{figure}

With this in mind, we now compare the cosmic voids used in, and expected from, different surveys over the years. This comparison is shown in \Cref{fig:Nvoids-years}, where different marker styles indicate standard, upper-limit and lower-limit estimates. For each survey, we report multiple void number counts when estimates for different sample types are available. The surveys presented in this figure are the same ones analyzed in the following sections, and their operational time spans are indicated by the corresponding range of years.
Despite the variety of estimates, it is noteworthy that the number of voids is expected to follow a trend of exponential growth, as future surveys will probe increasingly larger volumes of the Universe with ever greater detail. Notably absent from the result collection presented in \Cref{fig:Nvoids-years} is a forecast of void numbers for the Vera C. Rubin Observatory's Legacy Survey of Space and Time \citep[LSST,][]{Ivezic_2019}, along with those of other surveys described in \Cref{sec:other surveys}. Furthermore, we note that in this review we focus on cosmological surveys; however cosmic voids have been identified and studied in local galaxy surveys as well~\citep[e.g.][see also Appendix \ref{app:A}]{Courtois_2023,Hansen_2025,Malandrino_2025}.

In the following sections we provide a more detailed overview of the existing void counts and forecasts for past, ongoing, and future surveys (see \Cref{sec:past ongoing surveys,sec:future surveys}), and we also highlight other relevant surveys for which reliable void forecasts are not yet available (\Cref{sec:other surveys}).

\subsection{Past and ongoing surveys}\label{sec:past ongoing surveys}

\paragraph{VIPERS}
The VIMOS Public Extragalactic Redshift Survey \citep[VIPERS,][]{Scodeggio_2018} was a Large Program of the European Southern Observatory (ESO), conducted at its facilities in the Atacama Desert of northern Chile. It started in 2008 and completed in 2014, producing a catalog of $\sim90{,}000$ spectroscopic galaxies covering an area of $24 \, \mathrm{deg}^2$ in the redshift range $0.5<z<1.2$. Despite its relatively limited area, the exquisite redshift precision and high spectroscopic sampling made VIPERS one of the first effective test benches for identifying a statistically significant sample of cosmic voids. Using a finder based on the detection of empty, non-overlapping spheres, \citet{Micheletti_2014} identified 411 statistically significant underdensities in the range $0.55<z<0.9$, with radii $R>10.5 \ h^{-1} \, \mathrm{Mpc}$. Later, \citet{Hawken_2017} employed a less conservative selection from the same void catalog, consisting of 1263 voids, to constrain the growth rate of structure.

\paragraph{BOSS and eBOSS} 
The Baryon Oscillation Spectroscopic Survey \citep[BOSS,][]{Dawson_2013} and its successor, the extended BOSS \citep[eBOSS,][]{Dawson_2016}, were key components of the SDSS. Both surveys were conducted using the 2.5-meter Sloan Foundation Telescope at Apache Point Observatory in New Mexico, USA. BOSS was operational from 2009 to 2014 as part of SDSS-III \citep{Eisenstein_2011}, while eBOSS ran from 2014 to 2020 within SDSS-IV \citep{Blanton_2017}. Both surveys have played a pivotal role in the development of cosmic void studies, as shown in several works \citep[e.g.][]{Hamaus_2016,Achitouv_2019,Hamaus_2020, Aubert_2022, Contarini_2023, Contarini_2024} and thoroughly described in \Cref{sec:constraints_from_surveys}.
BOSS targeted over one million spectroscopic galaxies in the redshift range $0.2 < z < 0.75$ (LOWZ and CMASS samples), covering 9,493~$\mathrm{deg}^2$. From this data set, a total of 5,952 watershed voids were identified after applying purity cuts. When applying conservative purity cuts to spherical voids extracted from the same galaxy catalog, the number of voids is significantly reduced, yielding only a few hundred objects.
Subsequently, eBOSS expanded the redshift coverage by including multiple tracers: 174,816 luminous red galaxies (LRGs) in $0.6 < z < 1.0$ over 4,242~$\mathrm{deg}^2$, 73,736 emission line galaxies (ELGs) in $0.6 < z < 1.1$ over 1,170~$\mathrm{deg}^2$, and 343,708 quasars (QSOs) in $0.8 < z < 2.2$ over 4,808~$\mathrm{deg}^2$. The number of watershed voids detected in eBOSS were 4,228 for the combined LRG+CMASS sample, 2,097 for ELGs and 5,451 for QSOs. After applying purity cuts, these counts reduce to 2,814, 1,801 and 4,347, respectively.

\paragraph{DES}
The Dark Energy Survey \citep[DES,][]{DES_2016} was a six-year observational campaign that began in 2013 and mapped $5000 \ \mathrm{deg}^2$ of the southern sky using the 4-meter Blanco Telescope at Cerro Tololo Inter-American Observatory in Chile. It observed nearly 300 million galaxies and tens of thousands of clusters, reaching redshift $z = 1.4$.
Several studies have used 2D voids from DES to test the $\Lambda$CDM model, particularly by examining void-galaxy lensing, CMB lensing, and ISW \citep{Gruen_2016,Sanchez_2017,Kovacs_2019,Fang_2019,Vielzeuf_2021,Kovacs_2022c,Jeffrey_2021,Kovacs_2022a,Kovacs_2018,Demirbozan_2024}. In this review, we focus instead on estimates for 3D voids, which allow for a more direct comparison with other surveys (past, ongoing, and future). Their number counts are generally easier to model, as they mainly depend on the survey volume and the spatial resolution of tracers.
3D voids in DES data have been studied in four works: \citet{Pollina_2019}, \citet{Fang_2019} and \citet{Vielzeuf_2021}, based on Year 1 data (1,321~$\mathrm{deg}^2$), and \citet{Demirbozan_2024}, using the full Year~3 data set. These voids were identified both in the Red-sequence Matched-filter Galaxy Catalog \citep[redMaGiC,][]{Rozo_2016}---which includes photometrically selected luminous red galaxies \citep{Fang_2019,Vielzeuf_2021,Demirbozan_2024}---and in the galaxy cluster distribution from the Red-sequence Matched-filter Probabilistic Percolation algorithm \citep[redMaPPer,][]{Rykoff_2014} \citep{Pollina_2019}.
As a reference, we report here the number of 3D voids identified in DES~Y3 by \citet{Demirbozan_2024}, using the \textsc{Voxel} implementation of the \textsc{REVOLVER} watershed-based void finder \citep{Nadathur_2019a}. The galaxy sample was divided into three redshift bins, i.e. $0.2<z<0.43$, $0.43<z<0.59$, and $0.59<z<0.75$, yielding 6,821, 10,861, and 15,027 voids, respectively, for a total of 33,427 voids.

\paragraph{DESI}
The Dark Energy Spectroscopic Instrument \citep[DESI,][]{DESI_whitepaper_2016} started mapping the Universe in 2021 using ten spectrographs equipped with 5,000 robotic fibers mounted on the 4-meter Mayall Telescope at Kitt Peak National Observatory in Arizona, USA. Over the course of its program, DESI aims to measure redshifts for about 20 million luminous red galaxies (LRGs), emission line galaxies (ELGs), and quasars (QSOs), as well as 10 million galaxies from the Bright Galaxy Sample (BGS) at $z < 0.6$ \citep{Hahn_2003}. Covering 14,000~$\mathrm{deg}^2$, it has already delivered the largest spectroscopic data set of galaxies and quasars. The second data release alone comprises 14 million galaxies and has already enabled traditional galaxy clustering analyses, intriguingly hinting at the possibility of dynamical dark energy \citep{DESI_BAO_2025}.
From the perspective of void studies, a public void catalog has been released for the Year 1 BGS Bright Galaxy Sample, extending to $z=0.24$ and containing a few thousand voids\footnote{The relatively small number of voids arises from the limited and irregular Year 1 footprint, which strongly penalizes void finding, since voids intersecting the survey edge must be discarded.} \citep{Rincon_2024}. Cosmic voids have also been identified using the DESI photometric LRG sample, yielding a total of $140,712$ watershed voids in the range $0.35 < z < 0.95$, without imposing purity selections on their sizes \citep{Sartori_2025}.
 Theoretical expectations indicate that subsequent releases will yield on the order of $10^5$ voids across the full BGS ($0.05 < z < 0.4$), LRG ($0.4 < z < 1.0$), ELG ($0.6 < z < 1.6$), and QSO ($0.9 < z < 2.1$) samples\footnote{This estimate is based on theoretical models of the VSF, considers voids in the dark matter field, and excludes voids with radii smaller than twice the mean tracer separation.}. Such data sets will dramatically expand the landscape of void-based constraints on dark energy in modern surveys.

\paragraph{\textit{Euclid}} 
\textit{Euclid} is a medium-class mission of the European Space Agency launched on 1 July 2023 currently operating from the Sun–Earth L2 Lagrange point \citep{Mellier_2024}. Its main goal is to map the large-scale structure of the Universe in order to investigate the cause of its accelerated expansion. In particular, to probe the nature of dark energy, \textit{Euclid} aims at constraining the equation-of-state parameter $w(z)$ with sub-percent precision. It will survey approximately 14,000~$\mathrm{deg}^2$ of the extragalactic sky, obtaining spectroscopic redshifts for over 25 million galaxies in the range $0.9 \lesssim z \lesssim 1.8$, and photometric redshifts for about 1.5 billion galaxies up to $z \sim2.5$.

Various studies produced forecasts to assess the cosmological constraining power of cosmic voids in \textit{Euclid}. For instance, \citet{Contarini_2022} derived cosmological forecasts from the VSF, while \cite{Hamaus_2022} and \cite{Radinovic_2023} used the VGCF, and finally \citet{Bonici_2023} focused on the void-lensing cross-correlation. Depending on the chosen observable, the void size function can span different orders of magnitude and probe a wide range of spatial scales. For a catalog of voids extracted with a watershed-based algorithm like \textsc{VIDE} (see \Cref{sec: Void finders}), the number of voids will be of the order of $10^5$, which can be reduced to roughly 1/10th of the initial sample after applying a conservative cleaning procedure optimized for VSF modeling.

\begin{figure}[ht]
    \centering
    \includegraphics[width=0.5\linewidth]{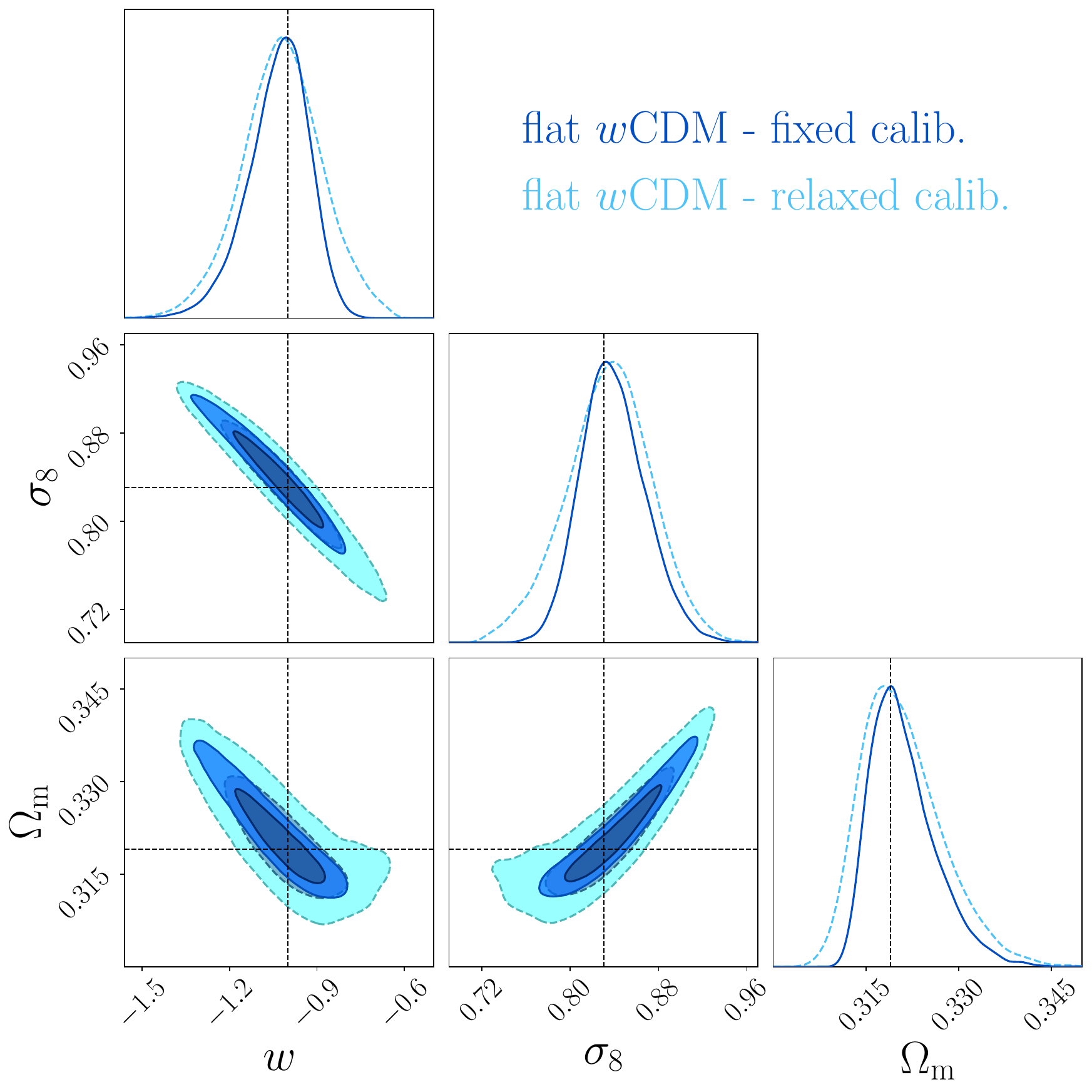}
    \includegraphics[width=0.48\linewidth]{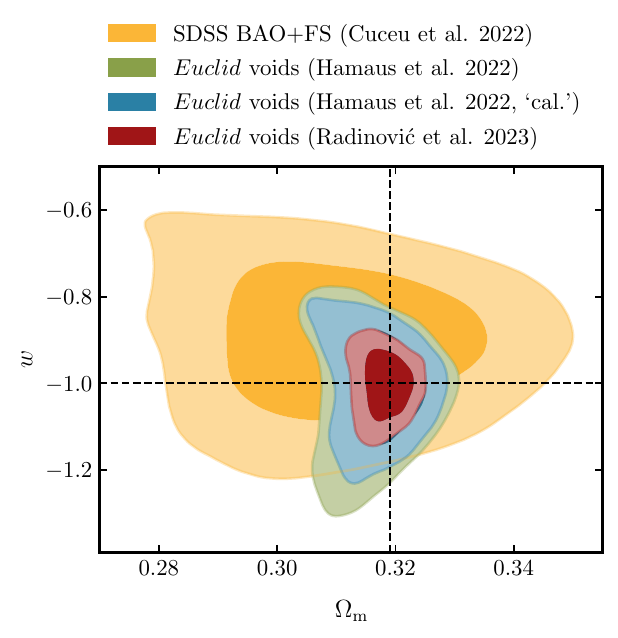}
    \caption{Confidence contours on the parameters of a flat $w$CDM model derived with different summary statistics. These cosmological forecasts have been derived assuming void samples consistent with those expected for the final \textit{Euclid} wide-field spectroscopic galaxy survey. \textit{Left}: cosmological forecasts from the VSF computed with two different model configurations (``fixed'' and ``relaxed'', in dark and light blue respectively). \textit{Right}: cosmological forecasts from the VGCF in different model configurations (``uncalibrated'' in green, ``calibrated'' in blue, see \citealt{Hamaus_2022}) and applying a hybrid-reconstruction technique (in red). As a comparison, this panel shows constraints derived from BAO and ‘full-shape’ measurements in the SDSS galaxy, quasar and Lyman-$\alpha$ data sets from the MGS, BOSS and eBOSS surveys (in yellow, from \citealt{Cuceu_2023}). For further details, we refer the reader to the original sources \citep{Contarini_2022,Radinovic_2023}.}
    \label{fig: Euclid constraints}
\end{figure}

The Euclid Collaboration papers mentioned above show that the void measurements derived from \textit{Euclid} data will provide high-precision constraints on the parameters of the standard model and its alternatives involving a dynamical dark energy. In \Cref{fig: Euclid constraints}, for example, we report some of the forecasts derived for a flat $w$CDM model by assuming a sample of 3D voids aligned with the one expected for the \textit{Euclid} final spectroscopic galaxy survey. According to these analyses, the VSF can constrain $\Omega_\mathrm{m}$ and $w$ with a precision of approximately $10\%$ and $16\%$, respectively, while the same parameters are constrained to $6\%$ [$1\%$] and $10\%$ [$6\%$] using the VGCF [extended with the hybrid-reconstruction and simulation-based template method].

\paragraph{SPHEREx} 
SPHEREx (the Spectro-Photometer for the History of the Universe, Epoch of Reionization, and Ices Explorer) is a NASA Medium Explorer mission launched on 11 March 2025 and currently operating from a Sun-synchronous polar orbit around Earth. During its two-year mission, SPHEREx will map the three-dimensional distribution of hundreds of millions of galaxies and stars to study cosmic inflation, galaxy evolution, and interstellar ices \citep{Crill_2020}. From the perspective of cosmic void studies, this survey stands out for its all-sky coverage and high number density at lower redshifts ($z<0.6$). According to \cite{Dore_2018}, SPHEREx will measure 200,000 voids\footnote{As is the case for the DESI forecast, these estimates are derived assuming voids identified in the underlying dark matter field.} in a redshift range so far poorly explored ($z<1$), and its VSF will provide constraints on dark energy (for example on the $w_0-w_a$ parameter space) strongly complementary to constraints from other surveys such as \textit{Euclid} or the Roman Space Telescope (see description below).

\paragraph{PFS}
The Subaru Prime Focus Spectrograph (PFS) is an optical and near-infrared spectrograph mounted on the Subaru telescope, located at the summit of Maunakea in Hawaii. It measures spectra in the range $380\, \mathrm{nm} \leq \lambda \leq 1260\, \mathrm{nm}$, with 2,400 fibers and a field of view with a diameter of 1.3 degree \citep{Takada_2014}. The Subaru telescope is characterized by an aperture of 8.2 meter, and the cosmology survey will target OII emission line galaxies, in a redshift range of $0.8 \leq z \leq 2.4$, covering a sky area of 1,200 square degrees. The survey has started in the Spring of 2025, and requires 300 observation nights.
The wide redshift range, high galaxy number density and galaxy population, different from other surveys exploring H-$\alpha$ emitters, will reflect in a unique void population from a dense sample counting around 60,000 watershed voids.

\subsection{Future surveys}\label{sec:future surveys}
\paragraph{CSST}
The China Space Station Telescope (CSST) is a next-generation Stage IV survey telescope consisting of a 2-meter space-based observatory that will operate in a co-orbit with the Chinese Space Station in low Earth orbit \citep{Gong_2025}. It is designed to carry out a 10-year wide-field survey, covering approximately 17,500~$\mathrm{deg}^2$ with high-resolution, multiband imaging and slitless spectroscopy over a wavelength range of 255--1,000~nm, targeting more than one billion galaxies up to redshift $z \sim2$. The combination of large area, high spatial resolution ($0.15 \, \mathrm{arcsec}$), and depth makes CSST a powerful tool for cosmology. Some studies \citep{Song_2024a,Song_2024c} have focused on forecasting the effectiveness of cosmic voids as cosmological probes for CSST, considering a realistic sample of watershed voids and modeling their VSF and VGCF. According to these works, the number of watershed voids expected to be identified with CSST is approximately $300,000$ (not accounting for any purity cuts).

\paragraph{Roman}

\begin{figure}[ht]
    \centering
    \includegraphics[width=1\linewidth]{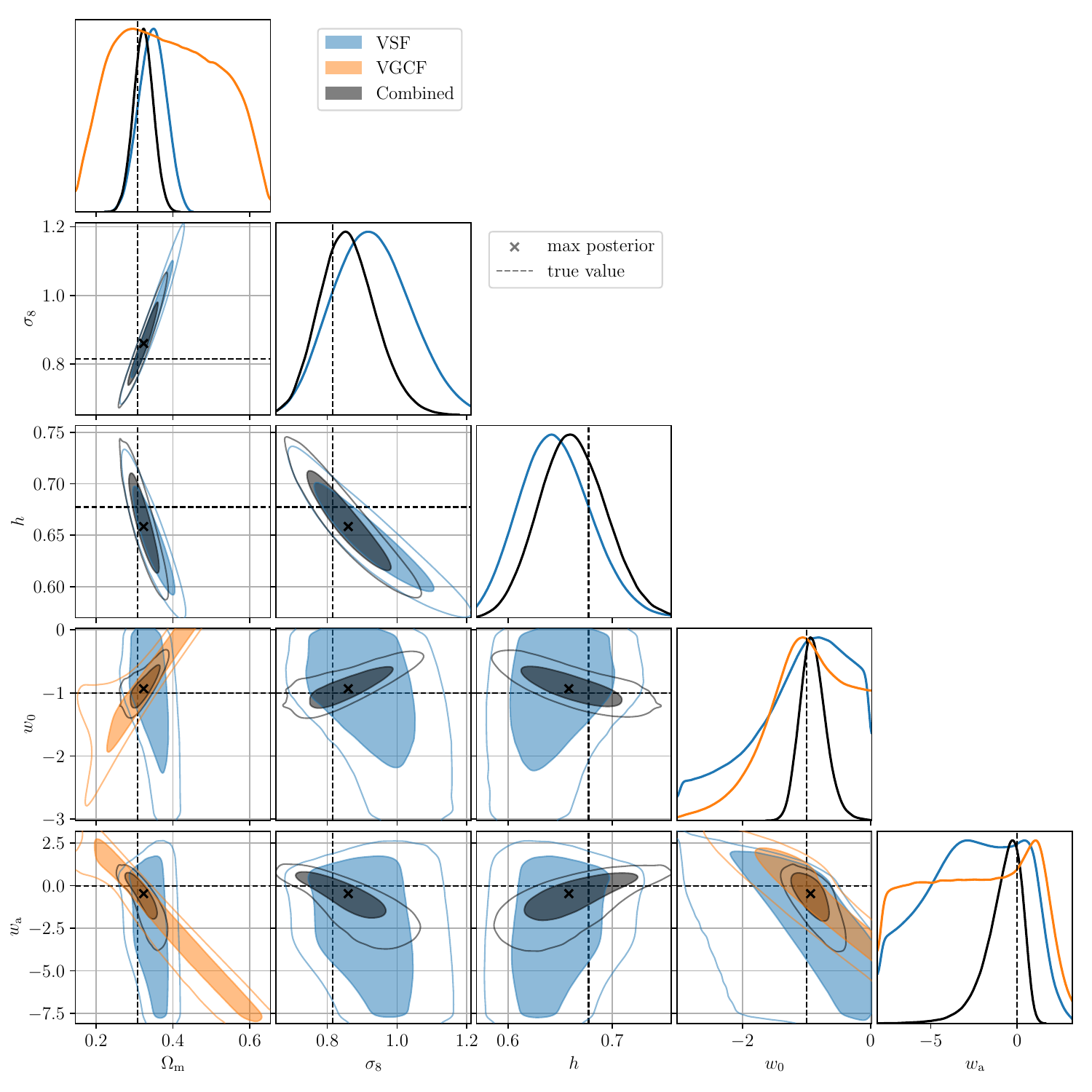}
    \caption{Cosmological forecasts from the VSF (blue), VGCF (orange), and their combination (dark gray) for the Roman survey.  The black dashed lines show the fiducial values of the cosmological parameters assumed in the analysis. For further details, we refer the reader to the original source \citep{Verza_2024b}.}
    \label{fig: Roman constraints}
\end{figure}

The Nancy Grace Roman Space Telescope, is a near infrared NASA mission expected to be launched no later than May 2027 \citep{Spergel_2015}. It will operate from a halo orbit around the Sun–Earth L2 Lagrange point, providing a stable thermal environment and continuous sky visibility. Besides the time allotted to settle essential questions in cosmology, Roman is an observatory with a wide range of research applications, including, but not limited to, the study of exoplanets and infrared astrophysics. 
The Roman telescope will conduct the High-Latitude Wide-Area Survey, combining both imaging and spectroscopy: the High Latitude Spectroscopic Survey \citep[HLSS,][]{Wang_2022} and the High Latitude Imaging Survey (HLIS). The survey area of the designed mission is 2,000 square degrees. Roman will allow us to reach an unprecedented high tracer number density in the redshift range $1<z<2$, but a sparser survey can extend up to $z=3$ using the OIII emission line \citep{Wang_2022}, overall unlocking a cosmic void sample of exceptional quality down to a few Mpcs. 

The expected high galaxy number density will allow a deep sampling of voids, making the target population of Roman voids complementary to other surveys from space, such as \textit{Euclid}, even if exploring a smaller area. In particular, in the HLSS, Roman is expected to observe $\sim82,000$ spectroscopic watershed voids with a minimum radius down to 5.8 $h^{-1} \,$Mpc at $z=1$ \citep{Verza_2024b}. The low-end side of the available void sample will allow tight constraints from the joint analysis of the VSF and the VGCF. The combination of the two void probes shows some complementarity, allowing Roman voids to provide tight constraints on standard cosmological parameters and, particularly, on dynamical dark energy. In the $\Lambda$CDM model, the joint constraining power of the VSF and VGCF is forecast to be up to $2.5\%$ for $\Omega_{\rm m}$, $3\%$ for $\sigma_8$, and $2.5\%$ for $H_0$. When considering the $w_0w_a$CDM model, the forecast constraints for Roman voids give an uncertainty of $8\%$ for $\Omega_{\rm m}$, $9\%$ for $\sigma_8$, $4\%$ for $H_0$, $20\%$ for $w_0$, and an absolute uncertainty of $\pm 1$ for $w_{a}$. It is important to note that the specific survey strategy, i.e. the trade-of between covered area and flux limit, is still under discussion and any modifications would lead to changes for these forecast constraints.

\paragraph{WST}

The Wide-field Spectroscopic Telescope (WST) is a proposed 12-meter class instrument, planned to be located at Cerro Paranal in northern Chile (alongside other major ESO facilities) and designed to carry out massive spectroscopic surveys of the sky \citep{Mainieri_2024}. With a 3.1~$\mathrm{deg}^2$ field of view and a multiplexing capability of at least 20,000 fibers, WST aims to provide over 150 million galaxy redshifts across a wide redshift range of $0<z<5$, extensible up to $z \sim7$ using Lyman-alpha emitters. The survey will cover about 18,000~$\mathrm{deg}^2$, enabling precision cosmology and deep insights into galaxy formation and evolution. Combining a multi-object spectrograph (MOS) with an integral field unit (IFU), WST could represent a cornerstone for multi-messenger and time-domain astrophysics in the coming decades.

Since the WST survey is still in its design phase and its first light is assumed around 2040, predicting the expected number of cosmic voids is non-trivial. Despite the fact that an accurate estimate would require specifically designed mock light-cones, a rough prediction can be obtained by rescaling the theoretical VSF based on the survey's expected volume and spatial resolution. This, however, requires the assumption of a galaxy bias model for the observed tracers. Taking as a reference a galaxy sample similar to that expected for the \textit{Euclid} mission, the expected number of voids is of the order of one million, potentially reduced to approximately $1/20$th of the initial void sample when applying conservative purity cuts\footnote{This reduction in number counts results from restricting the sample to voids larger than $2.5$ times the mean spatial separation expected for the galaxy catalog. Compared to \textit{Euclid}, WST has a smaller mean galaxy separation, so the same purity cut corresponds to smaller void radii. At these smaller scales, where the VSF is less steep, a larger reduction in void numbers is therefore expected relative to \textit{Euclid}.}. It is important to note that this estimate refers to spherical underdensities with an internal density contrast of $-0.7$; alternative void definitions may lead to significantly different void-count estimates.

\begin{figure}[ht]
    \centering
    \includegraphics[width=0.75\linewidth]{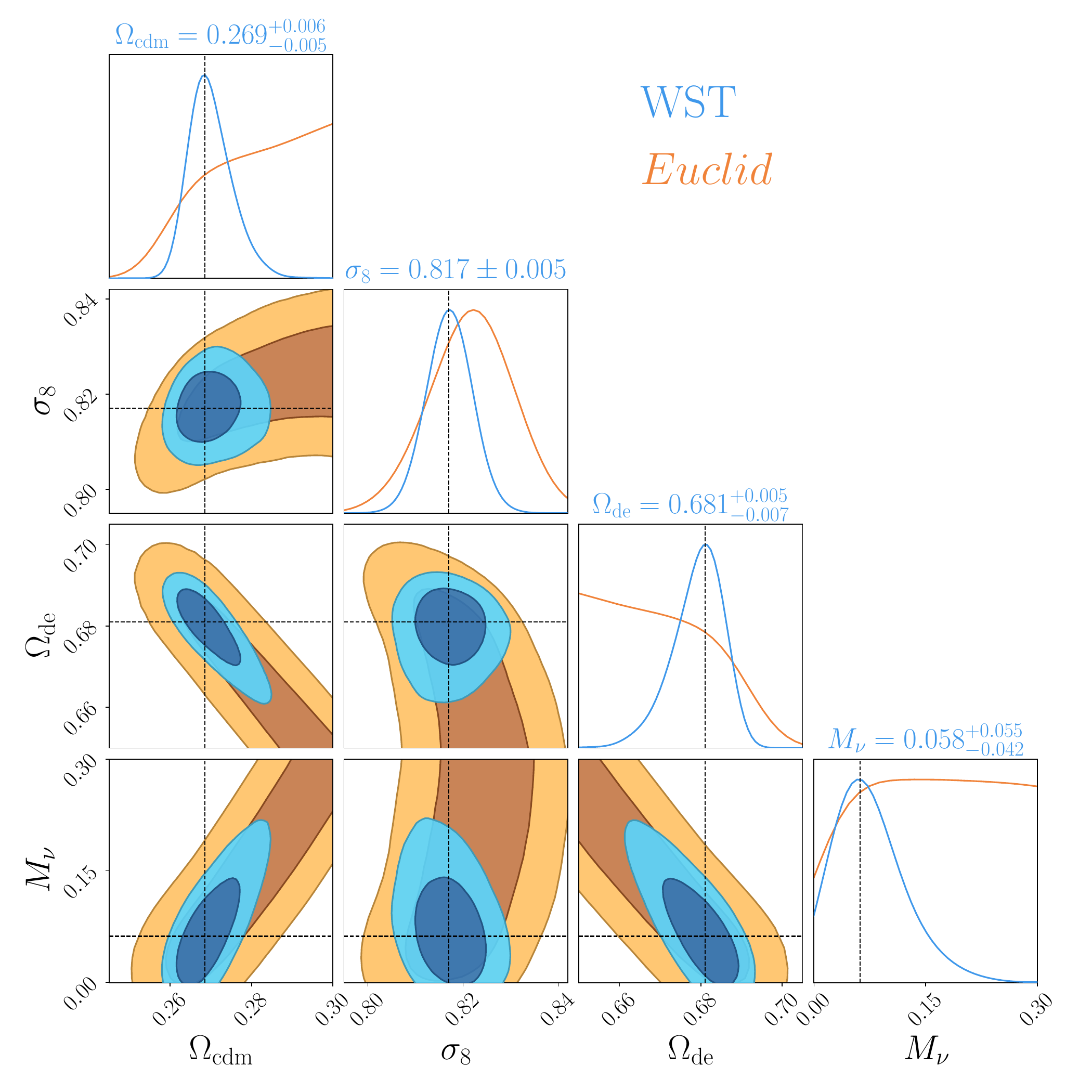}
    \caption{Cosmological forecasts from VSF computed for WST (blue), compared to those expected for a \textit{Euclid}-like survey (orange). The black dashed lines show the fiducial values of the cosmological parameters assumed in the analysis. The best-fit parameter values and their corresponding $1\sigma$ uncertainties are reported above the projected 1D constraints. For further details, we refer the reader to the original source \citep{Mainieri_2024}.}
    \label{fig: WST constraints}
\end{figure}

Thanks to the unique combination of large sky coverage, broad redshift range, and high spatial resolution, the cosmological constraints expected from voids identified with WST exhibit strong constraining power, enabling tests of alternative cosmological models. This is illustrated in \Cref{fig: WST constraints}, where cosmological forecasts based on the VSF expected from \textit{Euclid} and WST are compared under the assumption of a $\Lambda$CDM cosmology with massive neutrinos. The total neutrino mass $M_\nu$ is strongly degenerate with the other cosmological parameters ($\Omega_\mathrm{cdm}$, $\Omega_\mathrm{de}$, and $\sigma_8$), and the constraining power of \textit{Euclid} alone is not sufficient to disentangle these correlations. The expected improvement provided by WST is therefore particularly significant, enabling tight cosmological constraints within this extended parameter space.

\subsection{Other surveys} \label{sec:other surveys}

In this section, we describe surveys that are expected to deliver relevant results in the context of cosmic voids, but for which, at the time of writing this review, dedicated void number-count analyses or theoretical forecasts are not yet available. Future studies specifically aimed at forecasting void numbers for these missions will represent an important step towards quantifying their full cosmological potential in terms of constraints from voids. 

\paragraph{UNIONS}
The Ultraviolet Near-Infrared Optical Northern Survey \citep[UNIONS,][]{Gwyn_2025} is a collaboration between the Hawaiian observatories CFHT (Canada–France–Hawaii Telescope), Subaru, and Pan-STARRS (Panoramic Survey Telescope and Rapid Response System), designed to carry out a deep, multi-band imaging survey of the northern sky. Initiated in 2017, UNIONS is nearing completion at the time of writing this review, with a footprint of about 10,000~$\mathrm{deg}^2$ and a core high-quality region of approximately 6,250~$\mathrm{deg}^2$. The survey is expected to deliver precise photometry for hundreds of millions of galaxies, providing a foundational data set for studies of galaxy evolution and large-scale structure in the northern hemisphere. Although quantitative measurement or forecasts for void counts from the UNIONS survey are not yet available, \citet{Hunter_2025} have already employed its data for void studies, using the UNIONS galaxy catalog as a background source to measure the weak-lensing signal of voids identified in the overlapping BOSS region.

\paragraph{eROSITA}  
The extended ROentgen Survey with an Imaging Telescope Array \citep[eROSITA,][]{Predehl_2021} is a wide-field, high-throughput X-ray telescope on board the Russian-German Spectrum-Roentgen-Gamma (SRG) space mission. Launched on 12 December 2019 and currently operating at the Sun–Earth L2 Lagrange point, it was designed to detect large samples of galaxy clusters up to redshifts $z>1$ by performing eight full-sky scans over roughly four years. The first eROSITA All-Sky Survey Catalog (eRASS1) represents the largest collection of X-ray sources ever published, comprising about $900{,}000$ individual detections in the 0.2--2.3~keV band. 
Owing to its wide sky coverage, depth, and ability to identify both clusters and AGN, eROSITA holds strong potential for studies of the large-scale structure, including cosmic voids.

\paragraph{4MOST} 
The 4-meter Multi-Object Spectroscopic Telescope \citep[4MOST,][]{4most_2019,Verdier_2025} is a fiber-fed spectroscopic survey facility currently being installed on the Visible and Infrared Survey Telescope for Astronomy (VISTA) telescope at ESO's Paranal Observatory. It is designed to carry out large spectroscopic surveys of the southern sky over 5 years, with full science operations expected to begin in 2026. Thanks to its wide 4.2~$ \mathrm{deg}^2$ field of view and its ability to simultaneously collect about $2{,}400$ spectra, 4MOST will provide redshifts for millions of galaxies and quasars, enabling detailed mapping of the large-scale structure. This will make the survey highly relevant for cosmological studies, including void science, where spectroscopic information is crucial for accurate void identification and for measuring their dynamics.

\paragraph{Rubin} 
The Vera C. Rubin Observatory's Legacy Survey of Space and Time---also known as LSST---will be conducted with the 8.4-meter Simonyi Survey Telescope, located on Cerro Pachón in northern Chile, and will provide an unprecedented deep and wide photometric galaxy survey \citep{Ivezic_2019}. Over its planned ten years of operation, LSST is expected to cover about 18,000~$\mathrm{deg}^2$ of the southern sky, reaching a depth of $r \sim27.5$ for point sources and detecting tens of billions of galaxies. The combination of large sky coverage, high source density, and multi-epoch observations will open up unique opportunities for cosmology.  In particular, the survey will enable void studies over a cosmological volume far larger than any currently available, offering transformative potential for the field.

\paragraph{DESI-II}
The next phase of the Dark Energy Spectroscopic Instrument \citep[DESI-II,][]{Schlegel_2022}, will extend the current DESI program at the Mayall Telescope, targeting over 40 million galaxies and quasars to probe the evolution of dark energy and test models of inflation. This future campaign is scheduled to begin operations around 2029 and will focus on mapping large-scale structure at higher redshifts, with a particular interest in detecting the approximately 2.5 million LBGs expected in the range $2<z<4.5$ \citep{Payerne_2025}. 
The resulting spectroscopic sample will provide an unprecedented 3D map of the distant Universe, enabling new opportunities for cosmological studies, including the statistical characterization of cosmic voids across a significant volume and redshift baseline.

\paragraph{MUST}
The MUltiplexed Survey Telescope \citep[MUST,][]{Zhao_2024} is a 6.5-meter wide-field spectroscopic facility under development, designed to carry out highly multiplexed cosmological surveys in the 2030s. Located at Saishiteng Mountain in Qinghai, China, MUST will observe a field of view of approximately 5~$\mathrm{deg}^2$ and simultaneously acquire spectra for over 20,000 targets. The survey aims to map the 3D distribution of more than 100 million galaxies and quasars over roughly 13,000~$\mathrm{deg}^2$ of the northern sky, spanning cosmic epochs from the nearby Universe to redshifts of $z \sim 5.5$. With its large aperture, extreme multiplexing, and wide sky coverage, MUST will enable Stage-V spectroscopic studies of large-scale structure, providing unprecedented access to the high-redshift Universe and powerful constraints on dark energy and cosmic evolution.

\paragraph{Spec-S5}
The Stage-5 Spectroscopic Facility \citep[Spec-S5,][]{Besuner_2025} is a next-generation, all-sky spectroscopic experiment designed to map the 3D distribution of matter with unprecedented precision. The survey will be carried out using two upgraded 6-m wide-field telescopes located in the Northern and Southern hemispheres, in Arizona and Chile, enabling full-sky coverage. Spec-S5 will target approximately 25,000~$\mathrm{deg}^2$ of low-extinction extragalactic sky and obtain redshifts for about 110 million galaxies and quasars, an order-of-magnitude increase over current spectroscopic surveys. With its extreme multiplexing capability, measuring more than 13,000 spectra simultaneously, Spec-S5 will deliver a transformational gain in survey speed and statistical power. The combination of all-sky coverage and an unprecedented density of spectroscopic tracers will enable large-scale structure and void studies over cosmological volumes far exceeding those accessible to existing surveys.

\section{Conclusions}
Cosmic voids have undergone a remarkable transformation over the past decade, evolving from relatively unexplored structures in the cosmic web to powerful and versatile tools for precision cosmology. This review has provided a comprehensive overview of the current status of void science, presenting both the variety of void statistics and the theoretical frameworks developed to interpret them. We discussed how void cosmology consists of multiple statistics, including the VSF, density and velocity profiles, cross- and auto-correlation functions, ellipticity, and their imprints on the cosmic microwave background and gravitational lensing. Each of these observables opens a unique window on the evolution of structures and on the nature of gravity in underdense regions.

A substantial part of this review has focused on the theoretical modeling of void statistics, from early analytical frameworks such as the excursion-set formalism to more recent developments based on peak theory, bias expansions, and effective moving barrier. Beyond these topics, we also devote space to the so-called void model, to void dynamics based on linear theory and its extensions, as well as to gravitational lensing, the ISW effect in voids, and other theoretical approaches. Accurately describing voids in redshift space and accounting for observational systematic errors has motivated the introduction of semi-analytical model extensions and advanced reconstruction techniques, many of which have already been successfully applied to data. Analyses of voids in surveys such as BOSS, eBOSS, and DES have yielded cosmological constraints that not only validate results from traditional probes but, in some cases, provide competitive and complementary constraints, while also offering potential insights into physics beyond $\Lambda$CDM. Indeed, one of the key strengths of voids lies in their sensitivity to parameters that are typically difficult to constrain, including the growth rate of structure, the dark energy equation of state, neutrino mass, and fifth forces in modified gravity. Their underdense, unscreened environments and large spatial scales amplify subtle physical effects, making voids a powerful probe of physics beyond the standard $\Lambda$CDM model. 

The convergence of increasingly precise void measurements and robust theoretical predictions marks a series of important milestones, and positions the field to enter a new phase of rapid progress, fueled by the advent of next-generation galaxy surveys. Currently ongoing and upcoming missions---such as DESI, \textit{Euclid}, Rubin, SPHEREx, PFS, CSST, Roman, and WST---are expected to dramatically expand the surveyed volume and redshift coverage, opening unprecedented opportunities for void studies. These data sets will usher in the Big Data era for cosmic voids, enabling high-precision measurements of their statistical properties across a wide range of scales and cosmic epochs. To fully exploit this potential, further progress is however required in several directions: refining theoretical models, understanding effects such as the environmental dependence of tracer bias and the impact of galaxy properties, and developing fast and accurate tools for mock generation and data analysis.

In this rapidly evolving landscape, cosmic voids are poised to become a central component of modern cosmological analyses. Their unique sensitivity to the geometry and dynamics of the Universe, combined with the relative simplicity of the physical processes governing them, makes them an ideal complement to traditional high-density probes. As we enter this new phase, the framework presented in this review offers a solid foundation for interpreting current findings and charting future developments for cosmic void cosmology.


\backmatter





\bmhead{Acknowledgements}

SC acknowledges the use of computational resources from HPC system Raven of the Max Planck Computing and Data Facility (MPCDF) in
Garching, Germany. GV acknowledges support from the Simons Foundation to the Center for Computational Astrophysics at the Flatiron Institute. AP acknowledges support from the European Research Council (ERC) under the European Union's Horizon programme (COSMOBEST ERC funded project, grant agreement 101078174), and from the French government under the France 2030 investment plan, as part of the Initiative d'Excellence d'Aix- Marseille Universit{\'e} - A*MIDEX AMX-22-CEI- 03. 

\section*{Declarations}

\bmhead{Conflict of interest}
The authors declare no conflict of interest.



\clearpage

\begin{appendices}
\section{Publicly available analysis tools}\label{app:A}
In this Appendix, we provide a list of publicly available tools for void-related analyses, including algorithms for void identification, libraries for preparing and modeling void statistics, and existing public catalogs of cosmic voids. Each entry includes a concise description of the tool, key references, and a link to its public repository or webpage. Only tools whose repositories are active and accessible at the time of writing this review have been included, ensuring that all the listed resources can be readily obtained and used for research purposes.

\paragraph{Void finders} \smallskip
\begin{itemize}
    \item \textbf{\textsc{ZOBOV}} (ZOnes Bordering On Voidness)  is a topological void finder introduced in \cite{Neyrinck_2008}, available at \url{http://skysrv.pha.jhu.edu/~neyrinck/voboz/}. It performs a Voronoi tessellation of tracer particles to estimate a continuous density field, followed by a watershed transform~\citep{Schaap_2000,Platen_2007} to merge neighboring cells into zones and subsequently into voids. \textsc{ZOBOV} is written in C and interfaces with \textsc{Qhull}\footnote{\url{http://www.qhull.org/}} for constructing the Voronoi tessellation; it is designed to operate on periodic simulation boxes, as it does not include boundary treatments for survey geometries. \smallskip
    \item \textbf{\textsc{VIDE}} (Void IDentification and Examination toolkit) is a topological void finder introduced in \cite{Sutter_2014a}, available at \url{https://bitbucket.org/cosmicvoids/vide_public/}. VIDE is an enhanced implementation of \textsc{ZOBOV}, which constructs the density field through the Voronoi tessellation and applies a watershed transform to identify voids as relative minima surrounded by basins. With respect to \textsc{ZOBOV}, \textsc{VIDE} improves the efficiency of the tessellation and watershed procedures and provides extensive pre-processing tools, such as catalog subsampling and filtering. Crucially, it also generalizes the method to observational data by incorporating survey masks and dedicated boundary treatments. The output includes key void properties such as the center position, effective radius, ellipticity, density contrast, and central density. A Python API is also provided for post-processing and analysis, allowing the user to manipulate void catalogs and member particles, compute clustering statistics, and fit density profiles. \smallskip
    \item \textbf{\textsc{Sparkling}} is a spherical void finder introduced in \cite{Ruiz_2015} and \cite{Ruiz_2019}, available at \url{https://gitlab.com/andresruiz/Sparkling}. This C++ code can be applied to both N-body simulations and galaxy surveys. It first applies a Voronoi tessellation to the tracer distribution to locate local density minima, which serve as initial candidates for void centers. Around each candidate, a spherical region is expanded until the mean enclosed density reaches a user-defined density contrast threshold. A random walk is then performed around the preliminary center to refine its position and maximize the void size. When two voids overlap, only the largest one is retained. The resulting catalog provides the main geometric and physical properties of the voids, together with additional quantities useful for quality assessment and environmental characterization, and optionally includes the computation of integrated density profiles around the voids. \smallskip
    \item \textbf{\textsc{REVOLVER}} (REal-space VOid Locations from surVEy Reconstruction) is a topological void finder presented in \cite{Nadathur_2019c}, available at \url{https://github.com/seshnadathur/Revolver}. Similarly to \textsc{VIDE}, it can process both simulated and observational data sets, but it additionally offers the capability to identify voids in reconstructed real space by correcting RSD through a FFT-based method. \textsc{REVOLVER} provides two Python routines for void finding: the first relies on \textsc{ZOBOV} to reconstruct a continuous density field, while the second estimates the tracer density through particle-mesh interpolation. In both cases, voids are ultimately identified by applying a watershed transform that merges adjacent Voronoi cells or voxels into coherent underdense regions. The resulting catalogs undergo post-processing and quality-control steps and report, for each void, its main geometrical and physical properties. \smallskip
    \item \textbf{\textsc{Pylians}} \citep[Python libraries for the analysis of numerical simulations,][]{Villaescusa-Navarro_2018} is a collection of Python libraries aimed at providing fundamental tools for the analysis of numerical simulations. In particular, it implements a spherical void-finding algorithm based on the work of \cite{Banerjee_2016} and is available at \url{https://pylians3.readthedocs.io/en/master/voids.html}. This finder can be applied to cubic N-body simulations with periodic boundary conditions, using a 3D grid to construct a continuous density contrast field and locate local minima corresponding to candidate void centers. The algorithm identifies voids by progressively smoothing the density field on different spatial scales (from large to small) and selecting underdense regions whose mean density falls below a user-defined threshold. Each spherical region is tested against previously detected voids to prevent multiple detections of overlapping structures. The final output catalog provides a naturally hierarchical sample of voids, including their center positions and estimated radii. \smallskip
    \item \textbf{\textsc{PopCorn}} is an extension of \textsc{Sparkling}, presented in \cite{Paz_2023} and available at \url{https://gitlab.com/dante.paz/popcorn_void_finder}. It belongs to the category of spherical void finders as it defines voids as union of spheres of maximum volume, which collectively enclose isolated regions with a user-defined density contrast. This approach allows the formation of free-form, irregular structures (referred to as ``popcorn'' voids) resulting from the natural merging of neighboring spheres. As a consequence, overlapping voids are automatically handled by the algorithm, without the need for additional prescriptions. \smallskip
    \item \textbf{\textsc{VAST}} \citep[Void Analysis Software Toolkit,][]{Douglass_2022}, available at \url{https://vast.readthedocs.io/en/latest/}, is a Python toolkit that provides implementations of two widely used classes of void-finding algorithms for galaxy catalogs: spherical void finders and watershed-based methods. In particular, it includes Python3 versions of the \textsc{VoidFinder} and \textsc{V$^2$} algorithms, enabling the identification of cosmic voids either by growing spherical underdense regions or through topological segmentation of the density field. In addition to the void-finding routines, \textsc{VAST} offers \textsc{VoidRender}, an \textsc{OpenGL}-based module for interactive three-dimensional visualization of the identified voids overlaid on the reference galaxy distribution. Each algorithm implemented in \textsc{VAST} provides its own visualization interface, allowing users to inspect, compare, and analyze the structure of voids within the cosmic web. \smallskip
    \item \textbf{\textsc{PyTwinPeaks}} \citep{Maggiore_2025} is a Python pipeline for identifying and characterizing tunnel voids in weak-lensing convergence maps, available at \url{https://github.com/LeonardoMaggiore/PyTwinPeaks}. The algorithm operates directly on 2D convergence fields, with or without Gaussian shape noise, applying Gaussian smoothing to suppress pixel-level fluctuations and enhance extended underdensities. Candidate voids are identified through threshold-based segmentation of the convergence field, with centers defined as absolute minima traced across multiple signal-to-noise levels, and radii determined by circular growth until the mean signal reaches a user-specified threshold. The resulting catalog is cleaned to remove overlapping detections and includes derived statistical observables such as the probability density function, angular power spectrum, and void size function. \textsc{PyTwinPeaks} has been used to analyze weak-lensing maps from cosmological simulations, demonstrating its ability to enhance the tangential shear signal around 2D voids.
\end{itemize}
\paragraph{Post-processing and analysis tools} \smallskip
\begin{itemize} 
    \item \textbf{\textsc{CosmoBolognaLib}} \citep{Marulli_2016} is a comprehensive collection of \textit{free} C++/Python libraries, available at \url{https://gitlab.com/federicomarulli/CosmoBolognaLib}. It provides an efficient numerical environment for cosmological analyses of the large-scale structure of the Universe. Among its many functionalities are tools to compute the VSF models developed by \cite{Sheth_van_de_Weygaert_2004} and \cite{Jennings_2013}, which are also available with extensions that account for observational effects. The library is particularly well suited for handling catalogs of astronomical objects, including cosmic voids from both simulations and galaxy surveys, and provides routines to measure void-related statistics such as the VSF, the void–tracer and the void–auto correlation functions. It also implements the cleaning algorithm of \cite{Ronconi_2017}, designed to align the void samples with the theoretical Vdn model (see \Cref{sec:obs_VSF}). \smallskip
    \item \textbf{\textsc{Victor}} is a Python package available at \url{https://github.com/seshnadathur/victor}, which provides tools to measure the cross-correlation function between density regions (e.g., voids or density-split centers) and galaxies. The package also includes a functionality for modeling these statistics through Bayesian inference, and has been employed in several studies \citep{Nadathur_2020b,Woodfinden_2022,Radinovic_2023,Radinovic_2024}. \smallskip
    \item \textbf{\textsc{Voiager}} (Void dynamics and geometry explorer) is a Python library available at \url{https://voiager.readthedocs.io/en/latest/} that offers a comprehensive framework for the analysis of the VGCF, following the pipeline presented in several works \citep[e.g.][]{Hamaus_2020,Hamaus_2022,Beyond2pt_2025}. The code interfaces naturally with void catalogs extracted using \textsc{VIDE} and measures the VGCF multipoles through a stacking procedure. It then models the dynamical and geometrical distortions of the stacked voids and performs a Bayesian inference to derive cosmological constraints for different cosmological models ($\Lambda$CDM, $w$CDM, $w_0w_a$CDM). \smallskip
    \item \textbf{\textsc{ExcursionSetFunctions}} is a C++/Python library available at \url{https://github.com/Giovanni-Verza/excursion_set_functions}, which provides a collection of tools for computing excursion-set related quantities (see \Cref{subsec:excursion-set}). It includes analytical and numerical multiplicity functions, Lagrangian void density profiles, integration moments, and power-spectrum covariances. The library is particularly useful for efficiently computing the VSF model presented in \cite{Verza_2024a}. \smallskip
    \item \textbf{\textsc{Vorothreshold}} \citep{Verza_2024b}, available at \url{https://github.com/Giovanni-Verza/Voronoi_postprocess}, is a Python module for finding voids reaching a specific density threshold without assuming any symmetry, which relies on the Voronoi tessellation and Delaunay scheme of the discrete tracers distribution. It is mainly designed to post-process watershed voids, such as the ones from \textsc{VIDE}, \textsc{ZOBOV}, or \textsc{REVOLVER}, but it can also find voids if the Voronoi tesselation is provided. The code is implemented for simulation boxes and light-cones and can handle survey masks and weights.
    \item \textbf{\textsc{Void Profile Analysis Toolkit}} \citep{Schuster_2023,Schuster_2024,Schuster_2025} is a Python package available at \url{https://github.com/nicosmo/void_profile_analysis}. It provides tools for calculating, saving, and stacking radial profiles around cosmic voids using data from cosmological simulations. The toolkit measures physical quantities---such as tracer density and velocity---in spherical shells as a function of distance from the void center and includes robust methods for error estimation based on Jackknife resampling.
\end{itemize}
\paragraph{Void catalogs} \smallskip
\begin{itemize}
    \item \textbf{SDSS DR7 VoidFinder Catalog} \citep{Pan_2012} provides a publicly available catalog of cosmic voids and void galaxies identified in the SDSS DR7, through the \textsc{SIMBAD} database (\url{https://simbad.cds.unistra.fr/simbad/sim-ref?bibcode=2012MNRAS.421..926P&simbo=on}). Voids were identified using the \textsc{VoidFinder} algorithm, based on the method originally proposed by \cite{El-Ad_Piran_1997} and implemented by \cite{Hoyle_Vogeley_2002}. The catalog includes 1,054 statistically significant voids in the Northern Galactic Hemisphere with effective radii larger than $10~h^{-1} \, \mathrm{Mpc}$ and deep underdense cores $\Delta < -0.85$. The catalog also identifies $\sim$8,000 galaxies residing within the void regions, providing a valuable dataset for comparisons with simulations and studies of galaxy evolution in low-density environments. \smallskip
    \item \textbf{VIDE Public Void Catalogs} represents a collection of cosmic void catalogs extracted from galaxy redshift surveys and simulations using the finder \textsc{VIDE}, available through a dedicated link at \url{http://www.cosmicvoids.net}. It includes the catalogs produced for the analysis of voids in SDSS DR7 \citep{Sutter_2012b} and SDSS DR9 \citep{Sutter_2013}, as well as the mock catalogs constructed to match these two data sets \citep{Sutter_2014c}. Finally, it provides the SDSS DR10 void catalogs employed in the cosmological analysis presented in \citet{Sutter_2014a}. Taken together, these catalogs span a redshift range from $0.2$ to $0.7$, and have been used to test void number counts, ellipticity distributions, and density profiles against mock catalogs, assessing the impact of the survey mask, sparse sampling, and galaxy bias. \smallskip    
    \item \textbf{VIPERS Void Catalog} \citep{Micheletti_2014} presents the identification and characterization of cosmic voids in the VIMOS Public Extragalactic Redshift Survey (VIPERS) at redshifts $0.55<z<0.9$, and is available through the SIMBAD and VizieR databases (\url{https://simbad.cds.unistra.fr/simbad/sim-ref?bibcode=2014A%26A...570A.106M&simbo=on}; \url{http://cdsarc.u-strasbg.fr/viz-bin/qcat?J/A+A/570/A106}). A new void-finding method was developed for this analysis, based on identifying empty spheres that can be inscribed between galaxies while accounting for the complex survey geometry and internal gaps. The approach was validated against mock catalogs through measurements of the void size distribution and the void–galaxy correlation function, resulting in a catalog whose properties are in agreement with the simulations. \smallskip
    \item \textbf{SDSS DR7 Watershed Void Catalog} \citep{Nadathur_2014} provide publicly available catalogs of cosmic voids identified in the SDSS DR7 main galaxy and luminous red galaxy samples, and are accessible through the \textsc{VizieR} database (\url{https://cdsarc.cds.unistra.fr/viz-bin/cat/J/MNRAS/440/1248}). These catalogs build upon the earlier work of \cite{Sutter_2012}, adopting the same watershed-based void-finding principles while introducing significant methodological improvements. In particular, they use an enhanced version of the \textsc{ZOBOV} algorithm, featuring a more robust treatment of survey boundaries, masks, and the radially varying selection function. Two versions of the catalogs are available---in comoving and redshift coordinates---each including voids and corresponding super-clusters. For each structure, the catalog provides detailed information such as barycentric position, effective radius, volume, average and minimum density, and the number of member galaxies. \smallskip
    \item \textbf{\textsc{SDSS DR12 BOSS Void Catalog}} \citep{Mao_2017} provides a catalog of cosmic voids identified in the BOSS galaxy sample from the SDSS DR12, available at \url{https://cdsarc.cds.unistra.fr/viz-bin/cat/J/ApJ/835/161}. Voids were detected using the \textsc{ZOBOV} algorithm, accounting for survey boundaries, masks, and both angular and radial selection functions. The catalog contains 10,643 voids in total, with a high-quality subsample of 1,228 voids having effective radii in the range $20$–$100 \ h^{-1} \, \mathrm{Mpc}$ and central densities about $30\%$ of the mean galaxy density. The basic void statistics---such as size, redshift distribution, and stacked radial density profiles---are consistent with those obtained from 1,000 mock catalogs. The resulting BOSS and mock void samples provide a valuable resource for cosmological and environmental studies of galaxies. \smallskip
    \item \textbf{\textsc{GIGANTES}} \citep{Kreisch_2022} is an extensive suite of cosmic void catalogs extracted from the \textsc{QUIJOTE} Latin Hypercube simulations \citep{Villaescusa-Navarro_2020} using the void finder \textsc{VIDE}, and available at \url{https://gigantes.readthedocs.io/en/latest/}. It comprises 15,000 void catalogs for the fiducial cosmology and more than 9,000 for non-fiducial models, spanning a wide range of cosmological parameters and redshifts ($0\le z\le2$), for a total of over one billion voids. \textsc{GIGANTES} enables detailed studies of void statistics, including the VSF, VGCF, and auto-correlation functions, and provides a massive data set for machine-learning applications linking void properties to cosmological parameters. It has been employed for likelihood-free cosmological inference in \cite{Kreisch_2022} and \cite{Wang_2023}. \smallskip
    \item \textbf{SDSS DR7 Updated Void Catalogs} \citep{Douglass_2022} present several public catalogs of cosmic voids identified in a volume-limited subsample of the SDSS DR7, available at \url{https://cdsarc.cds.unistra.fr/viz-bin/cat/J/ApJS/265/7}. The authors applied three different void-finding algorithms—\textsc{VoidFinder}, \textsc{VIDE} and \textsc{REVOLVER}—to identify 1,163, 531, and 518 voids with radii larger than $10 \ h^{-1} \, \mathrm{Mpc}$. The catalog includes derived void properties such as centers, radial density profiles, volume fractions, and galaxy contents. Comparisons with 64 mock catalogs show good agreement between the observed and simulated void populations, providing a robust reference data set for cosmological and environmental studies.\smallskip
    \item \textbf{CAVITY Void Catalog} \citep{Garcia-Benito_2024} consists of sample of low-redshift voids derived from the DR1 of the Calar Alto Void Integral-field Treasury surveY (CAVITY), available at \url{https://cavity.caha.es/data/dr1/}. The project focuses on identifying and characterizing cosmic voids in the nearby Universe ($0.005<z<0.050$), drawing its parent sample from the SDSS catalog of \cite{Pan_2012}. Starting from 80 voids containing 19,732 galaxies, completeness and quality criteria were applied---requiring at least $20$ galaxies per void and excluding systems near the survey boundaries---yielding a refined sample of $42$ voids. From these, 15 representative voids were selected to form the CAVITY parent sample, comprising 4,866 galaxies distributed within well-characterized underdense regions. The catalog provides detailed spectroscopic data and is designed to enable studies of the physical and environmental properties of galaxies residing in cosmic voids. \smallskip
    \item \textbf{\textsc{DESIVAST}} \citep{Rincon_2024} is the value-added catalog of voids from the DESI Year 1 Bright Galaxy Survey, available at \url{https://data.desi.lbl.gov/doc/releases/dr1/vac/desivast/}. It provides three complementary void catalogs constructed with the \textsc{VoidFinder} and \textsc{V$^2$} algorithms, the latter implemented with both \textsc{VIDE} and \textsc{REVOLVER} pruning schemes. The catalogs are based on a volume-limited sample of bright galaxies extending to $z=0.24$ and contain between $\sim300$ and 1,500 interior voids, depending on the method. \textsc{DESIVAST} offers a new data set for studying galaxy evolution across environments and for cosmological analyses based on void statistics. \smallskip
    \item \textbf{Bayesian Void Catalog} \citep{Malandrino_2025} provides a collection of cosmic voids identified in the Local Universe and available at \url{https://github.com/RosaMalandrino/LocalVoids/}. The catalog is based on the \textsc{Manticore} project\footnote{\url{http://cosmictwin.org/}}, which includes a set of 50 constrained $N$-body simulations representing independent realizations of the local cosmic web. Voids were identified by running \textsc{VIDE} on the halo catalogs of each realization, and their centers were combined in three-dimensional space to locate statistically significant underdense regions. The resulting catalog contains $100$ voids with well-defined centers, shapes, and boundaries, offering a precise description of the local density environment. Produced within a Bayesian framework, this void catalog delivers a rigorous quantification of statistical uncertainties and high-quality data products for applications ranging from galaxy evolution and biasing to cosmological probes and high-energy astrophysics. \smallskip
    \item \textbf{Quaia High-z Void Catalog} \citep{Arsenov_2025} presents a catalog of cosmic voids identified in the distribution of 708,483 quasars from the \textsc{Quaia} data set, covering $24{,}372$~deg$^{2}$ in the redshift range $0.8<z<2.2$. The analysis aims to extend cosmic web mapping to high redshift using quasars as tracers of large-scale structure. Voids were identified with \textsc{REVOLVER}, resulting in 12,842 structures with effective radii up to $\sim250 \ h^{-1} \, \mathrm{Mpc}$. Comparisons with 50 mock realizations show $5–10\%$ agreement in void sizes, mean densities, and profiles, confirming the consistency between observations and simulations. The main products—including the density-field reconstruction, void catalog, and corresponding mock data—will be publicly released upon publication; for now, equivalent products for one mock realization are available at \url{https://zenodo.org/records/16359457}.
\end{itemize}





\end{appendices}


\phantomsection
\addcontentsline{toc}{section}{References}
\bibliography{bibliography}

\end{document}